\documentclass[10pt]{article}
\pdfoutput=1

\usepackage[utf8]{inputenc}
\usepackage[T2A]{fontenc}
\input{epsf}
\usepackage{epsfig}
\usepackage{amssymb}
\usepackage{amsfonts}
\usepackage{amsbsy}
\usepackage[cmtip,all]{xy}
\usepackage{amsmath}
\usepackage{esint}
\usepackage{collectbox}
\usepackage{amscd}
\usepackage{mathrsfs}
\usepackage{amsmath,amsthm}
\usepackage{enumitem}
\usepackage{dsfont}
\usepackage{soul}
\usepackage{comment}
\usepackage{cite} %To compress consecutive citations
\usepackage[most]{tcolorbox}
\usepackage{stmaryrd}
\usepackage{xcolor}
\definecolor{burgundy}{rgb}{0.5, 0.0, 0.13}
\definecolor{olive}{rgb}{0.50, 0.50, 0.0}
\usepackage[linktocpage=true,colorlinks=true,linkcolor=burgundy,citecolor=black!20!blue,urlcolor=violet]{hyperref}
\usepackage{accents}
\usepackage{xfrac}
\usepackage{bm}

\usepackage{tikz}
\usepackage{tikz-cd}
\usetikzlibrary{arrows}
\usetikzlibrary{arrows.meta}
\usetikzlibrary{positioning}
\usetikzlibrary{shapes}
\usetikzlibrary{fit}
\usetikzlibrary{decorations.pathmorphing,decorations.pathreplacing,decorations.markings}
\usetikzlibrary{calc}

\usepackage[font=footnotesize,labelfont=bf]{caption}

\theoremstyle{definition}

\newcommand{\sect}[1]{\setcounter{equation}{0}\section{#1}}
\DeclareMathAlphabet{\mathpzc}{OT1}{pzc}{m}{it}

\def\exp{{\rm exp}}
\def\Ext{{\rm Ext}}
\def\Hom{{\rm Hom}}
\def\I{{\rm i}}
\def\im{\mbox{Im }}
\renewcommand{\Im}{{\rm Im }}
\def\ker{{\rm ker}}
\def\log{{\rm log}}
\def\mod{{\rm mod}}

\def\re{\mbox{Re }}
\renewcommand{\Re}{{\rm Re }}
\def\sdet{{\rm sdet}}
\def\sgn{{\rm sgn\,}}
\def\STr{{\rm STr}}
\def\sym{{\rm Sym}}
\def\tr{{\rm tr\,}}
\def\Tr{{\rm Tr}}
\def\vol{{\rm vol\,}}

\def\la{\langle}
\def\ra{\rangle}
\def\half{\frac{1}{2}}
\def\p{\partial}
\def\prl{\parallel}

\def\rem{$\clubsuit$}

\def\pbar{\bar{\p}}

\def\rem{$\clubsuit$}
\def\bcom{{$\blacktriangleright$}}
\def\ecom{{$\blacktriangleleft$}}

\def\tM{\tilde{M}}
\def\tOmega{\tilde{\Omega}}
\def\tp{\tilde \phi}

\def\sx{{\bf x}}
\def\sy{{\bf y}}
\def\vx{{\vec{x}}}
\def\vy{{\vec{y}}}

\def\horr{{{\smallsmile}\atop{\smallfrown}}}

\def\bDelta{\boldsymbol{\Delta}}

\def\CA{{\cal A}}
\def\CB{{\cal B}}
\def\CC {{\cal C}}
\def\CD {{\cal D}}
\def\CE {{\cal E}}
\def\CF {{\cal F}}
\def\CG {{\cal G}}
\def\CH {{\cal H}}
\def\CI {{\cal I}}
\def\CJ {{\cal J}}
\def\CK {{\cal K}}
\def\CL {{\cal L}}
\def\CM {{\cal M}}
\def\CN {{\cal N}}
\def\CO {{\cal O}}
\def\CP {{\cal P}}
\def\CR {{\cal R}}
\def\CV {{\cal V}}
\def\CW {{\cal W}}
\def\CX {{\cal X}}
\def\CO {{\cal O}}
\def\CZ {{\cal Z}}
\def\CE {{\cal E}}
\def\CG {{\cal G}}
\def\CH {{\cal H}}
\def\CI {{{\cal I}}}
\def\CB {{\cal B}}
\def\CQ {{\cal Q}}
\def\CS {{\cal S}}
\def\CT {{\cal T}}
\def\CU {{\cal U}}
\def\CX {{\cal X}}
\def\CY {{\cal Y}}
\def\CZ{{\cal Z}}

\def\IA{\mathbb{A}}
\def\IB{\mathbb{B}}
\def\IC{\mathbb{C}}
\def\ID{\mathbb{D}}
\def\IE{\mathbb{E}}
\def\IF{\mathbb{Z}}
\def\IG{\mathbb{G}}
\def\IH{\mathbb{H}}
\def\II{\mathbb{I}}
\def\IJ{\mathbb{J}}
\def\IK{\mathbb{K}}
\def\IL{\mathbb{L}}
\def\IM{{\cal E}} % % % % MODIFIED % % %
\def\IN{\mathbb{N}}
\def\IO{\mathbb{O}}
\def\IP{\mathbb{P}}
\def\IQ{\mathbb{Q}} % % % % MODIFIED % % % %
\def\IR{{\mathbb{R}}}
\def\IS{{\mathbb{S}}}
\def\IT{{\mathbb{T}}}
\def\IV{{\mathbb{V}}}
\def\IW{{\mathbb{W}}}
\def\IZ{{\mathbb{Z}}}

\def\fa{\mathfrak{a}}
\def\fb{\mathfrak{b}}
\def\fc{\mathfrak{c}}
\def\fd{\mathfrak{d}}
\def\fD{\mathfrak{D}}
\def\fe{\mathfrak{e}}
\def\ff{\mathfrak{f}}
\def\lieg{\mathfrak{g}}
\def\fg{\mathfrak{g}}
\def\lieh{\mathfrak{h}}
\def\fh{\mathfrak{h}}
\def\fj{\mathfrak{j}}
\def\fk{\mathfrak{k}}
\def\fl{\mathfrak{l}}
\def\fm{\mathfrak{m}}
\def\fn{\mathfrak{n}}
\def\fo{\mathfrak{o}}
\def\fp{\mathfrak{p}}
\def\fq{\mathfrak{q}}
\def\fr{\mathfrak{r}}
\def\fs{\mathfrak{s}}
\def\ft{\mathfrak{t}}
\def\liet{\mathfrak{t}}
\def\fp{\mathfrak{p}}
\def\fq{\mathfrak{q}}
\def\fr{\mathfrak{r}}
\def\fs{\mathfrak{s}}
\def\ft{\mathfrak{t}}
\def\fu{\mathfrak{u}}
\def\fv{\mathfrak{v}}
\def\fw{\mathfrak{w}}
\def\fx{\mathfrak{x}}
\def\fy{\mathfrak{y}}
\def\fz{\mathfrak{z}}
\def\fA{\mathfrak{A}}
\def\fB{\mathfrak{B}}
\def\fC{\mathfrak{C}}
\def\fE{\mathfrak{E}}
\def\fH{\mathfrak{H}}
\def\fG{\mathfrak{G}}
\def\fI{\mathfrak{I}}
\def\fJ{\mathfrak{J}}
\def\fK{\mathfrak{K}}
\def\fL{\mathfrak{L}}
\def\fM{\mathfrak{M}}
\def\fN{\mathfrak{N}}
\def\fR{\mathfrak{R}}
\def\fS{\mathfrak{S}}
\def\fT{\mathfrak{T}}
\def\fP{\mathfrak{P}}
\def\fV{\mathfrak{V}}
\def\fQ{\mathfrak{Q}}
\def\fX{\mathfrak{X}}
\def\fY{\mathfrak{Y}}

\def\bPhi{{\boldsymbol{\Phi}}}
\def\bSigma{{\boldsymbol{\Sigma}}}
\def\bPsi{{\boldsymbol{\Psi}}}
\def\bphi{{\boldsymbol{\phi}}}
\def\bpsi{{\boldsymbol{\psi}}}
\def\bvphi{{\boldsymbol{\varphi}}}

\def\myA{{\mathsf{A}}}
\def\myB{{\mathsf{B}}}
\def\myC{{\mathsf{C}}}
\def\myF{{\mathsf{F}}}
\def\myG{{\mathsf{G}}}

\def\lm{\limits}
\def\be{\begin{eqnarray}}
\def\ee{\end{eqnarray}}
\def\nn{\nonumber}
\def\nb{\nabla}

\hoffset= - 1.0in

\numberwithin{equation}{section}
\tcbset{boxrule=0.6pt,colback=white!97!green}
\DeclareSymbolFont{bbsymbol}{U}{bbold}{m}{n}
\DeclareMathSymbol{\bbzero}{\mathbin}{bbsymbol}{"30}
\DeclareMathSymbol{\bbone}{\mathbin}{bbsymbol}{"31}
\DeclareMathSymbol{\bbtwo}{\mathbin}{bbsymbol}{"32}
\DeclareMathSymbol{\bbthree}{\mathbin}{bbsymbol}{"33}
\DeclareMathSymbol{\bbfour}{\mathbin}{bbsymbol}{"34}
\DeclareMathSymbol{\bbfive}{\mathbin}{bbsymbol}{"35}
\DeclareMathSymbol{\bbsix}{\mathbin}{bbsymbol}{"36}
\DeclareMathSymbol{\bbseven}{\mathbin}{bbsymbol}{"37}
\DeclareMathSymbol{\bbeight}{\mathbin}{bbsymbol}{"38}
\DeclareMathSymbol{\bbnine}{\mathbin}{bbsymbol}{"39}

\newcommand\Kappa{\mathrm{K}}
\def\myblue{white!40!blue}
\def\mygreen{black!40!green}
\def\myred{black!40!red}
\def\mygray{gray}

\definecolor{palette1}{rgb}{0.603922, 0.466667, 0.811765}
\definecolor{palette2}{rgb}{0.329412, 0.219608, 0.517647}
\definecolor{palette3}{rgb}{0.0156863, 0.282353, 0.333333}
\definecolor{palette4}{rgb}{0.631373, 0.211765, 0.439216}
\definecolor{palette5}{rgb}{0.92549, 0.254902, 0.462745}
\definecolor{palette6}{rgb}{1., 0.643137, 0.368627}
\definecolor{palette7}{rgb}{0.313725, 0.45098, 0.85098}

\definecolor{Xmagenta}{HTML}{D60270}
\definecolor{Xpurple}{HTML}{9B4F96}
\definecolor{Xblue}{HTML}{0038A8}

\newcommand\sqbox[1]{{
		\setbox0=\hbox{\mbox{$\Box$}}
		\setbox1=\hbox{\mbox{\raisebox{0.35ex}{\small #1}}}
		\mbox{\raisebox{-0.2ex}{\rlap{\hbox to \wd0{\hss{\box1}\hss}}\box0}}
}}
\newcommand\ssqbox[1]{{
		\setbox0=\hbox{\mbox{$\scriptstyle\Box$}}
		\setbox1=\hbox{\mbox{\raisebox{0.35ex}{\tiny #1}}}
		\mbox{\raisebox{-0.2ex}{\rlap{\hbox to \wd0{\hss{\box1}\hss}}\box0}}
}}

\begin{document}

\hfill MIPT/TH-02/26

\hfill ITEP/TH-02/26

\hfill IITP/TH-02/26

\vskip 1.5in
%\vskip 1cm
\begin{center}
	
    {\bf\Large Weyl Mutations in Quiver Yangians}

	\vskip 0.2in
	\renewcommand{\thefootnote}{\fnsymbol{footnote}}
	{Dmitry Galakhov$^{1,2,3,}$\footnote[2]{e-mail: d.galakhov.pion@gmail.com, galakhov@itep.ru},
		Alexei Gavshin$^{1,2,}$\footnote[3]{e-mail: gavshin.an@phystech.edu} and
		Alexei Morozov$^{1,2,3,}$\footnote[4]{e-mail: morozov@itep.ru}}
	\vskip 0.2in
	\renewcommand{\thefootnote}{\roman{footnote}}
	{\small{
			\textit{$^1$MIPT, 141701, Dolgoprudny, Russia}
			\vskip 0 cm
			\textit{$^2$NRC ``Kurchatov Institute'', 123182, Moscow, Russia}
			\vskip 0 cm
			\textit{$^3$IITP RAS, 127051, Moscow, Russia}
			\vskip 0 cm
			\textit{$^4$ITEP, Moscow, Russia}
	}}
\end{center}

\vskip 0.2in
\baselineskip 16pt

\centerline{ABSTRACT}

\bigskip

{\footnotesize
	The problem of solving non-linear equations would be considerably simplified
	by a possibility to convert known solutions into the new ones.
	This could seem an element of art, but in the context of ADHM-like equations describing quiver varieties
	there is a systematic approach.
    In this note we study moduli spaces and dualities of quiver gauge theories associated to effective dynamics of D-branes compactified on Calabi-Yau resolutions.
    We concentrate on a subfamily of quivers $\mathfrak{Q}_{\mathfrak{g}}$ covering Dynkin diagrams for simple Lie algebras $\mathfrak{g}$, where the respective BPS algebra is expected to be the Yangian algebra $Y(\mathfrak{g})$.
    For Yangians labeled by quivers their representations are described
    by solutions of ADHM-like equations.
    As quivers substitute Dynkin diagrams a generalization of the Weyl group $\mathcal{W}_{\mathfrak{g}}$
    acts on the ADHM solutions.
    Here we work with the case $\mathfrak{g}=\mathfrak{sl}_{n+1}$ and treat this group as a group of electro-magnetic Seiberg-like dualities (we call them Weyl mutations) on the respective quiver gauge theories.
    We lift it to the case of higher representations associated to rectangular Young diagrams.
    An action of Weyl mutations on the BPS Yangian algebra is also discussed.

}

\bigskip

\bigskip

\tableofcontents

\bigskip

\section{Introduction}

Among the most successful models mimicking strongly coupled QFTs and admitting analytic approaches one might select families embracing supersymmetry.
For such theories one of the basic questions about low energy effective excitation spectra could be partially resolved for BPS states preserving a part of supersymmetry.
Moreover under these circumstances BPS spectra exhibit a piece-wise dependence on parameters of the theory in question including vacuum moduli spaces etc.
The parameter space is divided into \emph{stability} chambers with constant or smoothly varying spectra.
Chambers are separated by \emph{marginal stability walls}.
Collectively processes of BPS spectrum transformations on and across marginal stability walls are called \emph{wall-crossing phenomena} \cite{Galakhov:2024foa,Andriyash:2010qv,Cecotti:2009uf,Aganagic:2009kf,Kontsevich:2008fj,FelixKlein,Aganagic:2010qr,Sulkowski:2009rw,Manschot:2010qz,Pioline:2013wta,Kontsevich:2013rda}.

Wall-crossing phenomena attract a lot of attention of physics community as a possible sandbox test environment mimicking phase transitions in full-fledged QFTs without supersymmetry.
Yet for this toy model supersymmetric ``critical phenomena'' in many cases it turns out to be possible to calculate a critical coupling constant exactly and analytically.

In this paper we consider families of quiver supersymmetric $\CN=4$ quantum mechanics (SQM) describing effective dynamics of D-branes on divisors of toric Calabi-Yau three-folds \cite{Galakhov:2021xum,Noshita:2021dgj,Nishinaka:2013mba}.
Quivers of this type with corresponding modifications for the superpotential were exploited in \cite{Bykov:2019cst,Yang:2024ubh} to propose a geometric description to higher spin rational $\fs\fl_2$ R-matrices.
More specifically we concentrate on quivers of $\myA_n$-family analogous to Dynkin diagrams of $\fs\fl_{n+1}$ Lie algebras generalizing Nakajima quiver varieties of the $\myA_n$-type \cite{2009arXiv0905.0686G}.
A canonical parameter space for such a model is a space of quiver stability moduli realized physically as FI parameters $\zeta_a$ associated to unitary gauge groups ``sitting'' in the quiver nodes.
A behavior of BPS spectra ($\frac{1}{2}$-supersymmetric ground states for SQM) is rather well-studied in a so called \emph{cyclic chamber} where all $\zeta_a>0$.
For the cyclic chamber one is able to construct a BPS algebra and its particular representation following 3 steps:
\begin{enumerate}[label=\alph*)]
	\item We consider a toric Calabi-Yau three-fold quiver with an $\Omega$-background turned on.
	Higgs branch classical vacua (equivariant fixed points) in SQM are in a 1-to-1 correspondence with molten crystals (for a review of the molten crystal construction see e.g.\cite{Yamazaki:2010fz,Li:2023zub,Yamazaki:2022cdg,Bao:2024ygr} and references therein), probably truncated \cite{Galakhov:2021xum} depending on a choice of extra terms in the superpotential.
	\item Matrix elements of raising and lowering BPS algebra operators shifting quiver dimensions $d_a\to d_a\pm 1$ are calculated mathematically via an equivariant Hecke correspondence \cite{Galakhov:2020vyb,NakajimaKacMoody,Rapcak:2018nsl,Rapcak:2020ueh} or physically via tunneling amplitudes \cite{Galakhov:2025fhd}.
	\item Calculated in the previous step matrix elements satisfy defining relations of an abstract algebra $\mathscr{A}$ in a specific representation $R$.
	We call $\mathscr{A}$ the respective BPS algebra, in our particular case this is a Yangian algebra $Y(\fs\fl_{n+1})$ \cite{Gavshin:2025akn}.
	For representation $R$ we obtain a canonically fixed basis where basic vectors are in a 1-to-1 correspondence with the classical vacua.
\end{enumerate}

So the main aim of the present paper is to consider a behavior of the above BPS algebra system on the whole moduli space beyond the cyclic chamber across the marginal stability walls.

Let us accumulate here our a priori expectations for this behavior.
Alternative estimates from the Coulomb branch \cite{Galakhov:2018lta,Kontsevich:2010px} predict for the Borel positive part of $\mathscr{A}$ another abstract algebra known as a cohomological Hall algebra.
The cohomological Hall algebra is independent of stability parameters by construction, so there is a good chance that $\mathscr{A}$ is a \emph{wall-crossing invariant}.
Therefore we would expect that wall-crossing would act on $\mathscr{A}$ by some automorphisms.
Quite a useful technique allowing one to extend applicability of a BPS spectrum construction in the cyclic chamber is a \emph{mutation}, or Seiberg duality \cite{Seiberg:1994pq,Intriligator:1995au}.
Seiberg duality (or Seiberg-Kutasov duality \cite{Kutasov:1995ve,Kutasov:1995np,Kutasov:1995ss} in more general cases) is duality of an electro-magnetic type connecting different chambers of moduli spaces of different quiver (say, $\fQ$ and $\check \fQ$) theories.
In the case when $\check \fQ=\fQ$ Seiberg duality maps the cyclic region to outer regions.
As we consider a theory family where quiver $\fQ=\myA_n$ Dynkin diagram, and Seiberg duality should act by automorphisms on $\mathscr{A}$, it is natural to expect that this is exactly our case $\check\fQ=\fQ=\myA_n$.
Yet mutations imply a rather specific action \cite{Berenstein:2002fi,Alim:2011kw,Benini:2014mia} on stability parameters $\zeta_a$ and dimensions $d_a$ reminiscent of the \emph{Weyl group} action on root and weight lattices of $\fs\fl_{n+1}$ correspondingly.

\begin{table}[ht!]
	\centering
	$\begin{array}{|c|c|c|c|c|}
		\hline
		\mbox{No.} & \mbox{Lie algebra} & \begin{array}{c}
			\mbox{Weyl}\\
			\mbox{group}
		\end{array} & \mbox{Quiver} & \begin{array}{c}
			\mbox{Weyl}\\
			\mbox{mutation}
		\end{array}\\
		\hline
		1 & \mbox{Dynkin diagram} & \mbox{\bf\color{black!40!green} Inv.}& \mbox{Unframed quiver} & \mbox{\bf\color{black!40!green} Inv.} \\
		\hline
		2 & \mbox{Roots }\alpha_i & \begin{array}{c}
			\mbox{\bf\color{black!40!red} Cov.}\\
			\eqref{Weyl_root}
		\end{array} & \mbox{FI (stability) parameters }\zeta_a &  \begin{array}{c}
			\mbox{\bf\color{black!40!red} Cov.}\\
			\eqref{non_Nak_du_sta}=\eqref{Weyl_root}
		\end{array}\\
		\hline
		3 & \mbox{Fundamental weights }\omega_i & \begin{array}{c}
			\mbox{\bf\color{black!40!red} Cov.}\\
			\eqref{Weyl_weight}
		\end{array} & \mbox{Dimensions }d_a & \begin{array}{c}
			\mbox{\bf\color{black!40!red} Cov.}\\
			\eqref{non_Nak_du_dim}\approx\eqref{Weyl_weight}
		\end{array}\\
		\hline
		4 & \begin{array}{c}
			\mbox{Representation}\\
			(\mbox{Young diagram }\Upsilon)
		\end{array} & \mbox{\bf\color{black!40!green} Inv.} & \begin{array}{c}
			\mbox{Framing+superpotential}\\
			(\mbox{rectangular }\Upsilon\mbox{ only})
		\end{array} & \mbox{\bf\color{black!40!green} Inv.} \\
		\hline
		5(a) & \begin{array}{c}
			\mbox{Rep vector as an element}\\
			\mbox{of rep vector space }v\in V_\Upsilon
		\end{array} & \mbox{\bf\color{black!40!green} Inv.} & \begin{array}{c}
			\mbox{Atomic plot in}\\
			\mbox{{\bf\color{burgundy}dual} theory, cyclic chamber}\\
			(=\mbox{Young diagram for }Y(\widehat{\fg\fl}_1))
		\end{array} & \mbox{\bf\color{black!40!green} Inv.}\\
		\hline
		5(b) & \begin{array}{c}
			\mbox{Rep vector as the highest}\\
			\mbox{weight module member}\\
			|v\rangle=f_af_bf_c\ldots|0\rangle
		\end{array}& \mbox{\bf\color{black!40!red} Cov.}& \begin{array}{c}
			\mbox{Atomic plot}\\
			\mbox{{\bf\color{burgundy}beyond} cyclic chamber}
		\end{array} & \mbox{\bf\color{black!40!red} Cov.}\\
		\hline
		6 & \begin{array}{c}
			\mbox{Chevalley basis of generators}\\
			(e_i,f_i,h_i)
		\end{array} & \begin{array}{c}
			\mbox{\bf\color{black!40!red} Cov.}\\ \eqref{Weyl_Chev} \end{array}& \begin{array}{c}
			\mbox{BPS algebra generators}\\
			\mbox{= Hecke operators}\\
			\mbox{= Yangian generators}
		\end{array} & \mbox{\bf\color{black!40!red} Cov.} \\
		\hline
		7 & \mbox{Character, Schur function} & \mbox{\bf\color{black!40!green} Inv.}& \mbox{DT/GW gen. function} & \mbox{\bf\color{black!40!green} Inv.} \\
		\hline
	\end{array}$
	\caption{Comparing actions of the Weyl group and of Weyl mutations.}\label{thetable}
\end{table}

Indeed in this note we will argue that the Seiberg-like dualities for the $\myA_n$-type quiver family form a group (of what we call \emph{Weyl mutations}) isomorphic to the Weyl group of $\fs\fl_{n+1}$ and describe its action on various elements of the construction of $\mathscr{A}=Y(\fs\fl_{n+1})$.
Similarities of actions of the Weyl group on $\fs\fl_{n+1}$ Lie algebras and their representations and of Weyl mutations on respective quiver BPS algebras and their representations (classical vacua) are summarized in Table \ref{thetable} (here {\bf\color{black!40!green} Inv.} implies that the characteristic remains invariant with respect to an action, and {\bf\color{black!40!red} Cov.} marks characteristics transforming covariantly).

Let us comment here on the structure of Table \ref{thetable}.

A framed quiver with a superpotential carries information about both the algebra and a representation.
We could separate approximately these two pieces into the unframed quiver defining underlying Dynkin diagram and framing together with superpotential extra terms fixing Dynkin labels of the irrep highest/lowest weight.
As both the objects are preserved by the Weyl group we put them in line 1 and 4 in the table.

Similarities of transformations for simple roots and FI parameters, as well as for fundamental weights and quiver dimensions have been noted earlier.
Similarly to the defining quadratic structure on roots and wights $(\omega_i,\alpha_j^{\vee})=\delta_{ij}$ remaining invariant with respect to the Weyl group for FI parameters and dimensions we could select stability slopes $\mu=\sum\lm_a d_a\zeta_a$ \cite{donaldson1983new,king1994moduli} characterizing fixed points.
So we put them respectively in lines 2 and 3.

Quiver dimensions define ranks of gauge groups $U(d_a)$ associated to nodes.
The very representations of the $\fs\fl_{n+1}$ Lie algebras admit two constructions: via Young projectors from the tensor powers of the $\Box$ representation or as a truncated Verma module of the lowest/highest weight vector.
As the Weyl group modifies the basis in the weight lattice this affects the notion of the lowest weight vector, so a representation of a vector as a word of raising/lowering Chevalley generators acting on the lowest/highest weight vector is modified as well.
In line 5(b) we associate this modification with a demonstrative modification of the fixed points depicted with atomic structure plots.
Moreover quiver dimensions defining numbers of elements of these plots and modifying by mutations count as well how many times distinct Chevalley generators are encountered in Verma words \eqref{weight}.

On the other hand weights as abstract vectors of the weight space are not affected by basis choice variations, similarly each fixed point outside the cyclic chamber is related to a fixed point in the cyclic chamber (labeled uniquely by a boxed 3d Young diagram) with a sequence of mutations.
Acting on a fixed point by a new mutation modifies the sequence, yet it does not modify the very pre-image in the cyclic chamber.
So we put these two entities in line 5(a).

It is well-known that the action of the Weyl group on the algebra generators, say, in the Chevalley basis, is performed by exponential maps \cite{fulton2013representation, reynoso2024standard}.
For quiver Yangians however we would find certain technical difficulties in a definition of the algebra outside the cyclic chamber.
Therefore following a strategy proposed in \cite{Galakhov:2024foa} we define quiver Yangian raising and lowering generators $e_a^{(k)}$ and $f_a^{(k)}$ associated to quiver node $a$ as Hecke modifications yet for pairs of fixed points whose pre-image 3d Young diagrams in the cyclic chamber differ by adding/subtracting a box correspondingly.
For generators defined in such a way we can no longer say that they shift quiver dimensions $\check d_a\to \check d_a\pm 1$ in the chosen stability chamber.
However we can transform them to a new basis of generators $\check e_a^{(k)}$ and $\check f_a^{(k)}$ acting on dimensions by unit shifts again by exponential maps.
So we put generators of the Lie and quiver Yangian algebras on similar footing to line 6.

Finally, a DT/GW invariant generating function for fixed points could be derived from ordinary Schur character functions of the associated representation and is not affected by the Weyl group.
It is put to line 7.

The paper is organized as follows.
In Sec.~\ref{sec:Weyl} we review the action of the Weyl group on Lie algebras of the $\myA$-family, and also on Nakajima quiver varieties of the $\myA$-family.
In Sec.~\ref{sec:quiv_Yang} we review a construction of quiver Yangian representations in terms of equivariant fixed points on respective quiver varieties.
Sec.~\ref{sec:Mutations} is devoted to the construction extension of Weyl mutations to quiver varieties in question and corresponding fixed points.
In Sec.~\ref{sec:MutYang} we consider an action of Weyl mutations on quiver Yangians $Y(\fs\fl_{n+1})$.
Finally, in Sec.~\ref{sec:Towards_Lie} we propose some ways to generalize the proposed machinery to Dynkin diagrams beyond the $\myA$ type.

%%%%%%%%%%%%%%%%%%%%%%%%%%%%%%%%%%%%%%%%%%%%%%%%%%%%%%%%%%%%%%%%%%%%%%%%%%%%%%%
%%%%%%%%%%%%%%%%%%%%%%%%%%%%%%%%%%%%%%%%%%%%%%%%%%%%%%%%%%%%%%%%%%%%%%%%%%%%%%%
%%%%%%%%%%%%%%%%%%%%%%%%%%%%%%%%%%%%%%%%%%%%%%%%%%%%%%%%%%%%%%%%%%%%%%%%%%%%%%%

\section{Weyl group}\label{sec:Weyl}

\subsection{Weyl group action on \texorpdfstring{$\myA_n$}{An} root and weight lattices}

Let us start with a short reminder of a few facts from the theory of Lie algebras \cite{humphreys2012introduction,DiFrancesco:1997nk}. Suppose, $\Delta$ is a root system of a semi-simple Lie algebra $\fg$. Then the Weyl group acting on a root lattice is defined as an orthogonal subgroup generated by all reflections $s_{\alpha}$ with respect to planes perpendicular to all roots $\alpha$:
\begin{equation}\label{Weyl_reflection}
	s_{\alpha}(v)=v-2\frac{(v,\alpha)}{(\alpha,\alpha)}\alpha\,.
\end{equation}
Therefore, the Weyl group represents an ambiguity in a choice of simple roots in a given root system $\Delta$.

Here we would like to concentrate on Lie algebras with Dynkin diagrams of $\myA_n$-type having the respective Cartan matrix and a nice embedding of the simple roots into $\IR^n$ with unit orthonormal vectors $e_i$:
\begin{equation}\label{sl(n)_data}
	\begin{aligned}
		&\begin{array}{c}
			\begin{tikzpicture}
				\draw[thick] (0,0) -- (3,0);
				\draw[thick, dashed] (3,0) -- (5,0);
				\draw[fill=\myblue] (0,0) circle (0.1) (1,0) circle (0.1) (2,0) circle (0.1) (3,0) circle (0.1) (5,0) circle (0.1);
				\draw (-0.1,0.1) to[out=90,in=180] (0.1,0.3) -- (2.3,0.3) to[out=0,in=270] (2.5,0.5) to[out=270,in=180] (2.7,0.3) -- (4.9,0.3) to[out=0,in=90] (5.1,0.1);
				\node[above] at (2.5,0.5) {\footnotesize $n$ nodes};
			\end{tikzpicture}
		\end{array}\\
		&\CA_{ij}=2\delta_{i,j}-\delta_{i+1,j}-\delta_{i,j+1},\quad 1\leq i,j\leq n\,,\\
		&\alpha_i=e_i-e_{i+1},\quad 1\leq i,j\leq n\,.
	\end{aligned}
\end{equation}
Apparently, it is sufficient to consider only the action on simple roots:
\begin{equation}\label{Weyl_root}
	s_i(\alpha_i)=-\alpha_i,\quad s_i(\alpha_{i\pm 1})=\alpha_{i\pm 1}+\alpha_i,\quad s_i(\alpha_j)=\alpha_j,\;\mbox{for }|i-j|>1\,.
\end{equation}
Respectively, this action induces the inverse transform on fundamental weights:
\begin{equation}\label{Weyl_weight}
	s_i(\omega_i)=\omega_{i-1}+\omega_{i+1}-\omega_i,\quad s_i(\omega_j)=\omega_j,\;\mbox{for }j\neq i\,.
\end{equation}
For $A_n$-root lattices the Weyl group is simply isomorphic to the permutation group $S_{n + 1}$:
\begin{equation}\label{sl(n)_Weyl_group_relations}
	s_i^2=1,\quad \left(s_is_{i+1}\right)^3=1,\quad s_is_j=s_js_i,\;\mbox{for }|i-j|>1\,.
\end{equation}

A semi-simple Lie algebra can be represented using generators $(e_i,f_i,h_i)$, $1\leq i\leq n$ in the Chevalley basis satisfying Chevalley-Serre relations:
\begin{equation}\label{Lie_algebra_relations}
	\left[h_i,h_j\right]=0,\quad [h_i,e_j]=\CA_{ij}e_j,\quad [h_i,f_j]=-\CA_{ij}f_j,\quad [e_i,f_j]=\delta_{ij}h_i,\quad {\rm ad}_{e_i}^{1-\CA_{ij}}e_j=0,\quad {\rm ad}_{f_i}^{1-\CA_{ij}}f_j=0\,.
\end{equation}
It is a natural desire to extend the Weyl group action from the root lattice preserved by it to the algebra automorphisms.
For example, one could verify easily that the transformations like:
\begin{equation}
	r(X) = \exp\bigl({\rm ad}(X)\bigr)\,, \quad X\in \fg\,,
\end{equation}
are automorphisms of Lie algebra $\fg$, since operators ${\rm ad}(X)$ are nilpotent. However, in this form they do not preserve the chosen Chevalley basis nor do they reflect roots according to the Weyl group action.

There is a canonical solution to the problem above \cite{fulton2013representation, reynoso2024standard}. 
The Weyl reflections $s_i$ induce Lie algebra automorphisms by the operators:
\begin{equation}\label{Weyl_Tits}
	\tau_{i} = e^{{\rm ad}(e_{i})}e^{-{\rm ad}(f_{i})}e^{{\rm ad}(e_{i})}\,.
\end{equation}
In our case (cf. App. \ref{App: Weyl Automorphisms} and \cite{2019SIGMA..15..020K}), the automorphisms take the following form:
\begin{equation}\label{Weyl_Chev}
	\begin{aligned}
		&\tau_i(e_i)= -f_i,\quad \tau_i(f_i) = -e_i,\quad \tau_i(h_{i})= -h_i\,,\\
		&\tau_i(e_{i\pm 1})=\left[e_i,e_{i\pm 1}\right],\quad \tau_i(f_{i\pm 1})=-\left[f_i,f_{i\pm 1}\right],\quad \tau_i(h_{i\pm 1})=h_i+h_{i\pm 1}\,,\\
		&\tau_i(e_j)=e_j,\quad \tau_i(f_j)=f_j,\quad \tau_i(h_{j})=h_j,\quad |i-j|>1\,.
	\end{aligned}
\end{equation}

%%%%%%%%%%%%%%%%%%%%%%%%%%%%%%%%%%%%%%%%%%%%%%%%%%%%%%%%%%%%%%%%%%%%%%%%%%%%%%%
%%%%%%%%%%%%%%%%%%%%%%%%%%%%%%%%%%%%%%%%%%%%%%%%%%%%%%%%%%%%%%%%%%%%%%%%%%%%%%%
%%%%%%%%%%%%%%%%%%%%%%%%%%%%%%%%%%%%%%%%%%%%%%%%%%%%%%%%%%%%%%%%%%%%%%%%%%%%%%%

\subsection{Weyl group action on \texorpdfstring{$\myA_n$}{An} Nakajima quiver varieties}\label{sec:Nak_quiv_Weyl}

Nakajima quiver varieties characterize self-dual connections on ALE spaces \cite{2009arXiv0905.0686G,kronheimer1990yang,nakajima1994instantons}.
Formally speaking \cite{Douglas:1996sw,Hanany:2003hp} instanton collective coordinates that could be mimicked also by strings stretched between D-branes establish effective dynamics analogous to a quiver gauge theory.
The particular case of $\myA_n$-type Nakajima quiver varieties are associated with the following framed chain quiver reminiscent of the $\myA_n$ Dynkin diagram and a superpotential function:
\begin{equation}\label{ALE_quiv}
	\begin{aligned}
	&\begin{array}{c}
		\begin{tikzpicture}
			\begin{scope}
				\draw[postaction=decorate, decoration={markings, mark= at position 0.7 with {\arrow{stealth}}}] (0,0) to[out=20,in=160] node[pos=0.5,above] {$\scriptstyle A_1$} (1.5,0);
				\draw[postaction=decorate, decoration={markings, mark= at position 0.7 with {\arrow{stealth}}}] (1.5,0) to[out=200,in=340] node[pos=0.5,below] {$\scriptstyle B_1$} (0,0);
				\draw[postaction=decorate, decoration={markings, mark= at position 0.8 with {\arrow{stealth}}}] (0,0) to[out=60,in=0] (0,0.6) to[out=180,in=120] (0,0);
				\node[above] at (0,0.6) {$\scriptstyle C_1$};
				\draw[postaction=decorate, decoration={markings, mark= at position 0.7 with {\arrow{stealth}}}] (0,-1.2) to[out=100,in=260] node[pos=0.3, left] {$\scriptstyle R_1$} (0,0);
				\draw[postaction=decorate, decoration={markings, mark= at position 0.7 with {\arrow{stealth}}}] (0,0) to[out=280,in=80] node[pos=0.7, right] {$\scriptstyle S_1$} (0,-1.2);
				\begin{scope}[shift={(0,-1.2)}]
					\draw[fill=burgundy] (-0.08,-0.08) -- (-0.08,0.08) -- (0.08,0.08) -- (0.08,-0.08) -- cycle;
				\end{scope}
			\end{scope}
			\begin{scope}[shift={(1.5,0)}]
				\draw[postaction=decorate, decoration={markings, mark= at position 0.7 with {\arrow{stealth}}}] (0,0) to[out=20,in=160] node[pos=0.5,above] {$\scriptstyle A_2$} (1.5,0);
				\draw[postaction=decorate, decoration={markings, mark= at position 0.7 with {\arrow{stealth}}}] (1.5,0) to[out=200,in=340] node[pos=0.5,below] {$\scriptstyle B_2$} (0,0);
				\draw[postaction=decorate, decoration={markings, mark= at position 0.8 with {\arrow{stealth}}}] (0,0) to[out=60,in=0] (0,0.6) to[out=180,in=120] (0,0);
				\node[above] at (0,0.6) {$\scriptstyle C_2$};
				\draw[postaction=decorate, decoration={markings, mark= at position 0.7 with {\arrow{stealth}}}] (0,-1.2) to[out=100,in=260] node[pos=0.3, left] {$\scriptstyle R_2$} (0,0);
				\draw[postaction=decorate, decoration={markings, mark= at position 0.7 with {\arrow{stealth}}}] (0,0) to[out=280,in=80] node[pos=0.7, right] {$\scriptstyle S_2$} (0,-1.2);
				\begin{scope}[shift={(0,-1.2)}]
					\draw[fill=burgundy] (-0.08,-0.08) -- (-0.08,0.08) -- (0.08,0.08) -- (0.08,-0.08) -- cycle;
				\end{scope}
			\end{scope}
			\begin{scope}[shift={(3,0)}]
				\draw[postaction=decorate, decoration={markings, mark= at position 0.8 with {\arrow{stealth}}}] (0,0) to[out=60,in=0] (0,0.6) to[out=180,in=120] (0,0);
				\node[above] at (0,0.6) {$\scriptstyle C_3$};
				\draw[postaction=decorate, decoration={markings, mark= at position 0.7 with {\arrow{stealth}}}] (0,-1.2) to[out=100,in=260] node[pos=0.3, left] {$\scriptstyle R_3$} (0,0);
				\draw[postaction=decorate, decoration={markings, mark= at position 0.7 with {\arrow{stealth}}}] (0,0) to[out=280,in=80] node[pos=0.7, right] {$\scriptstyle S_3$} (0,-1.2);
				\begin{scope}[shift={(0,-1.2)}]
					\draw[fill=burgundy] (-0.08,-0.08) -- (-0.08,0.08) -- (0.08,0.08) -- (0.08,-0.08) -- cycle;
				\end{scope}
			\end{scope}
			\begin{scope}[shift={(4.5,0)}]
				\draw[postaction=decorate, decoration={markings, mark= at position 0.8 with {\arrow{stealth}}}] (0,0) to[out=60,in=0] (0,0.6) to[out=180,in=120] (0,0);
				\node[above] at (0,0.6) {$\scriptstyle C_n$};
				\draw[postaction=decorate, decoration={markings, mark= at position 0.7 with {\arrow{stealth}}}] (0,-1.2) to[out=100,in=260] node[pos=0.3, left] {$\scriptstyle R_n$} (0,0);
				\draw[postaction=decorate, decoration={markings, mark= at position 0.7 with {\arrow{stealth}}}] (0,0) to[out=280,in=80] node[pos=0.7, right] {$\scriptstyle S_n$} (0,-1.2);
				\begin{scope}[shift={(0,-1.2)}]
					\draw[fill=burgundy] (-0.08,-0.08) -- (-0.08,0.08) -- (0.08,0.08) -- (0.08,-0.08) -- cycle;
				\end{scope}
			\end{scope}
			\draw[fill=\myblue] (0,0) circle (0.08) (1.5,0) circle (0.08) (3,0) circle (0.08) (4.5,0) circle (0.08);
			\draw[fill=black] (3.5,0) circle (0.03) (3.75,0) circle (0.03) (4,0) circle (0.03);
		\end{tikzpicture}
	\end{array},\; W=\Tr\left(\sum\lm_{a=1}^{n}C_a(A_aB_a-A_{a-1}B_{a-1}+R_aS_a)\right)\,,
	\end{aligned}
\end{equation}
here and in what follows fields $A_0=B_0=A_n=B_n=0$.

To vector spaces $V_a$ associated with gauge nodes and $W_a$ associated to framing nodes one assigns dimensions:
\begin{equation}
	{\rm dim}\,V_a=d_a,\quad {\rm dim}\,W_a=w_a,\quad 1\leq a\leq n\,.
\end{equation}

Vacua of this theory correspond to solutions of ADHM equations on an ALE space in the following way.
Potential terms of this theory consist of canonical D-terms and F-terms.
F-terms are generated by derivatives of the superpotential with respect to fields, and it is clear that assigning to all adjoint fields $C_a$ a zero vacuum expectation value would deliver a minimum.
Eventually D-terms and remaining superpotential derivatives with respect to $C_a$'s are exactly real and complex moment map ADHM equations:
\begin{equation}\label{ALE_ADHM}
	\begin{aligned}
		&B_aB_a^{\dagger}-A_a^{\dagger}A_a+A_{a-1}A_{a-1}^{\dagger}-B_{a-1}^{\dagger}B_{a-1}+R_aR_a^{\dagger}-S_{a}^{\dagger}S_a=\zeta_a\bbone_{V_a},\quad 1\leq a\leq n\,,\\
		&A_aB_a-A_{a-1}B_{a-1}+R_aS_a=\bbzero_{V_a}\,,
	\end{aligned}
\end{equation}
where $\zeta_a$ are canonical Fayet–Iliopoulos couplings (stability parameters) adding a non-commutative deformation to self-dual connections \cite{Nekrasov:1998ss,Szabo:2001kg,Hamanaka:2013vca}.

To work with quiver varieties by applying methods of algebraic geometry one eliminates canonically \cite{king1994moduli,NakajimaKacMoody} the first non-holomorphic equation in \eqref{ALE_ADHM} following the Narashiman-Shishadri-Hitchin-Kobayashi correspondence \cite{donaldson1983new}.
Instead one has to consider solely stable quiver representations and complexify the gauge group:
\begin{equation}
	G_{\IR}(\fQ):=\prod\lm_a U(d_a)\;\to \;\prod\lm_a GL(d_a,\IC)=:G_{\IC}(\fQ)\,.
\end{equation}

There is an action of the Weyl group on the Nakajima quiver varieties descending naturally from the action on the respective ALE space \cite{kronheimer1989construction,kronheimer1990yang,lusztig2000quiver,Maffei,NakajimaKacMoody,nakajima2003reflection}.
Here we follow an explicit action of the Weyl group presented in \cite{lusztig2000quiver} on quiver dimension and morphisms.
A reflection element of the Weyl group $s_a$ acts on a particular quiver node $a$ and maps a quiver of the $\myA_n$-type again to a quiver of the same type \eqref{ALE_quiv}.
Yet a stable representation with morphisms $(A_i,B_i,C_i,R_i,S_i)$, spaces $(V_i,W_i)$ and stability parameters $\zeta_i$ is mapped to a new representation with new parameters we denote by ``checks''.
This map is subjected to the following rules for generator $s_a$:
\begin{enumerate}
	\item Stability parameters are acted upon by the Weyl reflection:
	\begin{equation}\label{Nak_du_sta}
		\check \zeta_a=-\zeta_a,\quad \check\zeta_{a\pm 1}=\zeta_{a\pm 1}+\zeta_a,\quad \check\zeta_b=\zeta_b,\;\mbox{for }|a-b|>1\,.
	\end{equation}
	\item Dimensions behave as fundamental weights:
	\begin{equation}\label{Nak_du_dim}
		\check d_a=w_a+d_{a+1}+d_{a-1}-d_a,\quad \check d_b=d_b,\;\mbox{for }b\neq a,\quad \check w_b=w_b\mbox{ for all nodes}\,.
	\end{equation}
	\item Morphisms not adjacent to node $a$ are unmodified:
	\begin{equation}
		\check R_b=R_b,\;\check S_b=S_b, \;\mbox{for }b\neq a;\quad \check A_b=A_b,\;\check B_b=B_b,\;\mbox{for }b\neq a,\;b\neq a-1\,.
	\end{equation}
	\item Maps $\alpha$ and $\beta$ constructed as:
	\begin{equation}
		\alpha=\left(\begin{array}{ccc}
			\check S_a & \check B_{a-1} & \check A_{a}
		\end{array}\right)^T,\quad \beta=\left(\begin{array}{ccc}
		R_a & -A_{a-1} & B_{a}
		\end{array}\right)\,,
\	\end{equation}
	constitute a short exact sequence:
	\begin{equation}\label{exact_du}
		\begin{array}{c}
			\begin{tikzpicture}
				\node (A) at (-4,0) {$0$};
				\node (B) at (-3,0) {$\check V_a$};
				\node (C) at (0,0) {$W_a\oplus V_{a-1}\oplus V_{a+1}$};
				\node (D) at (3,0) {$V_a$};
				\node (E) at (4,0) {$0$};
				\path (A) edge[->] (B) (B) edge[->] node[above] {$\scriptstyle \alpha$} (C) (C) edge[->] node[above] {$\scriptstyle \beta$} (D) (D) edge[->] (E);
			\end{tikzpicture}
		\end{array}\,.
	\end{equation}
	Note that the space in the middle is invariant with respect to $s_a$: $W_a\oplus V_{a-1}\oplus V_{a+1}=\check W_a\oplus \check V_{a-1}\oplus \check V_{a+1}$.
	\item Duality relations for all two-arrow paths passing through node $a$ read:
	\begin{equation}\label{quad_du}
		\begin{aligned}
			&A_{a}A_{a-1}=\check A_{a}\check A_{a-1},\; B_{a-1}B_{a}=\check B_{a-1}\check B_{a},\; B_{a-1}A_{a-1}=\check B_{a-1}\check A_{a-1},\; A_aB_a=\check A_a\check B_a\,,\\ 
			&S_aA_{a-1}=\check S_a\check A_{a-1},\;S_aB_{a}=\check S_a\check B_{a},\;A_aR_a=\check A_a\check R_a,\;B_{a-1}R_a=\check B_{a-1}\check R_a,\;S_aR_a=\check S_a\check R_a\,.
		\end{aligned}
	\end{equation}
\end{enumerate}

It is natural to compare a mathematical equivalence of quiver representations with physical equivalence of corresponding vacuum varieties.
Indeed IR behavior of different quiver theories is equivalent if theories are dual to each other.
Transition rules between parameters of the original theory and the dual, ``checked'' one are reminiscent in this case of \emph{Seiberg electro-magnetic duality} \cite{Seiberg:1994pq,Intriligator:1995au}, a.k.a. a quiver \emph{mutation}.
In particular, transition rules for FI stability parameters \eqref{Nak_du_sta} and relations between electric gauge groups $U(d_b)$ and magnetic gauge groups $U(\check d_b)$ \eqref{Nak_du_dim} coincide with those of Seiberg duality.

As another argument in favor of a treatment of Weyl group elements as Seiberg-like dualities we could point out the Weyl groupoid \cite{2023arXiv230614598S,2025arXiv251004221V} associated with affine Lie superalgebras $\widehat{\fg\fl}(m|n)$.
For sufficiently large $m$ and $n$ these algebras admit substantially different choices of simple roots and respectively more than one Dynkin diagrams.
Some of the Weyl reflections extending the Weyl group to the Weyl groupoid in this case deliver transitions between different Dynkin diagrams of the same $\widehat{\fg\fl}(m|n)$ \cite{bezerra2021braid}.
It is simple to derive those transition on quiver varieties directly as Seiberg mutations \cite[Appendix A]{Galakhov:2024foa}.

However despite we set adjoint fields $C_b=0$ in the vacuum, they perform a non-trivial contribution on the quantum level, in particular, they contribute to duality.
A presence of adjoint matter makes duality relations more involved \cite{Kutasov:1995ve,Kutasov:1995np,Kutasov:1995ss} allowing to construct composite mesonic operators of potentially infinite length.
In this case certain terms in the superpotential might suppress this growth making Kutasov duality less universal than Seiberg duality without adjoint matter.

Nevertheless even taking into account possibilities of the adjoint matter influence we can not avoid noting certain discrepancies of the Weyl reflection actions with the Seiberg-Kutasov duality canonical action.
In the dual description pairs of fields adjacent to the node to be dualized are substituted by a composite meson field $M$, so the relations between fields take a form where a polynomial of original fields is equated to a polynomial of dual fields.
For example, if we would like to dualize, say, node $a$, in addition to others higher Kutasov mesons there will be operator $M=B_{a-1}A_{a-1}$ corresponding to a path starting at node $a-1$ passing through $a$ and returning back.
We should invert this path into two new dual fields $\check B_{a-1}\check A_{a-1}$, as a result the superpotential term containing these fields should be dualized as well:
\begin{equation}
	W=\Tr\left(C_aB_{a-1}A_{a-1}\right)+\ldots\;\to\;\check W=\Tr\left(C_aM-M\check B_{a-1}\check A_{a-1}\right)+\ldots\,,
\end{equation}
where we substituted $B_{a-1}A_{a-1}$ by $M$ in the superpotential expression.
Now a derivative of $\check W$ with respect to $M$ delivers a new vacuum constraint:
\begin{equation}
	C_a=\check B_{a-1}\check A_{a-1}\,,
\end{equation}
however as we will see in Sec.~\ref{sec:BoxYoung} fields $A$, $B$ and $C$ behave as independent coordinates in the field space, so there should not be a constraint equating values of $A$, $B$ with $C$, and such a theory becomes not self-dual as we expected.
Instead self-duality implies the following natural transform of the superpotential:
\begin{equation}
	W=\Tr\left(C_aB_{a-1}A_{a-1}\right)+\ldots\;\to\;\check W=\Tr\left(C_a\check B_{a-1}\check A_{a-1}\right)+\ldots\,.
\end{equation}

Similarly, if we ease exactness constraint \eqref{exact_du} to ${\rm Im}\,\alpha\subset{\rm Ker}\,\beta$, this would lead to a quadratic polynomial constraint $\beta\alpha=0$ where fields of the original theory and dual fields are mixed.
So for safety precautions we call this duality of $\myA_n$-type quiver varieties with adjoint matter a Seiberg-like \emph{Weyl mutation}.

%%%%%%%%%%%%%%%%%%%%%%%%%%%%%%%%%%%%%%%%%%%%%%%%%%%%%%%%%%%%%%%%%%%%%%%%%%%%%%%%%%%%%%%%%%%%%%%%%%%
%%%%%%%%%%%%%%%%%%%%%%%%%%%%%%%%%%%%%%%%%%%%%%%%%%%%%%%%%%%%%%%%%%%%%%%%%%%%%%%%%%%%%%%%%%%%%%%%%%%
%%%%%%%%%%%%%%%%%%%%%%%%%%%%%%%%%%%%%%%%%%%%%%%%%%%%%%%%%%%%%%%%%%%%%%%%%%%%%%%%%%%%%%%%%%%%%%%%%%%

\section{Modeling Yangian \texorpdfstring{$Y(\fs\fl_{n+1})$}{Y(su(n))} representations with quivers}\label{sec:quiv_Yang}

\subsection{Framed quiver diagram}

In what follows we will be interested in a specific family of framed quivers $\fQ_{n,u,h}$ with superpotentials parameterized by a triplet of positive integers $n$, $u$, $h$:
\begin{equation}\label{quiver}
	\fQ_{n,u,h}=\left\{\begin{array}{c}
		\begin{tikzpicture}
			\draw[postaction=decorate, decoration={markings, mark= at position 0.7 with {\arrow{stealth}}}] (0,0) to[out=20,in=160] node[pos=0.5,above] {$\scriptstyle A_1$} (1.5,0);
			\draw[postaction=decorate, decoration={markings, mark= at position 0.7 with {\arrow{stealth}}}] (1.5,0) to[out=200,in=340] node[pos=0.5,below] {$\scriptstyle B_1$} (0,0);
			\draw[postaction=decorate, decoration={markings, mark= at position 0.8 with {\arrow{stealth}}}] (0,0) to[out=60,in=0] (0,0.6) to[out=180,in=120] (0,0);
			\node[above] at (0,0.6) {$\scriptstyle C_1$};
			\begin{scope}[shift={(1.5,0)}]
				\draw[postaction=decorate, decoration={markings, mark= at position 0.8 with {\arrow{stealth}}}] (0,0) to[out=60,in=0] (0,0.6) to[out=180,in=120] (0,0);
				\node[above] at (0,0.6) {$\scriptstyle C_2$};
			\end{scope}
			\draw[fill=\myblue] (0,0) circle (0.08) (1.5,0) circle (0.08);
			%%%%%%%%%%%%%%%%%%%%%%%%%%%%%%%%%%%%%%%%%%%%%%%%%%%%%%%%%%%%%
			\begin{scope}[shift = {(4,0)}]
				\draw[postaction=decorate, decoration={markings, mark= at position 0.7 with {\arrow{stealth}}}] (0,-1.2) to[out=100,in=260] node[pos=0.3, left] {$\scriptstyle R$} (0,0);
				\draw[postaction=decorate, decoration={markings, mark= at position 0.7 with {\arrow{stealth}}}] (0,0) to[out=280,in=80] node[pos=0.7, right] {$\scriptstyle S$} (0,-1.2);
				%%%%%%%%%%%%%%%%%%%%%%%%%%%%%%%%%%%%%%%%%%%%%%%%%%%%5
				\draw[postaction=decorate, decoration={markings, mark= at position 0.7 with {\arrow{stealth}}}] (0,0) to[out=20,in=160] node[pos=0.5,above] {$\scriptstyle A_u$} (1.5,0);
				\draw[postaction=decorate, decoration={markings, mark= at position 0.7 with {\arrow{stealth}}}] (1.5,0) to[out=200,in=340] node[pos=0.5,below] {$\scriptstyle B_u$} (0,0);
				\begin{scope}[shift={(-1.5,0)}]
					\draw[postaction=decorate, decoration={markings, mark= at position 0.7 with {\arrow{stealth}}}] (0,0) to[out=20,in=160] node[pos=0.5,above] {$\scriptstyle A_{u-1}$} (1.5,0);
					\draw[postaction=decorate, decoration={markings, mark= at position 0.7 with {\arrow{stealth}}}] (1.5,0) to[out=200,in=340] node[pos=0.5,below] {$\scriptstyle B_{u-1}$} (0,0);
				\end{scope}
				\draw[postaction=decorate, decoration={markings, mark= at position 0.8 with {\arrow{stealth}}}] (0,0) to[out=60,in=0] (0,0.6) to[out=180,in=120] (0,0);
				\node[above] at (0,0.6) {$\scriptstyle C_u$};
				\begin{scope}[shift={(1.5,0)}]
					\draw[postaction=decorate, decoration={markings, mark= at position 0.8 with {\arrow{stealth}}}] (0,0) to[out=60,in=0] (0,0.6) to[out=180,in=120] (0,0);
					\node[above] at (0,0.6) {$\scriptstyle C_{u+1}$};
				\end{scope}
				\begin{scope}[shift={(-1.5,0)}]
					\draw[postaction=decorate, decoration={markings, mark= at position 0.8 with {\arrow{stealth}}}] (0,0) to[out=60,in=0] (0,0.6) to[out=180,in=120] (0,0);
					\node[above] at (0,0.6) {$\scriptstyle C_{u-1}$};
				\end{scope}
				\draw[fill=\myblue] (-1.5,0) circle (0.08) (0,0) circle (0.08) (1.5,0) circle (0.08);
				%%%%%%%%%%%%%%%%%%%%%%%%%%%%%%%%
				\begin{scope}[shift={(0,-1.2)}]
					\draw[fill=burgundy] (-0.08,-0.08) -- (-0.08,0.08) -- (0.08,0.08) -- (0.08,-0.08) -- cycle;
				\end{scope}
			\end{scope}
			%%%%%%%%%%%%%%%%%%%%%%%%%%%%%%%%%%%%%%%%%%%%%%%%%%%%%%%%%%%%%%%
			\begin{scope}[shift = {(8,0)}]
				\begin{scope}[shift={(-1.5,0)}]
					\draw[postaction=decorate, decoration={markings, mark= at position 0.7 with {\arrow{stealth}}}] (0,0) to[out=20,in=160] node[pos=0.5,above] {$\scriptstyle A_{n-1}$} (1.5,0);
					\draw[postaction=decorate, decoration={markings, mark= at position 0.7 with {\arrow{stealth}}}] (1.5,0) to[out=200,in=340] node[pos=0.5,below] {$\scriptstyle B_{n-1}$} (0,0);
				\end{scope}
				\draw[postaction=decorate, decoration={markings, mark= at position 0.8 with {\arrow{stealth}}}] (0,0) to[out=60,in=0] (0,0.6) to[out=180,in=120] (0,0);
				\node[above] at (0,0.6) {$\scriptstyle C_{n}$};
				\begin{scope}[shift={(-1.5,0)}]
					\draw[postaction=decorate, decoration={markings, mark= at position 0.8 with {\arrow{stealth}}}] (0,0) to[out=60,in=0] (0,0.6) to[out=180,in=120] (0,0);
					\node[above] at (0,0.6) {$\scriptstyle C_{n-1}$};
				\end{scope}
				\draw[fill=\myblue] (-1.5,0) circle (0.08) (0,0) circle (0.08);
			\end{scope}
			\draw[fill=black] (1.75,0) circle (0.03) (2,0) circle (0.03) (2.25,0) circle (0.03) (5.75,0) circle (0.03) (6,0) circle (0.03) (6.25,0) circle (0.03);
		\end{tikzpicture}\\
		W=\Tr\left[A_1C_1B_1+\sum\lm_{a=2}^{n-1}\left(A_aC_aB_a-B_{a-1}C_aA_{a-1}\right)-B_{n-1}C_{n}A_{n-1}+{\color{burgundy}S C_u^{h}R}\right]
	\end{array}\right\}\,.
\end{equation}
In addition we would like to introduce an equivariant toric action on these quiver varieties by introducing $\Omega$-background parameters.
Those parameters enter quiver theory as chiral flavors charging fields representing quiver arrows.
We assign the following values of the equivariant charges to our quiver arrows:
\begin{equation}\label{fcharges}
	\begin{array}{c|c|c|c|c|c}
		\mbox{Field} & A_a & B_a & C_a & R & S\\
		\hline
		\mbox{Equiv. ch.} & \epsilon_1 & \epsilon_2 & -\epsilon_1-\epsilon_2 & 0 & h(\epsilon_1+\epsilon_2)
	\end{array}\,.
\end{equation}

From now on we assume that the framing vector space is one-dimensional and the corresponding flavor charge is set to $\mu$.

As classical vacua for $\fQ_{n,u,h}$ one should count equivariant fixed points on the respective quiver variety \cite{Galakhov:2020vyb} modulo gauge transformations.
Let us summarize all the constraints on quiver morphisms imposed by D-term, F-term and the equivariant fixed point condition:
\begin{equation}\label{fp_eq}
	\begin{aligned}
		&B_aB_a^{\dagger}-A_a^{\dagger}A_a+A_{a-1}A_{a-1}^{\dagger}-B_{a-1}^{\dagger}B_{a-1}+\left[C_a,C_a^{\dagger}\right]+\delta_{a,u}\left(RR^{\dagger}-S^{\dagger}S\right)=\zeta_a\bbone_{d_a\times d_a},\quad 1\leq a\leq n\,;\\
		&A_aB_a-A_{a-1}B_{a-1}+\delta_{a,u}\sum\lm_{k=0}^{h-1}C_u^kRSC_u^{h-1-k}=\bbzero_{d_a\times d_a},\quad 1\leq a\leq n\,;\\
		& C_aB_a=B_aC_{a+1},\quad A_aC_a=C_{a+1}A_a,\quad 1\leq a\leq n-1;\quad C_u^kR=\bbzero_{d_u\times 1},\quad S C_u^k=\bbzero_{1\times d_u}\,;\\
		& \left[\Phi_a,C_a\right]=(\epsilon_1+\epsilon_2)C_a,\quad \Phi_{a+1}A_a-A_a\Phi_a=\epsilon_1 A_a,\quad \Phi_{a}B_a-B_a\Phi_{a+1}=\epsilon_2 B_a,\quad 1\leq a\leq n\,,\\
		& \Phi_u R =\mu R,\quad \mu S-S\Phi_u=h(\epsilon_1+\epsilon_2)S\,,
	\end{aligned}
\end{equation}
where $\Phi_a$ are scalars belonging to the gauge multiplets developing a vev in vacua.

For $\fQ_{n,u,h}$ it is known that the respective BPS algebra is isomorphic to Yangian algebra $Y(\fs\fl_{n+1})$ \cite{Galakhov:2024bzs,Gavshin:2025akn} (see also \cite{Bao:2023ece,Bao:2025dqs}).
Moreover fixed points in the cyclic chamber (solutions to \eqref{fp_eq} modulo gauge transforms for $\zeta_a>0$) form a module with respect to $Y(\fs\fl_{n+1})$ that is isomorphic to an irreducible representation of $\fs\fl_{n+1}$ labeled by a rectangular Young diagram:
\begin{equation}
	\Upsilon_{u,h}=\underbrace{[h,h,h,\ldots,h]}_{n+1-u\;{\rm times}}\;=\!\!\!\begin{array}{c}
		\begin{tikzpicture}[scale=0.4]
			\foreach \i in {0,1,2,3}
			{
				\draw[thick] (0,\i) -- (4,\i);
			}
			\foreach \i in {0,1,2,3,4}
			{
				\draw[thick] (\i,0) -- (\i,3);
			}
			\draw (0,0) to[out=270, in=180] (0.25,-0.25) -- (1.75,-0.25) to[out=0,in=90] (2,-0.5) to[out=90,in=180] (2.25,-0.25) -- (3.75,-0.25) to[out=0,in=270] (4,0);
			\draw (4,0) to[out=0,in=270] (4.25,0.25) -- (4.25,1.25) to[out=90,in=180] (4.5,1.5) to[out=180,in=270] (4.25,1.75) -- (4.25,2.75) to[out=90,in=0] (4,3);
			\node[below] at (2,-0.5) {$\scriptstyle n+1-u$ \scriptsize columns};
			\node[right] at (4.5,1.5) {$\scriptstyle h$ \scriptsize rows};
		\end{tikzpicture}
	\end{array}\,.
\end{equation}
This particular type of representations of $\fs\fl_{n+1}$ is such that the lowest weight $\lambda$ is proportional to one of the fundamental weights $\omega_i$, in particular, $\lambda=-h\omega_u$ \cite[Sec.13.3.1]{DiFrancesco:1997nk}.
It is known from the literature \cite{drinfel1985hopf} that such irreducible representations of $\fs\fl_{n+1}$ could be extended to representations of the respective Yangian algebra $Y(\fs\fl_{n+1})$.

The Yangian algebra $Y(\fs\fl_{n+1})$ realized as a BPS algebra in the \emph{cyclic chamber} delivers raising Chevalley generators $e_a^{(k)}$ where index $a$ runs over nodes of quiver/Dynkin diagram and zero modes $e_a^{(0)}$ belong to subalgebra $\fs\fl_{n+1}$.
Physically, $e_a^{(k)}$ shift respective quiver dimensions $d_a\to d_a+1$.
Representation $\Upsilon_{u,h}$ is irreducible, and therefore it could be constructed as a truncated Verma module starting with a vacuum lowest weight vector $|0\rangle$, then quiver dimensions $d_a$ for fixed point solutions to \eqref{fp_eq} could be treated naturally as how many times one should act on $|0\rangle$ by $e_a$ to arrive to a desired vector in the Verma module.
This reasoning sets a natural relation between quiver dimensions of a solution to \eqref{fp_eq} and weight $w$ of a vector of irrep $\Upsilon_{u,h}$ it corresponds to:
\begin{tcolorbox}
\begin{equation}\label{weight}
	w=-h\omega_u+\sum\lm_{a=1}^nd_a\alpha_a\,.
\end{equation}
\end{tcolorbox}

In practice we could have considered a more generic framing for quivers $\fQ_{n,u,h}$ similar to one used in \eqref{ALE_quiv}: attach a quiver node to each gauge node and modify the superpotential by respective terms $\Tr\left(R_aC_a^{h_a}S_a\right)$.
In this case as it was discussed in e.g. \cite{Galakhov:2022uyu} fixed points of the quiver variety \eqref{fp_eq} would correspond to a tensor sum of $Y(\fs\fl_{n+1})$ irreps:\footnote{In practice these irreps carry also spectral parameters defined by respective flavor charges $\mu$ provided by framing nodes, see \cite{Galakhov:2022uyu}.}
\begin{equation}
	R_{\fQ}=\bigoplus\lm_{a=1}^nW_a\otimes \Upsilon_{a,h_a}\,,
\end{equation}
where flavor spaces $W_a$ play the role of invariant subspaces of $\fs\fl_{n+1}$. 

A short way to describe solutions to \eqref{fp_eq} is to construct a generating function for solution numbers $N(d_a)$ depending on quiver dimensions $d_a$.
This generating function is analogous to one counting charged D-branes, or Donaldson-Thomas/Gromov-Witten invariants \cite{maulik2006gromov} on Calabi-Yau resolutions:
\begin{equation}
	{\bf DT}_{\fQ,\vec \zeta}\left(\{q_k\}\right):=\sum\lm_{d_a\geq 0} N_{\vec \zeta}(d_a)\times\prod\lm_{a=1}^n q_a^{d_a}\,.
\end{equation}
In general solution numbers, and generating functions $\bf DT$ respectively, depend piece-wise on a ray in the FI parameter space $\zeta_a$.
This piece-wise dependence is known as wall-crossing phenomena \cite{Galakhov:2024foa,Andriyash:2010qv,Cecotti:2009uf,Aganagic:2009kf,Kontsevich:2008fj,FelixKlein,Aganagic:2010qr,Sulkowski:2009rw}.

As we mentioned for our quiver $\fQ_{n,u,h}$ in the cyclic chamber (all $\zeta_a>0$) solutions to \eqref{fp_eq} correspond to vectors of representation $\Upsilon_{u,h}$, therefore the generating function coincides with the character (the Schur function $\chi_{\Upsilon}\left(\left\{p_k\right\}\right)$) of representation $\Upsilon_{u,h}$ of $\fs\fl_{n+1}$:
\begin{equation}\label{DTSchur}
	{\bf DT}_{\fQ,{\rm cyc}}\left(\{q_k\}\right)=\prod\lm_{i=u}^nq_i^{-h(u-i)}\times\chi_{\Upsilon_{u,h}}\left(\{p_k\}\right)\big|_{\fs\fl_{n+1}}\,,
\end{equation}
where
\begin{equation}
	p_k \big|_{\fs\fl_{n+1}}:=\sum\lm_{i=1}^{n+1}\left(\prod\lm_{j=n+2-i}^{n}q_j^k\right)\,.
\end{equation}
For example:
\begin{equation}
	\begin{aligned}
		&\chi_{[1,1]}=\frac{p_1^2}{2}-\frac{p_2}{2}\;\to\;(n,u,h)\left[\fs\fl_3,[1,1]\right]=(2,1,1)\;\to\;{\bf DT}_{2,1,1}=1+q_1+q_1q_2\,,\\
		&\chi_{[1]}=p_1\;\to\;(n,u,h)\left[\fs\fl_3,[1]\right]=(2,2,1)\;\to\;{\bf DT}_{2,2,1}=1+q_2+q_1q_2\,,\\
		&\chi_{[2,2]}=\frac{p_1^4}{12}-\frac{p_3 p_1}{3}+\frac{p_2^2}{4}\;\to\;(n,u,h)\left[\fs\fl_3,[2,2]\right]=(2,1,2)\;\to\;{\bf DT}_{2,1,2}=1+q_1+q_1^2+q_1 q_2+q_1^2q_2 +q_1^2q_2^2 \,,\\
		&\chi_{[1,1,1]}=\frac{p_1^3}{6}-\frac{p_2 p_1}{2}+\frac{p_3}{3}\;\to\;(n,u,h)\left[\fs\fl_4,[1,1,1]\right]=(3,1,1)\;\to\;{\bf DT}_{3,1,1}=1+q_1+q_1q_2+q_1q_2q_3\,.
	\end{aligned}
\end{equation}

As we will see in what follows the generating function ${\bf DT}_{\fQ_{n,u,h}}$ remains the same across the walls of marginal stability, and is in fact a \emph{wall-crossing invariant}.
So that subscript cyc. in \eqref{DTSchur} could be omitted.

%%%%%%%%%%%%%%%%%%%%%%%%%%%%%%%%%%%%%%%%%%%%%%%%%%%%%%%%%%%%%%%%%%%%%%%%%%%%%%%%%%%%%%%%%%%%%%%%%%%
%%%%%%%%%%%%%%%%%%%%%%%%%%%%%%%%%%%%%%%%%%%%%%%%%%%%%%%%%%%%%%%%%%%%%%%%%%%%%%%%%%%%%%%%%%%%%%%%%%%
%%%%%%%%%%%%%%%%%%%%%%%%%%%%%%%%%%%%%%%%%%%%%%%%%%%%%%%%%%%%%%%%%%%%%%%%%%%%%%%%%%%%%%%%%%%%%%%%%%%

\subsection{Depicting fixed points with atomic structure plots}\label{sec:atomic}

Here we would like to exploit a diagrammatic way of denoting fixed points on quiver varieties.
In particular, we apply atomic structure plots suggested in \cite{Galakhov:2024foa}.
Atomic structure plots are a mere brief depiction of characteristic features of an apriori known solution to \eqref{fp_eq}.
Unlike molten crystals (for a review of the molten crystal construction see e.g.\cite{Yamazaki:2010fz,Li:2023zub,Yamazaki:2022cdg,Bao:2024ygr} and references therein) corresponding to various 3d partitions there are no construction rules for these plots making sure that a given plot corresponds to a valid fixed point on a quiver variety.
Yet depictions of atomic structure plots and molten crystals are quite similar, and basic rules for reconstructing actual matrices of quiver morphisms from such a depiction are the same.

We define an atomic picture plot as a connected oriented graph $G$ constructed from a solution to \eqref{fp_eq} subjected to the following rules:
\begin{enumerate}
	\item The bases in quiver vector spaces $V_a$ are fixed in such a way that all $\Phi_a$ are diagonal.
	\item An injective map $\iota$ sends all distinct basic vectors $v\in V_a$ for all $1\leq a\leq n$ and the basic vector of a single $W_u$ to vertices of $G$, also called ``atoms''.
	Atoms have a characteristic ``color'' corresponding to which quiver node $a$ belongs its pre-image $\iota^{-1}$.
	\item For all quiver morphisms $M:\;V_a\to V_b$ consider two basic vectors $v_1\in V_a$ and $v_2\in V_b$, if there is a non-zero matrix element $\left|\left\langle v_2|M|v_1\right\rangle\right|^2>0$ then in $G$ respective atoms are connected by an arrow with label $M$:
	\begin{equation}
		\left|\left\langle v_2|M|v_1\right\rangle\right|^2>0\;\Rightarrow\;\hspace{-0.35cm}\begin{array}{c}
			\begin{tikzpicture}
				\draw[thick, postaction=decorate, decoration={markings, mark= at position 0.7 with {\arrow{stealth}}}] (0,0) -- (1.5,0) node[pos=0.5,above] {$\scriptstyle M$};
				\draw[fill=\myblue] (0,0) circle (0.08) (1.5,0) circle (0.08);
				\node(A)[left] at (-0.08,0) {$\iota(v_1)$};
				\node(B)[right] at (1.58,0) {$\iota(v_2)$};
			\end{tikzpicture}
		\end{array}\,.
	\end{equation}
	\item Eigen value $\phi(\nu)$ of $\Phi_a$ for basic vector $\nu\in V_a$ is constructed in the following way.
	Consider basic vector $w_0\in W_u$ and construct a path $\wp$ along edges of $G$ connecting $\iota(w_0)$ to $\iota(\nu)$.
	A direction of edge $e$ in $\wp$ might be along $\wp$ or opposite, in the first case we say $\sigma_e=+1$, in the other case we say $\sigma_e=-1$.
	Then the $\Phi_a$ eigen value reads:
	\begin{equation}\label{flavor}
		\phi(\nu)=\mu+\sum\lm_{e\in \wp}\sigma_e\epsilon_e\,,
	\end{equation}
	where $\epsilon_e$ is the respective flavor charge of the field labeling $e$ from table \eqref{fcharges}.
\end{enumerate}

An inverse procedure of reconstructing a solution to \eqref{fp_eq} form an atomic structure plot is also simple.
First of all we count amounts of atoms of different colors, these will be quiver dimensions:
\begin{equation}
	d_a=\mbox{number of atoms of color }a\,.
\end{equation}
Further we enumerate atoms in arbitrary order within each color group.
From directed arrows with labels we construct an ansatz for quiver morphisms in the following way: if an arrow with label $M$ connects atoms with internal group numbers $a$ and $b$ then $(b,a)$-matrix element of $M$ acquires a non-zero expectation value:
\begin{equation}
	M={\color{white!80!black}\left(\begin{array}{cccccc}
			* & * & * & * & * & *\\
			* & * & {\color{black} m_{ba}\neq 0} & * & * & *\\
			* & * & * & * & * & *\\
			* & * & * & * & * & *\\
		\end{array}\right)}\,.
\end{equation}
Other elements of quiver morphisms that do not correspond to edges of an atomic  structure plot are set to zero.
Particular values $m_{ba}$ for non-zero matrix elements are not determined from the atomic structure plots.
To fix them for a $G_{\IR}(\fQ)$-orbit element (fixing those elements up to phases) one should substitute this ansatz into \eqref{fp_eq} and solve remaining equations for $m_{ba}$. 
When atomic structure plots incorporate many loops it turns out equations \eqref{fp_eq} are of rather high order and can not be solved in terms of radicals.
In those cases one could solve these equations numerically to find a single real solution starting with a rather generic seed point by the usual steepest descend method.

We present more explicit examples in Sec.~\ref{sec:mut_asp}.
Now to illustrate the procedure described above let us consider the following atomic structure plot for a fixed point of $\fQ_{3,2,2}$ \eqref{sl(4), [2,2], quiv} in the cyclic chamber:
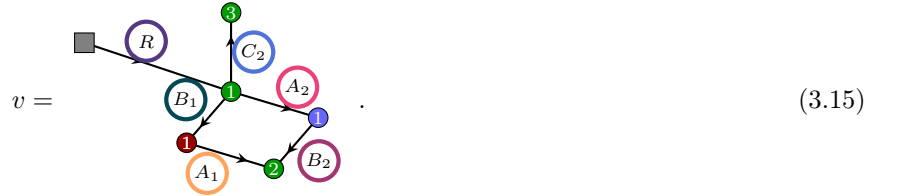
\begin{equation}\label{aps_examp}
v=
\begin{array}{c}
	\begin{tikzpicture}[scale=1.3]
		\tikzset{col1/.style={fill=black!40!red}}
		\tikzset{col2/.style={fill=black!40!green}}
		\tikzset{col3/.style={fill=\myblue}}
		\draw[thick, postaction=decorate, decoration={markings, mark= at position 0.4 with {\arrow{stealth}}}] (-1.5,0.5) -- (0,0) node(A)[pos=0.3,above right] {$\scriptstyle R$};
		\draw[thick, postaction=decorate, decoration={markings, mark= at position 0.7 with {\arrow{stealth}}}] (0.,0.) -- (-0.45399,-0.52372) node(B)[pos=0.5,above left] {$\scriptstyle B_1$};
		\draw[thick, postaction=decorate, decoration={markings, mark= at position 0.7 with {\arrow{stealth}}}] (0.891007,-0.266849) -- (0.437016,-0.790569)  node(C)[pos=0.5,below right] {$\scriptstyle B_2$};
		\draw[thick, postaction=decorate, decoration={markings, mark= at position 0.7 with {\arrow{stealth}}}] (0.,0.) -- (0.891007,-0.266849) node(D)[pos=0.5,above right] {$\scriptstyle A_2$};
		\draw[thick, postaction=decorate, decoration={markings, mark= at position 0.7 with {\arrow{stealth}}}] (-0.45399,-0.52372) -- (0.437016,-0.790569)   node(E)[pos=0.5,below left] {$\scriptstyle A_1$};
		\draw[thick, postaction=decorate, decoration={markings, mark= at position 0.7 with {\arrow{stealth}}}] (0.,0.) -- (0.,0.809017) node(F)[pos=0.5,right] {$\scriptstyle C_2$};
		\foreach \a/\b/\c in {0./0./col2,0./0.809017/col2,-0.45399/-0.52372/col1,0.891007/-0.266849/col3,0.437016/-0.790569/col2}
		{
			\draw[\c] (\a,\b) circle (0.1);
		}
		\node[white] at (0,0) {$\scriptstyle 1$};
		\node[white] at (-0.45399,-0.52372) {$\scriptstyle 1$};
		\node[white] at (-0.45399,-0.52372) {$\scriptstyle 1$};
		\node[white] at (0.891007,-0.266849) {$\scriptstyle 1$};
		\node[white] at (0.,0.809017) {$\scriptstyle 3$};
		\node[white] at (0.437016,-0.790569) {$\scriptstyle 2$};
		\begin{scope}[shift={(-1.5,0.5)}]
			\draw[fill=gray] (-0.1,-0.1) -- (-0.1,0.1) -- (0.1,0.1) -- (0.1,-0.1) -- cycle;
		\end{scope}
		\draw[ultra thick,palette2] (A) circle (0.2);
		\draw[ultra thick,palette3] (B) circle (0.2);
		\draw[ultra thick,palette4] (C) circle (0.2);
		\draw[ultra thick,palette5] (D) circle (0.2);
		\draw[ultra thick,palette6] (E) circle (0.2);
		\draw[ultra thick,palette7] (F) circle (0.2);
	\end{tikzpicture}
\end{array}\,.
\end{equation}
Atom colors correspond to those of quiver \eqref{sl(4), [2,2], quiv} nodes: there is 1 {\bf\color{black!40!red}red} atom corresponding to node {\bf\color{black!40!red}1}, 3 {\bf\color{black!40!green}green} atoms corresponding to node {\bf\color{black!40!green}2} and 1 {\bf\color{\myblue}blue} atom corresponding to node {\bf\color{\myblue}3}.
Numbers of atoms count dimensions of respective vector spaces of the quiver representation:
\begin{equation}
	{\rm dim}\, V_{\bf\color{black!40!red}1}=1,\quad {\rm dim}\, V_{\bf\color{black!40!green}2}=3,\quad {\rm dim}\, V_{\bf\color{\myblue}3}=1\,.
\end{equation}
Atoms correspond to basic vectors in those spaces, and by enumerating them we choose some ordering on the basic vectors.
Different order choices are isomorphic by $G_{\IR}(\fQ)$ transformations permuting basic vectors, so it is sufficient to choose any.
We enumerated atoms as it is in \eqref{aps_examp}.
Enumerations are independent between different color groups and denote order only within a specific atom color group, so in \eqref{aps_examp} there are three atoms with number 1 of different colors.
Further we re-construct non-zero matrix elements of quiver morphisms from edges in the atomic structure plots.
An arrow with label ``$A_1$'' connecting atoms with numbers 1 and 2 corresponds to a non-zero vev of matrix field $A_1$ at position $21$ in the matrix.
Similarly, an arrow with label ``$C_2$'' connecting atoms with numbers 1 and 3 corresponds to a non-zero vev of matrix field $C_2$ at position $31$ in the matrix.
Continuing with this rule we would arrive to the following ansatz for quiver morphisms, where we used the same color code as in \eqref{aps_examp} to mark which vevs correspond to which arrows:
\begin{subequations}
\begin{equation}\label{asp_exmp1}
		C_1=\left(
		\begin{array}{c}
			0 \\
		\end{array}
		\right)\in{\rm Hom}(V_{\bf\color{black!40!red}1}, V_{\bf\color{black!40!red}1}),\; C_2=\hspace{-0.35cm}\begin{array}{c}
			\begin{tikzpicture}
				\node at (0,0) {$\left(
					\begin{array}{ccc}
						0 & 0 & 0 \\
						0 & 0 & 0 \\
						x_1 & 0 & 0 \\
					\end{array}
					\right)$};
				\draw[ultra thick, palette7] (-0.6,-0.45) circle (0.25);
			\end{tikzpicture}
		\end{array}\hspace{-0.35cm}\in{\rm Hom}(V_{\bf\color{black!40!green}2}, V_{\bf\color{black!40!green}2}),\;C_3=\left(
		\begin{array}{c}
		0 \\
		\end{array}
		\right)\in{\rm Hom}(V_{\bf\color{\myblue}3}, V_{\bf\color{\myblue}3})\,,
\end{equation}
\begin{equation}
	A_1=\hspace{-0.35cm}\begin{array}{c}
		\begin{tikzpicture}
			%%%%%%%%%%%%%%%%%%%%%%%%%%%%%%%%%%%%%%
			%%%%%%%%%%%%%%%%%%%%%%%%%%%%%%%%%%%%%%
			\node at (0,0) {$\left(
				\begin{array}{c}
					0 \\
					x_2 \\
					0 \\
				\end{array}
				\right)$};
			\draw[ultra thick, palette6] (0.,0.) circle (0.25);
		\end{tikzpicture}
	\end{array}\hspace{-0.35cm}\in{\rm Hom}(V_{\bf\color{black!40!red}1}, V_{\bf\color{black!40!green}2}),\;B_1=\hspace{-0.35cm}\begin{array}{c}
	\begin{tikzpicture}
		%%%%%%%%%%%%%%%%%%%%%%%%%%%%%%%%%%%%%%
		%%%%%%%%%%%%%%%%%%%%%%%%%%%%%%%%%%%%%%
		\node at (0,0) {$\left(
			\begin{array}{ccc}
				x_3 & 0 & 0 \\
			\end{array}
			\right)$};
		\draw[ultra thick, palette3] (-0.55,0.) circle (0.25);
	\end{tikzpicture}
	\end{array}\hspace{-0.35cm}\in{\rm Hom}(V_{\bf\color{black!40!green}2},V_{\bf\color{black!40!red}1}),\;A_2=\hspace{-0.35cm}\begin{array}{c}
	\begin{tikzpicture}
		%%%%%%%%%%%%%%%%%%%%%%%%%%%%%%%%%%%%%%
		%%%%%%%%%%%%%%%%%%%%%%%%%%%%%%%%%%%%%%
		\node at (0,0) {$\left(
			\begin{array}{ccc}
				x_4 & 0 & 0 \\
			\end{array}
			\right)$};
		\draw[ultra thick, palette5] (-0.55,0.) circle (0.25);
	\end{tikzpicture}
	\end{array}\hspace{-0.35cm}\in{\rm Hom}(V_{\bf\color{black!40!green}2},V_{\bf\color{\myblue}3})\,,
\end{equation}
\begin{equation}\label{asp_exmp3}
	B_2=\hspace{-0.35cm}\begin{array}{c}
		\begin{tikzpicture}
			%%%%%%%%%%%%%%%%%%%%%%%%%%%%%%%%%%%%%%
			%%%%%%%%%%%%%%%%%%%%%%%%%%%%%%%%%%%%%%
			\node at (0,0) {$\left(
				\begin{array}{c}
					0 \\
					x_5 \\
					0 \\
				\end{array}
				\right)$};
			\draw[ultra thick, palette4] (0.,0.) circle (0.25);
		\end{tikzpicture}
	\end{array}\hspace{-0.35cm}\in{\rm Hom}(V_{\bf\color{\myblue}3}, V_{\bf\color{black!40!green}2}),\;R=\hspace{-0.35cm}\begin{array}{c}
	\begin{tikzpicture}
		%%%%%%%%%%%%%%%%%%%%%%%%%%%%%%%%%%%%%%
		%%%%%%%%%%%%%%%%%%%%%%%%%%%%%%%%%%%%%%
		\node at (0,0) {$\left(
			\begin{array}{c}
				x_6 \\
				0 \\
				0 \\
			\end{array}
			\right)$};
		\draw[ultra thick, palette2] (0.,0.45) circle (0.25);
	\end{tikzpicture}
	\end{array}\hspace{-0.35cm}\in{\rm Hom}(W, V_{\bf\color{black!40!green}2}),\;S=\left(
	\begin{array}{ccc}
	0 & 0 & 0 \\
	\end{array}
	\right)\in{\rm Hom}(V_{\bf\color{black!40!green}2},W)\,.
\end{equation}
\end{subequations}
In the chosen basis fields $\Phi_a$ are diagonal, and it is easy to calculate their eigenvalues from the path formula \eqref{flavor} by drawing paths from the framing atom to respective atoms inside the plot edges:
\begin{equation}
	\begin{aligned}
	&\phi(\!\!\!\raisebox{-0.05cm}{$\begin{array}{c}
		\begin{tikzpicture}
			\draw[fill=black!40!red] (0,0) circle (0.15);
			\node[white] at (0,0) {$1$};
		\end{tikzpicture}
	\end{array}$}\!\!\!)=\mu+\epsilon_R+\epsilon_{B_1}=\mu+\epsilon_2\,,\\
	&\phi(\!\!\!\raisebox{-0.05cm}{$\begin{array}{c}
		\begin{tikzpicture}
			\draw[fill=black!40!green] (0,0) circle (0.15);
			\node[white] at (0,0) {$1$};
		\end{tikzpicture}
	\end{array}$}\!\!\!)=\mu+\epsilon_R=\mu\,,\\
	&\phi(\!\!\!\raisebox{-0.05cm}{$\begin{array}{c}
		\begin{tikzpicture}
			\draw[fill=black!40!green] (0,0) circle (0.15);
			\node[white] at (0,0) {$2$};
		\end{tikzpicture}
	\end{array}$}\!\!\!)=\mu+\epsilon_R+\epsilon_{A_1}+\epsilon_{B_1}=\mu+\epsilon_R+\epsilon_{A_2}+\epsilon_{B_2}=\mu+\epsilon_{1}+\epsilon_{2}\,,\\
	&\phi(\!\!\!\raisebox{-0.05cm}{$\begin{array}{c}
		\begin{tikzpicture}
			\draw[fill=black!40!green] (0,0) circle (0.15);
			\node[white] at (0,0) {$3$};
		\end{tikzpicture}
	\end{array}$}\!\!\!)=\mu+\epsilon_R+\epsilon_{C_2}=\mu-\epsilon_{1}-\epsilon_{2}\,,\\
	&\phi(\!\!\!\raisebox{-0.05cm}{$\begin{array}{c}
		\begin{tikzpicture}
			\draw[fill=\myblue] (0,0) circle (0.15);
			\node[white] at (0,0) {$1$};
		\end{tikzpicture}
	\end{array}$}\!\!\!)=\mu+\epsilon_R+\epsilon_{B_1}=\mu+\epsilon_{1}\,.
	\end{aligned}
\end{equation}
Summarizing expectation values of fields $\Phi_a$ read:
\begin{equation}
	\Phi_{\bf\color{black!40!red}1}=\left(\begin{array}{c}
		\mu+\epsilon_{2}
	\end{array}\right),\quad \Phi_{\bf\color{black!40!green}2}=\left(\begin{array}{ccc}
	\mu& 0 & 0\\
	0 & \mu+\epsilon_{1}+\epsilon_{2}& 0\\
	0 & 0 & \mu-\epsilon_{1}-\epsilon_{2}
	\end{array}\right),\quad \Phi_{\bf\color{\myblue}3}=\left(\begin{array}{c}
	\mu+\epsilon_{1}
	\end{array}\right)\,.
\end{equation}

To fix absolute values of variables $x_i$ on the $G_{\IR}(\fQ)$-orbit one should substitute ansatz \eqref{asp_exmp1}-\eqref{asp_exmp3} into \eqref{fp_eq}.
We should note that equations of \eqref{fp_eq} containing fields $\Phi_a$ are satisfied automatically.
For $x_i$ one arrives to the following set of equations:
\begin{equation}
	x_4 x_5-x_2 x_3=0,\;-x_2^2+x_3^2=\zeta _1,\; x_1^2=\zeta _2,\;x_2^2+x_5^2=\zeta _2,\;-x_1^2-x_3^2-x_4^2+x_6^2=\zeta _2,\;x_4^2-x_5^2=\zeta _3\,,
\end{equation}
having the following real solution:
\begin{equation}
	\begin{aligned}
	&x_1=\sqrt{\zeta _2},\;x_2=\sqrt{ \frac{\zeta _2 \left(\zeta _2+\zeta _3\right)}{\zeta _1+2 \zeta _2+\zeta _3}},\;x_3=\sqrt{\frac{\left(\zeta _1+\zeta _2\right) \left(\zeta _1+\zeta _2+\zeta _3\right)}{\zeta _1+2 \zeta _2+\zeta _3}},\,\\
	&x_4=\sqrt{ \frac{\left(\zeta _2+\zeta _3\right) \left(\zeta _1+\zeta _2+\zeta _3\right)}{\zeta _1+2 \zeta _2+\zeta _3}},\;x_5=\sqrt{ \frac{\zeta _2 \left(\zeta _1+\zeta _2\right)}{\zeta _1+2 \zeta _2+\zeta _3}},\;x_6=\sqrt{\zeta _1+3 \zeta _2+\zeta _3}\,.
	\end{aligned}
\end{equation}
valid in the cyclic locus $\zeta_a>0$.

%%%%%%%%%%%%%%%%%%%%%%%%%%%%%%%%%%%%%%%%%%%%%%%%%%%%%%%%%%%%%%%%%%%%%%%%%%%%%%%
%%%%%%%%%%%%%%%%%%%%%%%%%%%%%%%%%%%%%%%%%%%%%%%%%%%%%%%%%%%%%%%%%%%%%%%%%%%%%%%
%%%%%%%%%%%%%%%%%%%%%%%%%%%%%%%%%%%%%%%%%%%%%%%%%%%%%%%%%%%%%%%%%%%%%%%%%%%%%%%

\subsection{Cyclic chamber and plane partitions}\label{sec:BoxYoung}

A problem of counting fixed points for quiver varieties descending from D-branes on toric Calabi-Yau three-folds in the \emph{cyclic chamber} (all $\zeta_a>0$) has an elegant solution in terms of molten crystals (see \cite{Yamazaki:2010fz,Li:2023zub,Yamazaki:2022cdg,Bao:2024ygr} and references therein).
A derivation of this solution is rather involved, however the final answer represents a simple combinatorial construction generalizing 3d partitions.
For the purposes of the present paper we translate this rather generic construction to a requirement that atomic picture plots of fixed points (actual solutions to \eqref{fp_eq}) for $\fQ_{n,u,h}$ in the cyclic chamber are in a 1-to-1 correspondence to plane partitions (3d Young diagrams) bounded by a $(n+1-u)\times u\times h$ box.

This correspondence goes as follows (cf. Fig.~\ref{fig:boxedYoung}).
Consider a 3d space with axes $(\vec B,\vec A,\vec C)$ and embed a 3d Young diagram $Y$ into a rectangular box of dimensions $(n+1-u)\times u\times h$ so that all the cells of $Y$ have integral coordinates, and the cell at the corner is at the center of coordinates $(0,0,0)$.
Each cell of $Y$ corresponds to a round (``gauge'') atom of the respective atomic picture plot $P$.
Color $\bf col$ of an atom $\Box$ in $P$ corresponds to a node of $\fQ_{n,u,h}$ and is calculated from coordinates of the respective cell $\Box$ as:
\begin{equation}
	{\bf col}(\Box)=A_{\Box}-B_{\Box}+u\,.
\end{equation}
To each common face shared by two cells of $Y$ we associate an arrow of $P$ connecting the atoms, the label for this arrow is defined by to which axis this arrow is parallel $\vec B$, $\vec A$ or $\vec C$.
An atom at the center $(0,0,0)$ is connected to a square atom by an arrow $R$.

So in general if atom colors are neglected $q_a=q$ the generating function for solution numbers of \eqref{fp_eq} is given by the Macmahon formula for boxed plane partitions:
\begin{equation}
	{\bf DT}_{\fQ_{n,u,h}}(q_a=q)=\prod\lm_{i=1}^{n+1-u}\prod\lm_{j=1}^u\frac{1-q^{i+j+h-1}}{1-q^{i+j-1}}\,.
\end{equation}  

\begin{figure}[ht!]
	\centering
	\begin{tikzpicture}
		\tikzset{col1/.style={fill=palette1}}
		\tikzset{col2/.style={fill=palette3}}
		\tikzset{col3/.style={fill=palette4}}
		\tikzset{col4/.style={fill=palette7}}
		\tikzset{col5/.style={fill=palette6}}
		\tikzset{col6/.style={fill=palette2}}
		\tikzset{col7/.style={fill=palette5}}
		\draw[-stealth] (0,0) -- (-1.72516,-1.99014) node[pos=1,left] {$\scriptstyle B$};
		\draw[-stealth] (0,0) -- (3.38582, -1.01403) node[pos=1,right] {$\scriptstyle A$};
		\draw[-stealth] (0,0) -- (0., 3.07426)  node[pos=1,left] {$\scriptstyle C$};
		\draw[thick, postaction=decorate, decoration={markings, mark= at position 0.5 with {\arrow{stealth}}}] (-2,0.5) -- (0,0) node[pos=0.3, above right] {$\scriptstyle R$};
		\begin{scope}[shift={(-2,0.5)}]
			\draw[fill=gray] (-0.1,-0.1) -- (-0.1,0.1) -- (0.1,0.1) -- (0.1,-0.1) -- cycle;
		\end{scope}
		\foreach \a/\b/\c/\d in {0./0./-0.45399/-0.52372,0./0.809017/-0.45399/0.285296,0./1.61803/-0.45399/1.09431,0./2.42705/-0.45399/1.90333,-0.45399/-0.52372/-0.907981/-1.04744,-0.45399/0.285296/-0.907981/-0.238424,-0.907981/-1.04744/-1.36197/-1.57116,0.891007/-0.266849/0.437016/-0.790569,0.891007/0.542168/0.437016/0.0184476,0.891007/1.35119/0.437016/0.827465,0.437016/-0.790569/-0.0169745/-1.31429,-0.0169745/-1.31429/-0.470965/-1.83801,1.78201/-0.533698/1.32802/-1.05742,1.78201/0.275319/1.32802/-0.248401,1.32802/-1.05742/0.874032/-1.58114,2.67302/-0.800547/2.21903/-1.32427,0./0./0.891007/-0.266849,0./0.809017/0.891007/0.542168,0./1.61803/0.891007/1.35119,0./2.42705/0.891007/2.1602,-0.45399/-0.52372/0.437016/-0.790569,-0.45399/0.285296/0.437016/0.0184476,-0.45399/1.09431/0.437016/0.827465,-0.907981/-1.04744/-0.0169745/-1.31429,-1.36197/-1.57116/-0.470965/-1.83801,0.891007/-0.266849/1.78201/-0.533698,0.891007/0.542168/1.78201/0.275319,0.437016/-0.790569/1.32802/-1.05742,0.437016/0.0184476/1.32802/-0.248401,-0.0169745/-1.31429/0.874032/-1.58114,1.78201/-0.533698/2.67302/-0.800547,1.32802/-1.05742/2.21903/-1.32427,0./0./0./0.809017,0./0.809017/0./1.61803,0./1.61803/0./2.42705,-0.45399/-0.52372/-0.45399/0.285296,-0.45399/0.285296/-0.45399/1.09431,-0.45399/1.09431/-0.45399/1.90333,-0.907981/-1.04744/-0.907981/-0.238424,0.891007/-0.266849/0.891007/0.542168,0.891007/0.542168/0.891007/1.35119,0.891007/1.35119/0.891007/2.1602,0.437016/-0.790569/0.437016/0.0184476,0.437016/0.0184476/0.437016/0.827465,1.78201/-0.533698/1.78201/0.275319,1.32802/-1.05742/1.32802/-0.248401}
		{
			\draw[thick, postaction=decorate, decoration={markings, mark= at position 0.7 with {\arrow{stealth}}}] (\a,\b) -- (\c,\d);
		}
		\foreach \a/\b/\c in {0./0./col4,0./0.809017/col4,0./1.61803/col4,0./2.42705/col4,-0.45399/-0.52372/col3,-0.45399/0.285296/col3,-0.45399/1.09431/col3,-0.45399/1.90333/col3,-0.907981/-1.04744/col2,-0.907981/-0.238424/col2,-1.36197/-1.57116/col1,0.891007/-0.266849/col5,0.891007/0.542168/col5,0.891007/1.35119/col5,0.891007/2.1602/col5,0.437016/-0.790569/col4,0.437016/0.0184476/col4,0.437016/0.827465/col4,-0.0169745/-1.31429/col3,-0.470965/-1.83801/col2,1.78201/-0.533698/col6,1.78201/0.275319/col6,1.32802/-1.05742/col5,1.32802/-0.248401/col5,0.874032/-1.58114/col4,2.67302/-0.800547/col7,2.21903/-1.32427/col6}
		{
			\draw[\c] (\a,\b) circle (0.1);
		}
		%%%%%%%%%%%%%%%%%%%%%%%%%%%%%%%%%%%%%%%%%%%%%%%%%%%%%%%%%%%%%%%%%%%%%%%%%%%%%%
		\begin{scope}[shift = {(8,0)},scale=0.7]
			\draw[fill=black!15!white] (-2.03447, -2.10411) -- (-2.03447, 1.13196) -- (-0.218508, 3.22684)  -- (3.34552, 2.15945)  -- (3.34552, -1.07662) -- (1.52956, -3.1715)  -- cycle;
			\foreach \a/\b/\c/\d/\e/\f/\g/\h/\k in {0.218508/1.62726/0.218508/2.43627/1.10951/2.16943/1.10951/1.36041/col5,-1.12649/0.561369/-1.12649/1.37039/-0.235482/1.10354/-0.235482/0.29452/col3,-1.12649/1.37039/-1.12649/2.1794/-0.235482/1.91255/-0.235482/1.10354/col3,-0.235482/-0.514497/-0.235482/0.29452/0.655524/0.0276714/0.655524/-0.781346/col4,-0.235482/0.29452/-0.235482/1.10354/0.655524/0.836688/0.655524/0.0276714/col4,0.655524/-0.781346/0.655524/0.0276714/1.54653/-0.239178/1.54653/-1.04819/col5,1.54653/-1.85721/1.54653/-1.04819/2.43754/-1.31504/2.43754/-2.12406/col6,-1.58048/-0.771368/-1.58048/0.0376487/-0.689473/-0.2292/-0.689473/-1.03822/col2,0.201534/-2.11408/0.201534/-1.30507/1.09254/-1.57192/1.09254/-2.38093/col4,-2.03447/-2.10411/-2.03447/-1.29509/-1.14346/-1.56194/-1.14346/-2.37095/col1,-1.14346/-2.37095/-1.14346/-1.56194/-0.252457/-1.82879/-0.252457/-2.6378/col2,1.56351/1.07511/1.56351/1.88413/1.10951/1.36041/1.10951/0.551392/col5,1.56351/1.88413/1.56351/2.69315/1.10951/2.16943/1.10951/1.36041/col5,2.45451/-0.000753557/2.45451/0.808263/2.00052/0.284543/2.00052/-0.524474/col6,3.34552/-1.07662/3.34552/-0.267602/2.89153/-0.791323/2.89153/-1.60034/col7,0.218508/1.62726/0.218508/2.43627/-0.235482/1.91255/-0.235482/1.10354/col3,1.10951/0.551392/1.10951/1.36041/0.655524/0.836688/0.655524/0.0276714/col4,2.00052/-0.524474/2.00052/0.284543/1.54653/-0.239178/1.54653/-1.04819/col5,2.89153/-1.60034/2.89153/-0.791323/2.43754/-1.31504/2.43754/-2.12406/col6,-0.235482/-0.514497/-0.235482/0.29452/-0.689473/-0.2292/-0.689473/-1.03822/col2,1.54653/-1.85721/1.54653/-1.04819/1.09254/-1.57192/1.09254/-2.38093/col4,0.201534/-2.11408/0.201534/-1.30507/-0.252457/-1.82879/-0.252457/-2.6378/col2,-0.218508/3.22684/0.672499/2.96/0.218508/2.43627/-0.672499/2.70312/col4,0.672499/2.96/1.56351/2.69315/1.10951/2.16943/0.218508/2.43627/col5,1.56351/1.07511/2.45451/0.808263/2.00052/0.284543/1.10951/0.551392/col6,2.45451/-0.000753557/3.34552/-0.267602/2.89153/-0.791323/2.00052/-0.524474/col7,-0.672499/2.70312/0.218508/2.43627/-0.235482/1.91255/-1.12649/2.1794/col3,0.218508/1.62726/1.10951/1.36041/0.655524/0.836688/-0.235482/1.10354/col4,1.10951/0.551392/2.00052/0.284543/1.54653/-0.239178/0.655524/0.0276714/col5,2.00052/-0.524474/2.89153/-0.791323/2.43754/-1.31504/1.54653/-1.04819/col6,-1.12649/0.561369/-0.235482/0.29452/-0.689473/-0.2292/-1.58048/0.0376487/col2,-0.235482/-0.514497/0.655524/-0.781346/0.201534/-1.30507/-0.689473/-1.03822/col3,0.655524/-0.781346/1.54653/-1.04819/1.09254/-1.57192/0.201534/-1.30507/col4,-1.58048/-0.771368/-0.689473/-1.03822/-1.14346/-1.56194/-2.03447/-1.29509/col1,-0.689473/-1.03822/0.201534/-1.30507/-0.252457/-1.82879/-1.14346/-1.56194/col2}
			{
				\draw[\k, rounded corners=1] (\a,\b) -- (\c,\d) -- (\e,\f) -- (\g,\h) -- cycle;
			}
			\draw[ultra thick] (-2.03447, -2.10411) -- (-2.03447, 1.13196) (-2.03447, 1.13196) -- (-0.218508, 3.22684) (-0.218508, 3.22684) -- (3.34552, 2.15945) (3.34552, 2.15945) -- (3.34552, -1.07662) node[pos=0.5, right] {\scriptsize $h$ cells} (3.34552, -1.07662) -- (1.52956, -3.1715) node[pos=0.5, below right] {\scriptsize $u$ cells} (1.52956, -3.1715) -- (-2.03447, -2.10411) node[pos=0.5, below left] {\scriptsize $n+1-u$ cells} (1.52956, 0.0645665) -- (-2.03447, 1.13196) (1.52956, 0.0645665) -- (3.34552, 2.15945) (1.52956, 0.0645665) -- (1.52956, -3.1715);
		\end{scope}
		%%%%%%%%%%%%%%%%%%%%%%%%%%%%%%%%%%%%%%%%%%%%%%%%%%%%%%%%%%%%%%%%%%%%%%%
		\draw[thick, -stealth] (3,0) -- (5,0);
		\node at (4,3) {$\myA_n: \begin{array}{c}
				\begin{tikzpicture}[scale=0.7]
					\draw[thick] (0,0) -- (6,0);
					\foreach \a/\b in {col1/0,col2/1,col3/2,col4/3,col5/4,col6/5,col7/6}
					{
						\draw[\a] (\b,0) circle(0.15);
					}
				\end{tikzpicture}
			\end{array}$};
	\end{tikzpicture}
	\caption{\emph{Box-bounded} 3d Young diagrams (plane partitions) count fixed points of $\fQ_{n,u,h}$ quiver theory in the cyclic chamber.}\label{fig:boxedYoung}
\end{figure}

The structure standing behind the boxed 3d Young diagrams is absolutely the same as the one behind unrestricted colored 3d Young diagrams counting fixed points on quiver varieties associated to affine Dynkin diagrams of $\widehat{\fg\fl}_{n+1}$ (see e.g. \cite[Sec.~3.3]{Galakhov:2020vyb}).
In practice, affine quiver $\hat\myA_n$ is obtained from $\hat\myA_1$ by the McKay procedure \cite{ito2000mckay}, and the colored 3d Young diagrams are analogous to colorless ones, where diagrams depict ideals in the quiver path algebra spanned by operators $A^xB^yC^zR$.
The only effect of a transition from $\hat\myA_1$ to $\hat\myA_n$ in the above picture is captured by the subscripts of fields $A$, $B$, $C$, $R$ and by respective coloring.
When passing from $\hat\myA_n$ to $\myA_n$ we should break one of the edges in the Dynkin diagram, respectively delete fields $A_n$, $B_n$, then in monomials $A^xB^yC^zR$ we could act by $A$ and by $B$ as many times as this action does not bring us beyond nodes $1$ and $n$.
This restriction induces side walls of the restriction box in Fig.~\ref{fig:boxedYoung}.
And the marked term in \eqref{quiver} induces an ideal $C^hR=0$ in the path algebra that corresponds to the restriction box roof.

As an example let us consider the following case:
\begin{equation}\label{su(3), [2]}
	\fQ_{\fs\fl_3,[2,2]}=\left\{\begin{array}{c}
		\begin{tikzpicture}
			%%%%%%%%%%%%%%%%%%%%%%%%%%%%%%%%
			\begin{scope}[rotate=-90]
				\draw[postaction=decorate, decoration={markings, mark= at position 0.7 with {\arrow{stealth}}}] (0,-1.2) to[out=100,in=260] node[pos=0.3, above] {$\scriptstyle R$} (0,0);
				\draw[postaction=decorate, decoration={markings, mark= at position 0.7 with {\arrow{stealth}}}] (0,0) to[out=280,in=80] node[pos=0.7, below] {$\scriptstyle S$} (0,-1.2);
				\begin{scope}[shift={(0,-1.2)}]
					\draw[fill=gray] (-0.08,-0.08) -- (-0.08,0.08) -- (0.08,0.08) -- (0.08,-0.08) -- cycle;
				\end{scope}
			\end{scope}
			%%%%%%%%%%%%%%%%%%%%%%%%%%%%%%%%%%%%%%%%%%%%%%%%%%%%5
			\draw[postaction=decorate, decoration={markings, mark= at position 0.7 with {\arrow{stealth}}}] (0,0) to[out=20,in=160] node[pos=0.5,above] {$\scriptstyle A_1$} (1.5,0);
			\draw[postaction=decorate, decoration={markings, mark= at position 0.7 with {\arrow{stealth}}}] (1.5,0) to[out=200,in=340] node[pos=0.5,below] {$\scriptstyle B_1$} (0,0);
			\draw[postaction=decorate, decoration={markings, mark= at position 0.8 with {\arrow{stealth}}}] (0,0) to[out=60,in=0] (0,0.6) to[out=180,in=120] (0,0);
			\node[above] at (0,0.6) {$\scriptstyle C_1$};
			\begin{scope}[shift={(1.5,0)}]
				\draw[postaction=decorate, decoration={markings, mark= at position 0.8 with {\arrow{stealth}}}] (0,0) to[out=60,in=0] (0,0.6) to[out=180,in=120] (0,0);
				\node[above] at (0,0.6) {$\scriptstyle C_{2}$};
			\end{scope}
			\draw[fill=black!40!red] (0,0) circle (0.08);
			\draw[fill=black!40!green] (1.5,0) circle (0.08);
			%%%%%%%%%%%%%%%%%%%%%%%%%%%%%%%%%%%%%%%%%%%%%%%%%%%%%%%%%%%%%%%
			\node[below] at (0,-0.08) {$\scriptstyle \zeta_1$};
			\node[below] at (1.5,-0.08) {$\scriptstyle \zeta_2$};
		\end{tikzpicture}
	\end{array},\quad W=\Tr\left[A_1C_1B_1-B_{1}C_{2}A_{1}+{\color{burgundy}S C_1^{2}R}\right]\right\}\,.
\end{equation}
In this case the solution number generating function reads:
\begin{equation}
{\bf DT}_{\fs\fl_3,[2,2]}=1+q_1+q_1^2+q_1 q_2+q_1^2q_2 +q_1^2q_2^2\,,
\end{equation}
and it represents a generating function of plane partitions in a $1\times 2\times 2$ box, so that actually these are ordinary partitions in a $2\times 2$ box.
In the light of the correspondence Table \ref{thetable} we should compare these fixed points to vectors of irrep $[2,2]$ of $\fs\fu_3$.
This is rather simple since weights do not have multiplicities so that for each weight there is a single vector.
In this case the comparison is straightforward, we present it in Table \ref{tab2}. 
To quiver dimensions $d_a$ we associate a vector of type $e_2^{d_2}e_1^{d_1}|0\rangle$ constructed in the Verma module of the lowest weight vector $|0\rangle$.

\begin{table}[ht!]
	\centering
	$\begin{array}{c|c|c|c|c}
		\begin{array}{c} \mbox{Vector in}\\ \mbox{Verma module} \end{array} & \begin{array}{c} \mbox{Weight}\\ \mbox{Dynkin}\\ \mbox{labels} \end{array}  & \begin{array}{c} \mbox{Generating}\\ \mbox{function}\\ \mbox{weight} \end{array}  & \begin{array}{c} \mbox{Boxed}\\ \mbox{Young}\\ \mbox{diagram} \end{array} & \begin{array}{c} \mbox{Atomic}\\ \mbox{picture}\\ \mbox{plot} \end{array}\\
		\hline
		|0\rangle & (-2,0) & 1 & \varnothing & \begin{array}{c}
			\begin{tikzpicture}
				\begin{scope}[shift={(-0.8,0)}]
					\draw[fill=gray] (-0.08,-0.08) -- (-0.08,0.08) -- (0.08,0.08) -- (0.08,-0.08) -- cycle;
				\end{scope}
			\end{tikzpicture}
		\end{array}\\
		\hline
		e_1|0\rangle & (0,-1) & q_1 & \begin{array}{c}
			\begin{tikzpicture}[scale=0.2]
				\foreach \a/\b/\c/\d in {0/0/1/0, 0/0/0/1, 1/0/1/1, 0/1/1/1}
				{\draw[thick] (\a,\b) -- (\c,\d);}
			\end{tikzpicture}
		\end{array} & \begin{array}{c}
			\begin{tikzpicture}
				\draw[postaction=decorate, decoration={markings, mark= at position 0.7 with {\arrow{stealth}}}] (-0.8,0) -- (0,0) node[pos=0.5,below] {$\scriptstyle R$};
				\draw[fill=black!40!red] (0,0) circle (0.08);
				\begin{scope}[shift={(-0.8,0)}]
					\draw[fill=gray] (-0.08,-0.08) -- (-0.08,0.08) -- (0.08,0.08) -- (0.08,-0.08) -- cycle;
				\end{scope}
			\end{tikzpicture}
		\end{array}\\
		\hline
		e_1^2|0\rangle & (2,-2) & q_1^2 & \begin{array}{c}
			\begin{tikzpicture}[scale=0.2]
				\foreach \a/\b/\c/\d in {0/0/1/0, 0/0/0/2, 1/0/1/2, 0/1/1/1, 0/2/1/2}
				{\draw[thick] (\a,\b) -- (\c,\d);}
			\end{tikzpicture}
		\end{array} & \begin{array}{c}
			\begin{tikzpicture}
				\draw[postaction=decorate, decoration={markings, mark= at position 0.7 with {\arrow{stealth}}}] (-0.8,0) -- (0,0) node[pos=0.5,below] {$\scriptstyle R$};
				\draw[postaction=decorate, decoration={markings, mark= at position 0.7 with {\arrow{stealth}}}] (0,0) -- (0,0.8) node[pos=0.5,left] {$\scriptstyle C_1$};
				\draw[fill=black!40!red] (0,0) circle (0.08) (0,0.8) circle (0.08);
				\begin{scope}[shift={(-0.8,0)}]
					\draw[fill=gray] (-0.08,-0.08) -- (-0.08,0.08) -- (0.08,0.08) -- (0.08,-0.08) -- cycle;
				\end{scope}
			\end{tikzpicture}
		\end{array}\\
		\hline
		e_2e_1|0\rangle & (-1,1) & q_1q_2 & \begin{array}{c}
			\begin{tikzpicture}[scale=0.2,rotate=90]
				\foreach \a/\b/\c/\d in {0/0/1/0, 0/0/0/2, 1/0/1/2, 0/1/1/1, 0/2/1/2}
				{\draw[thick] (\a,\b) -- (\c,\d);}
			\end{tikzpicture}
		\end{array} & \begin{array}{c}
			\begin{tikzpicture}
				\draw[postaction=decorate, decoration={markings, mark= at position 0.7 with {\arrow{stealth}}}] (-0.8,0) -- (0,0) node[pos=0.5,below] {$\scriptstyle R$};
				\draw[postaction=decorate, decoration={markings, mark= at position 0.7 with {\arrow{stealth}}}] (0,0) -- (0.8,0) node[pos=0.5,above] {$\scriptstyle A_1$};
				\draw[fill=black!40!red] (0,0) circle (0.08);
				\draw[fill=black!40!green] (0.8,0) circle (0.08);
				\begin{scope}[shift={(-0.8,0)}]
					\draw[fill=gray] (-0.08,-0.08) -- (-0.08,0.08) -- (0.08,0.08) -- (0.08,-0.08) -- cycle;
				\end{scope}
			\end{tikzpicture}
		\end{array}\\
		\hline
		e_2e_1^2|0\rangle\sim e_1e_2e_1|0\rangle & (1,0) & q_1^2q_2 & \begin{array}{c}
			\begin{tikzpicture}[scale=0.2]
				\foreach \a/\b/\c/\d in {0/0/2/0, 0/1/2/1, 0/0/0/2, 1/0/1/2, 2/0/2/1, 0/2/1/2}
				{\draw[thick] (\a,\b) -- (\c,\d);}
			\end{tikzpicture}
		\end{array} &\begin{array}{c}
			\begin{tikzpicture}
				\draw[postaction=decorate, decoration={markings, mark= at position 0.7 with {\arrow{stealth}}}] (-0.8,0) -- (0,0) node[pos=0.5,below] {$\scriptstyle R$};
				\draw[postaction=decorate, decoration={markings, mark= at position 0.7 with {\arrow{stealth}}}] (0,0) -- (0,0.8) node[pos=0.5,left] {$\scriptstyle C_1$};
				\draw[postaction=decorate, decoration={markings, mark= at position 0.7 with {\arrow{stealth}}}] (0,0) -- (0.8,0) node[pos=0.5,above] {$\scriptstyle A_1$};
				\draw[fill=black!40!red] (0,0) circle (0.08) (0,0.8) circle (0.08);
				\draw[fill=black!40!green] (0.8,0) circle (0.08);
				\begin{scope}[shift={(-0.8,0)}]
					\draw[fill=gray] (-0.08,-0.08) -- (-0.08,0.08) -- (0.08,0.08) -- (0.08,-0.08) -- cycle;
				\end{scope}
			\end{tikzpicture}
		\end{array}\\
		\hline
		e_2^2e_1^2|0\rangle\sim \left(e_2e_1\right)^2|0\rangle & (0,2) & q_1^2q_2^2 & \begin{array}{c}
			\begin{tikzpicture}[scale=0.2]
				\foreach \a/\b/\c/\d in {0/0/2/0, 0/0/0/2, 0/2/2/2, 2/0/2/2, 1/0/1/2, 0/1/2/1}
				{\draw[thick] (\a,\b) -- (\c,\d);}
			\end{tikzpicture}
		\end{array} & \begin{array}{c}
			\begin{tikzpicture}
				\draw[postaction=decorate, decoration={markings, mark= at position 0.7 with {\arrow{stealth}}}] (-0.8,0) -- (0,0) node[pos=0.5,below] {$\scriptstyle R$};
				\draw[postaction=decorate, decoration={markings, mark= at position 0.7 with {\arrow{stealth}}}] (0,0) -- (0,0.8) node[pos=0.5,left] {$\scriptstyle C_1$};
				\draw[postaction=decorate, decoration={markings, mark= at position 0.7 with {\arrow{stealth}}}] (0,0) -- (0.8,0) node[pos=0.5,above] {$\scriptstyle A_1$};
				\draw[postaction=decorate, decoration={markings, mark= at position 0.7 with {\arrow{stealth}}}] (0,0.8) -- (0.8,0.8) node[pos=0.5,above] {$\scriptstyle A_1$};
				\draw[postaction=decorate, decoration={markings, mark= at position 0.7 with {\arrow{stealth}}}] (0.8,0) -- (0.8,0.8) node[pos=0.5,right] {$\scriptstyle C_2$};
				\draw[fill=black!40!red] (0,0) circle (0.08) (0,0.8) circle (0.08);
				\draw[fill=black!40!green] (0.8,0) circle (0.08) (0.8,0.8) circle (0.08);
				\begin{scope}[shift={(-0.8,0)}]
					\draw[fill=gray] (-0.08,-0.08) -- (-0.08,0.08) -- (0.08,0.08) -- (0.08,-0.08) -- cycle;
				\end{scope}
			\end{tikzpicture}
		\end{array}
	\end{array}$
	\caption{Comparing vectors with atomic structure plots.}\label{tab2}
\end{table}

%%%%%%%%%%%%%%%%%%%%%%%%%%%%%%%%%%%%%%%%%%%%%%%%%%%%%%%%%%%%%%%%%%%%%%%%%%%%%%%%%%%%%%%%%%%%%%%%%%%
%%%%%%%%%%%%%%%%%%%%%%%%%%%%%%%%%%%%%%%%%%%%%%%%%%%%%%%%%%%%%%%%%%%%%%%%%%%%%%%%%%%%%%%%%%%%%%%%%%%
%%%%%%%%%%%%%%%%%%%%%%%%%%%%%%%%%%%%%%%%%%%%%%%%%%%%%%%%%%%%%%%%%%%%%%%%%%%%%%%%%%%%%%%%%%%%%%%%%%%

\section{Weyl mutations}\label{sec:Mutations}

\subsection{Mutating quiver varieties}\label{sec:Weyl_mut}

We defined Weyl mutations as electro-magnetic self-dualities of quivers $\fQ_{n,u,h}$ generalizing Nakajima quiver dualities discussed in Sec.~\ref{sec:Nak_quiv_Weyl}, so that the dual quiver $\check\fQ_{n,u,h}=\fQ_{n,u,h}$ and the dual superpotential $\check W=W$.
However stability parameters $\zeta_a$, dimensions $d_a$ and fixed point quiver morphisms $(A_a,B_a,C_a,R,S)$ must transform accordingly into $\check\zeta_a$, $\check d_a$ and $(\check A_a,\check B_a,\check C_a,\check R,\check S)$.
As for the Weyl group Weyl mutations are generated by elementary reflections $s_a$.
Subscript $a$ runs in this case over quiver $\fQ_{n,u,h}$ nodes.

By considering various examples in App.~\ref{app:examples} we propose a \emph{hypothesis} that mutation $s_a$ is subjected to the following rules:
\begin{enumerate}
	\item Stability parameters are acted upon by the Weyl reflection:
	\begin{equation}\label{non_Nak_du_sta}
		\check \zeta_a=-\zeta_a,\quad \check\zeta_{a\pm 1}=\zeta_{a\pm 1}+\zeta_a,\quad \check\zeta_b=\zeta_b,\;\mbox{for }|a-b|>1\,.
	\end{equation}
	\item Dimensions behave as fundamental weights:
	\begin{equation}\label{non_Nak_du_dim}
		\check d_a=h\delta_{a,u}+d_{a+1}+d_{a-1}-d_a,\quad \check d_b=d_b,\;\mbox{for }b\neq a\,.
	\end{equation}
	\item Morphisms not adjacent to node $a$ are unmodified:
	\begin{equation}
		\check R=R,\;\check S=S, \;\mbox{if }u\neq a;\quad \check A_b=A_b,\;\check B_b=B_b,\;\mbox{for }b\neq a,\;b\neq a-1\,.
	\end{equation}
	\item \label{majoritem} Maps $\alpha$ and $\beta$ constructed as:
	\begin{equation}
	\begin{aligned}
		&\alpha=\left(\begin{array}{ccc}
			\check B_{a-1} & \check A_{a}
		\end{array}\right)^T,\quad \beta=\left(\begin{array}{ccc}
			-A_{a-1} & B_{a}
		\end{array}\right),\quad\mbox{if }u\neq a\,;\\
		&\alpha=\left(\begin{array}{cccccc}
			\check S & \check S\check C_u &\ldots & \check S\check C_u^{h-1} & \check B_{u-1} & \check A_{u}
		\end{array}\right)^T\,,\\
		&\beta=\left(\begin{array}{cccccc}
			C_u^{h-1}R & C_u^{h-2}R &\ldots & R&-A_{u-1} & B_{u}
		\end{array}\right),\;\mbox{if }u= a;
	\end{aligned}
	\end{equation}
	define a \emph{short exact sequence}:
	\begin{equation}
		\begin{array}{c}
			\begin{tikzpicture}
				\node (A) at (-4,0) {$0$};
				\node (B) at (-3,0) {$\check V_a$};
				\node (C) at (0,0) {$\IC^{h\delta_{u,a}}\oplus V_{a-1}\oplus V_{a+1}$};
				\node (D) at (3,0) {$V_a$};
				\node (E) at (4,0) {$0$};
				\path (A) edge[->] (B) (B) edge[->] node[above] {$\scriptstyle \alpha$} (C) (C) edge[->] node[above] {$\scriptstyle \beta$} (D) (D) edge[->] (E);
			\end{tikzpicture}
		\end{array}\,.
	\end{equation}
	\item\label{rule 5} Duality relations for all two-arrow paths passing through node $a$ read:
	\begin{equation}
		\begin{aligned}
			&A_{a}A_{a-1}=\check A_{a}\check A_{a-1},\; B_{a-1}B_{a}=\check B_{a-1}\check B_{a},\; B_{a-1}A_{a-1}=\check B_{a-1}\check A_{a-1},\; A_aB_a=\check A_a\check B_a\,,\\ 
			&\mbox{and }SA_{u-1}=\check S\check A_{u-1},\;SB_{u}=\check S\check B_{u},\;A_uR=\check A_u\check R,\;B_{u-1}R =\check B_{u-1}\check R_u,\;SR=\check S\check R\mbox{ if }a=u\,.
		\end{aligned}
	\end{equation}
\end{enumerate}

A few remarks are in order:
\begin{enumerate}
	\item All the rules are equivariant with respect to the $\Omega$-background torus action.
	\item A relation for dual dimensions \eqref{non_Nak_du_dim} acquires a correction in comparison to \eqref{Nak_du_dim}.
	For our choice of framing respective framing dimensions read $w_a=\delta_{u,a}$, so this term is rescaled by $h$ in \eqref{non_Nak_du_dim}.
	\item Traditionally quiver dimensions transform dually to spectral parameters so that a stability slope $\mu:=\sum\lm_{a=1}^n d_a\zeta_a$ \cite{donaldson1983new} remains invariant.
	Shift $h\delta_{u,a}$ in \eqref{non_Nak_du_dim} might contradict this principle at the first glance.
	However to apply the stability constraints to \emph{framed} quivers we suggest to extend rules for \emph{unframed} quivers in \cite{king1994moduli} by adding a dummy stability parameter $\zeta_0$ to the framing node.
	We believe the duality forces to transform $\zeta_0$ accordingly so that the slope is invariant.
	\item The major modification of the duality rules in comparison to Sec.\ref{sec:Nak_quiv_Weyl} is in item \ref{majoritem}.
	In Sec.\ref{sec:Nak_quiv_Weyl} we have already noted that item \ref{majoritem} is the most unusual in the framework of Seiberg-like dualities.
	When $u=a$, in other words when we mutate the framed node, there are $h$ copies of the framing vector space $\IC^{h\delta_{u,a}}$.
	Also we modify maps $\alpha$ and $\beta$ by adding dual maps $\check S \check C_u^{p}$ and $C_u^{h-1-p}R$ to this space.
	\item Let us note that the rule in item \ref{majoritem} is \emph{approximate}.
	In the case of Nakajima quiver varieties \cite{lusztig2000quiver} it was applied to resolved complex moment maps, in other words if one adds an extra correction $\sum\lm_{a=1}^n \nu_a\Tr\, C_a$ to the superpotential.
	Parameters $\nu_a$ are complex in general and behave under duality as real stability parameters \eqref{non_Nak_du_sta}, so one might call $\nu_a$ complex stability parameters.
	Universal transformation rules for 3d vectors $(\zeta_a,\nu_a)$ refer to an enhancement of an ordinary K\"ahler structure of $\CN=4$ SQM to a hyper-K\"ahler structure of $\CN=8$ SQM in the case of Nakajima quiver varieties.
	Here we work with the case $\nu_a=0$, and as we see in examples (see e.g. App.~\ref{sec:exp_sl_2}) rule \ref{majoritem} \emph{does not} work always exactly.
	Yet we are able to check how it works in examples.
	All the atomic picture plots could be \emph{checked} explicitly by substituting respective quiver morphisms into \eqref{fp_eq} numerically.
	After considering examples we suggest that to apply rule \ref{majoritem} one should decide which of two theories is $\fQ_{n,u,h}$ and which is $\check\fQ_{n,u,h}$.
	\begin{tcolorbox}
		For mutation $s_a$ and rule \ref{majoritem} theory $\fQ_{n,u,h}$ corresponds to $\zeta_a>0$, and $\check\fQ_{n,u,h}$ corresponds to $\check\zeta_a=-\zeta_a<0$.
	\end{tcolorbox}
	\item 
	It is natural to assume that mutation maps a \emph{single} solution of \eqref{fp_eq} to a single solution in a new chamber.
	As in the case of Lie algebras the Weyl group orbit of the fundamental chamber covers the whole weight space, similarly the Weyl mutation orbit of the cyclic chamber covers the whole moduli space of $\zeta_a$, and \emph{any} solution outside the cyclic chamber could be derived from a known solution in the cyclic chamber by a sequence of mutations.
	Respectively, we should stress that dimension transformation rule \eqref{non_Nak_du_dim} is such that vector weights in \eqref{weight} remain invariant as abstract vectors, so a mutation leaves generating function ${\bf DT}_{\fQ_{n,u,h}}$ \emph{invariant} and independent of values of $\zeta_a$.
\end{enumerate}

%%%%%%%%%%%%%%%%%%%%%%%%%%%%%%%%%%%%%%%%%%%%%%%%%%%%%%%%%%%%%%%%%%%%%%%%%%%%%%%%%
%%%%%%%%%%%%%%%%%%%%%%%%%%%%%%%%%%%%%%%%%%%%%%%%%%%%%%%%%%%%%%%%%%%%%%%%%%%%%%%%%
%%%%%%%%%%%%%%%%%%%%%%%%%%%%%%%%%%%%%%%%%%%%%%%%%%%%%%%%%%%%%%%%%%%%%%%%%%%%%%%%%

\subsection{Mutating atomic structure plots}\label{sec:mut_asp}

Atomic structure plots describe fixed points on quiver varieties.
The action of the Weyl group could be pulled back from rules discussed in Sec.~\ref{sec:Weyl_mut} to modify atomic structure plots.
As we have noted the quiver theory is self-dual under Weyl mutations, yet the latter acts on the stability parameters, so that the cyclic chamber could be mapped to any other chamber by a sequence of Weyl reflections.
So knowing the spectrum of fixed points in the cyclic chamber and acting on it by a Weyl reflection we are able to reconstruct spectra on the whole moduli space.

We consider numerous examples of spectra in App.~\ref{app:examples}.

Here we discuss in detail the case of $\fs\fl_4$ quiver and irrep $[2,2]$ with a quiver given by \eqref{sl(4), [2,2], quiv}.
The respective generating function for solution numbers \eqref{sl(4), [2,2], DT} has multiplier 2 in front of term $q_1q_2^2q_3$ implying that vectors with such weight contribute twice.
These are two fixed points reconstructed from respective partitions following Sec.~\ref{sec:BoxYoung}:
\begin{equation}
	v_1=\begin{array}{c}
		\begin{tikzpicture}[scale=0.5]
			\tikzset{col1/.style={fill=black!40!red}}
			\tikzset{col2/.style={fill=black!40!green}}
			\tikzset{col3/.style={fill=\myblue}}
			\foreach \a/\b/\c/\d/\e/\f/\g/\h/\k in {-1.12649/-1.05666/-1.12649/-0.247648/-0.235482/-0.514497/-0.235482/-1.32351/col1,-0.235482/-1.32351/-0.235482/-0.514497/0.655524/-0.781346/0.655524/-1.59036/col2,1.56351/-0.542922/1.56351/0.266095/1.10951/-0.257625/1.10951/-1.06664/col3,1.10951/-1.06664/1.10951/-0.257625/0.655524/-0.781346/0.655524/-1.59036/col2,-0.218508/0.799793/0.672499/0.532944/0.218508/0.00922379/-0.672499/0.276073/col2,0.672499/0.532944/1.56351/0.266095/1.10951/-0.257625/0.218508/0.00922379/col3,-0.672499/0.276073/0.218508/0.00922379/-0.235482/-0.514497/-1.12649/-0.247648/col1,0.218508/0.00922379/1.10951/-0.257625/0.655524/-0.781346/-0.235482/-0.514497/col2}
			{
				\draw[\k, rounded corners=1] (\a,\b) -- (\c,\d) -- (\e,\f) -- (\g,\h) -- cycle;
			}
		\end{tikzpicture}
	\end{array}=\begin{array}{c}
	\begin{tikzpicture}
		\tikzset{col1/.style={fill=black!40!red}}
		\tikzset{col2/.style={fill=black!40!green}}
		\tikzset{col3/.style={fill=\myblue}}
		\draw[postaction=decorate, decoration={markings, mark= at position 0.7 with {\arrow{stealth}}}] (-1,0.5) -- (0,0) node[pos=0.3,above right] {$\scriptstyle R$};
		\draw[postaction=decorate, decoration={markings, mark= at position 0.7 with {\arrow{stealth}}}] (0.,0.) -- (-0.45399,-0.52372) node[pos=0.5,above left] {$\scriptstyle B_1$};
		\draw[postaction=decorate, decoration={markings, mark= at position 0.7 with {\arrow{stealth}}}] (0.891007,-0.266849) -- (0.437016,-0.790569)  node[pos=0.5,below right] {$\scriptstyle B_2$};
		\draw[postaction=decorate, decoration={markings, mark= at position 0.7 with {\arrow{stealth}}}] (0.,0.) -- (0.891007,-0.266849) node[pos=0.5,above right] {$\scriptstyle A_2$};
		\draw[postaction=decorate, decoration={markings, mark= at position 0.7 with {\arrow{stealth}}}] (-0.45399,-0.52372) -- (0.437016,-0.790569) node[pos=0.5,below left] {$\scriptstyle A_1$};
		\foreach \a/\b/\c in {0./0./col2,-0.45399/-0.52372/col1,0.891007/-0.266849/col3,0.437016/-0.790569/col2}
		{
			\draw[\c] (\a,\b) circle (0.1);
		}
		\begin{scope}[shift={(-1,0.5)}]
			\draw[fill=gray] (-0.1,-0.1) -- (-0.1,0.1) -- (0.1,0.1) -- (0.1,-0.1) -- cycle;
		\end{scope}
	\end{tikzpicture}
	\end{array},\quad v_2=\begin{array}{c}
	\begin{tikzpicture}[scale=0.5]
		\tikzset{col1/.style={fill=black!40!red}}
		\tikzset{col2/.style={fill=black!40!green}}
		\tikzset{col3/.style={fill=\myblue}}
		\foreach \a/\b/\c/\d/\e/\f/\g/\h/\k in {-0.672499/0.276073/-0.672499/1.08509/0.218508/0.818241/0.218508/0.00922379/col2,0.218508/-0.799793/0.218508/0.00922379/1.10951/-0.257625/1.10951/-1.06664/col3,-1.12649/-1.05666/-1.12649/-0.247648/-0.235482/-0.514497/-0.235482/-1.32351/col1,0.672499/0.532944/0.672499/1.34196/0.218508/0.818241/0.218508/0.00922379/col2,1.56351/-0.542922/1.56351/0.266095/1.10951/-0.257625/1.10951/-1.06664/col3,0.218508/-0.799793/0.218508/0.00922379/-0.235482/-0.514497/-0.235482/-1.32351/col1,-0.218508/1.60881/0.672499/1.34196/0.218508/0.818241/-0.672499/1.08509/col2,0.672499/0.532944/1.56351/0.266095/1.10951/-0.257625/0.218508/0.00922379/col3,-0.672499/0.276073/0.218508/0.00922379/-0.235482/-0.514497/-1.12649/-0.247648/col1}
		{
			\draw[\k, rounded corners=1] (\a,\b) -- (\c,\d) -- (\e,\f) -- (\g,\h) -- cycle;
		}
	\end{tikzpicture}
	\end{array}=\begin{array}{c}
	\begin{tikzpicture}
		\tikzset{col1/.style={fill=black!40!red}}
		\tikzset{col2/.style={fill=black!40!green}}
		\tikzset{col3/.style={fill=\myblue}}
		\draw[postaction=decorate, decoration={markings, mark= at position 0.7 with {\arrow{stealth}}}] (-1,0.5) -- (0,0) node[pos=0.3,above right] {$\scriptstyle R$};
		\draw[postaction=decorate, decoration={markings, mark= at position 0.7 with {\arrow{stealth}}}] (0.,0.) -- (-0.45399,-0.52372) node[pos=0.5,above left] {$\scriptstyle B_1$};
		\draw[postaction=decorate, decoration={markings, mark= at position 0.7 with {\arrow{stealth}}}] (0.,0.) -- (0.891007,-0.266849) node[pos=0.8,below left] {$\scriptstyle A_1$};
		\draw[postaction=decorate, decoration={markings, mark= at position 0.7 with {\arrow{stealth}}}] (0.,0.) -- (0.,0.809017) node[pos=0.5,right] {$\scriptstyle C_1$};
		\foreach \a/\b/\c in {0./0./col2,0./0.809017/col2,-0.45399/-0.52372/col1,0.891007/-0.266849/col3}
		{
			\draw[\c] (\a,\b) circle (0.1);
		}
		\begin{scope}[shift={(-1,0.5)}]
			\draw[fill=gray] (-0.1,-0.1) -- (-0.1,0.1) -- (0.1,0.1) -- (0.1,-0.1) -- cycle;
		\end{scope}
	\end{tikzpicture}
	\end{array}\,.
\end{equation}

Using these atomic picture plots we can easily describe an ansatz for \eqref{fp_eq}.
For example, for $v_1$ we have:
\begin{equation}
	\begin{aligned}
	&v_1:\;
	C_1=\left(
	\begin{array}{c}
		0 \\
	\end{array}
	\right),\;C_2=\left(
	\begin{array}{cc}
		0 & 0 \\
		0 & 0 \\
	\end{array}
	\right),\;C_3=\left(
	\begin{array}{c}
		0 \\
	\end{array}
	\right),\; A_1=\left(
	\begin{array}{c}
		0 \\
		x_1 \\
	\end{array}
	\right),\; B_1=\left(
	\begin{array}{cc}
		x_2 & 0 \\
	\end{array}
	\right)\,,\\
	& A_2=\left(
	\begin{array}{cc}
		x_3 & 0 \\
	\end{array}
	\right),\;B_2=\left(
	\begin{array}{c}
		0 \\
		x_4 \\
	\end{array}
	\right),\;
	R=\left(
	\begin{array}{c}
		x_5 \\
		0 \\
	\end{array}
	\right),\;S=\left(
	\begin{array}{cc}
		0 & 0 \\
	\end{array}
	\right)\,,\\
	& \Phi_1=\left(\begin{array}{c}
		\mu+\epsilon_2
	\end{array}\right),\quad \Phi_2=\left(\begin{array}{cc}
	\mu & 0\\ 0 & \mu+\epsilon_1+\epsilon_2
	\end{array}\right),\quad \Phi_3=\left(\begin{array}{c}
	\mu+\epsilon_1
	\end{array}\right)\,.
	\end{aligned}
\end{equation} 
By substituting these expressions to \eqref{fp_eq} we arrive to the following system of equations for $x_i$:
\begin{equation}
	x_3 x_4-x_1 x_2=0,\;-x_1^2+x_2^2=\zeta _1,\;x_1^2+x_4^2=\zeta _2,\;-x_2^2-x_3^2+x_5^2=\zeta _2,\;x_3^2-x_4^2=\zeta _3\,.
\end{equation}
It is easy to present a solution to $x_i$ when all $x_i>0$ and respectively an element of the $G_{\IR}(\fQ)$-orbit (solutions with other sign choices lie in the same $G_{\IR}(\fQ)$-orbit):
\begin{equation}
	\begin{aligned}
	&x_1=\sqrt{ \frac{\zeta _2 \left(\zeta _2+\zeta _3\right)}{\zeta _1+2 \zeta _2+\zeta _3}},\; x_2=\sqrt{ \frac{\left(\zeta _1+\zeta _2\right) \left(\zeta _1+\zeta _2+\zeta _3\right)}{\zeta _1+2 \zeta _2+\zeta _3}},\; x_3=\sqrt{ \frac{\left(\zeta _2+\zeta _3\right) \left(\zeta _1+\zeta _2+\zeta _3\right)}{\zeta _1+2 \zeta _2+\zeta _3}}\,,\\ &x_4=\sqrt{ \frac{\zeta _2 \left(\zeta _1+\zeta _2\right)}{\zeta _1+2 \zeta _2+\zeta _3}},\; x_5=\sqrt{ \zeta _1+2 \zeta _2+\zeta _3}\,.
	\end{aligned}
\end{equation}
Similarly, for $v_2$ we derive:
\begin{equation}
	\begin{aligned}
		&v_2:\;
		C_1=\left(
		\begin{array}{c}
			0 \\
		\end{array}
		\right),\;C_2=\left(
		\begin{array}{cc}
			0 & 0 \\
			x_1 & 0 \\
		\end{array}
		\right),\;C_3=\left(
		\begin{array}{c}
			0 \\
		\end{array}
		\right),\; A_1=\left(
		\begin{array}{c}
			0 \\
			0 \\
		\end{array}
		\right),\; B_1=\left(
		\begin{array}{cc}
			x_2 & 0 \\
		\end{array}
		\right)\,,\\
		& A_2=\left(
		\begin{array}{cc}
			x_3 & 0 \\
		\end{array}
		\right),\;B_2=\left(
		\begin{array}{c}
			0 \\
			0\\
		\end{array}
		\right),\;
		R=\left(
		\begin{array}{c}
			x_4 \\
			0 \\
		\end{array}
		\right),\;S=\left(
		\begin{array}{cc}
			0 & 0 \\
		\end{array}
		\right)\,,\\
		& \Phi_1=\left(\begin{array}{c}
			\mu+\epsilon_2
		\end{array}\right),\quad \Phi_2=\left(\begin{array}{cc}
			\mu & 0\\ 0 & \mu-\epsilon_1-\epsilon_2
		\end{array}\right),\quad \Phi_3=\left(\begin{array}{c}
			\mu+\epsilon_1
		\end{array}\right)\,, 
	\end{aligned}
\end{equation} 
with $x_i$ taking the following values on the $G_{\IR}(\fQ)$-orbit:
\begin{equation}
	x_1=\sqrt{ \zeta _2},\quad x_2=\sqrt{ \zeta _1},\quad x_3=\sqrt{ \zeta _3},\quad x_4=\sqrt{ \zeta _1+2 \zeta _2+\zeta _3}\,.
\end{equation}

We would like to describe an action of mutation $s_2$ on these solutions.
According to the $\fs\fl_4$ quiver moduli space map depicted in Fig.~\ref{fig:su_4_moduli} mutation $s_2$ maps the cyclic chamber ${\bf Ph}_1$ ($\zeta_a>0$) to ${\bf Ph}_3$ following \eqref{non_Nak_du_sta}:
\begin{equation}
	\left(\check \zeta_1,\check \zeta_2,\check \zeta_3\right)=(\zeta_1+\zeta_2,-\zeta_2,\zeta_2+\zeta_3)\,.
\end{equation}
For mutated dimensions we find following \eqref{non_Nak_du_dim}:
\begin{equation}
	\left(\check d_1,\check d_2,\check d_3\right)=(1,2+1+1-2,1)=(1,2,1)\,.
\end{equation}
So the number of atoms and their coloring do not change in both plots under a mutation.

Vector spaces $V_1$ and $V_3$ are not modified during a mutation, these spaces are represented in the atomic picture plots by red and blue atoms, so these atoms remain intact.
According to rule \ref{rule 5} in the Sec.~\ref{sec:Weyl_mut} list $A_2R=\check A_2\check R$, $B_1R=\check B_1\check R$, these maps are represented by the shortest paths connecting the gray framing atom with blue and red atoms, so these paths remain invariant preserving the green atom they pass through.
Finally, the position of the remaining green atom is modified.
To determine a new position one should use rule \ref{majoritem}.

According to rule \ref{majoritem} to constrain maps $\check\fQ$ we should find possible $\alpha$ as a null space of $\beta$ (for this calculation we could neglect the real moment map and choose all $x_i=1$ form the $G_{\IC}(\fQ)$-orbit):
\begin{equation}\label{abv_1}
	\beta(v_1)=\left(
	\begin{array}{cccc}
		0 & 1 & 0 & 0 \\
		0 & 0 & -1 & 1 \\
	\end{array}
	\right)\;\Longrightarrow \; \alpha=\left(
	\begin{array}{cc}
		0 & {\bf\color{black!40!red}1} \\
		0 & 0 \\
		{\bf\color{black!40!green}1} & 0 \\
		{\bf\color{black!40!green}1} & 0 \\
	\end{array}
	\right)\,.
\end{equation}
The term marked by red color corresponds to the image of $\check S$, so we conclude that in the mutated plot the green atom is connected to the framing gray atom by an $\check S$ arrow:
\begin{equation}
	\check v_1=s_{\bf\color{black!40!green}2}(v_1)=s_{\bf\color{black!40!green}2}\left(\begin{array}{c}
		\begin{tikzpicture}
			\tikzset{col1/.style={fill=black!40!red}}
			\tikzset{col2/.style={fill=black!40!green}}
			\tikzset{col3/.style={fill=\myblue}}
			\draw[postaction=decorate, decoration={markings, mark= at position 0.7 with {\arrow{stealth}}}] (-1,0.5) -- (0,0) node[pos=0.3,above right] {$\scriptstyle R$};
			\draw[postaction=decorate, decoration={markings, mark= at position 0.7 with {\arrow{stealth}}}] (0.,0.) -- (-0.45399,-0.52372) node[pos=0.5,above left] {$\scriptstyle B_1$};
			\draw[postaction=decorate, decoration={markings, mark= at position 0.7 with {\arrow{stealth}}}] (0.891007,-0.266849) -- (0.437016,-0.790569)  node[pos=0.5,below right] {$\scriptstyle B_2$};
			\draw[postaction=decorate, decoration={markings, mark= at position 0.7 with {\arrow{stealth}}}] (0.,0.) -- (0.891007,-0.266849) node[pos=0.5,above right] {$\scriptstyle A_2$};
			\draw[postaction=decorate, decoration={markings, mark= at position 0.7 with {\arrow{stealth}}}] (-0.45399,-0.52372) -- (0.437016,-0.790569) node[pos=0.5,below left] {$\scriptstyle A_1$};
			\foreach \a/\b/\c in {0./0./col2,-0.45399/-0.52372/col1,0.891007/-0.266849/col3,0.437016/-0.790569/col2}
			{
				\draw[\c] (\a,\b) circle (0.1);
			}
			\begin{scope}[shift={(-1,0.5)}]
				\draw[fill=gray] (-0.1,-0.1) -- (-0.1,0.1) -- (0.1,0.1) -- (0.1,-0.1) -- cycle;
			\end{scope}
		\end{tikzpicture}
	\end{array}\right)=\begin{array}{c}
		\begin{tikzpicture}
			\tikzset{col1/.style={fill=black!40!red}}
			\tikzset{col2/.style={fill=black!40!green}}
			\tikzset{col3/.style={fill=\myblue}}
			\draw[postaction=decorate, decoration={markings, mark= at position 0.7 with {\arrow{stealth}}}] (-1,0.5) -- (0,0) node[pos=0.3,above right] {$\scriptstyle R$};
			\draw[postaction=decorate, decoration={markings, mark= at position 0.7 with {\arrow{stealth}}}] (0.,0.) -- (-0.45399,-0.52372) node[pos=0.5,above left] {$\scriptstyle B_1$};
			\draw[postaction=decorate, decoration={markings, mark= at position 0.7 with {\arrow{stealth}}}] (0.,0.) -- (0.891007,-0.266849) node[pos=0.8,below left] {$\scriptstyle A_1$};
			\draw[postaction=decorate, decoration={markings, mark= at position 0.7 with {\arrow{stealth}}}] (-2,1) -- (-1,0.5) node[pos=0.5,above right] {$\scriptstyle S$};
			\foreach \a/\b/\c in {0./0./col2,-2/1/col2,-0.45399/-0.52372/col1,0.891007/-0.266849/col3}
			{
				\draw[\c] (\a,\b) circle (0.1);
			}
			\begin{scope}[shift={(-1,0.5)}]
				\draw[fill=gray] (-0.1,-0.1) -- (-0.1,0.1) -- (0.1,0.1) -- (0.1,-0.1) -- cycle;
			\end{scope}
		\end{tikzpicture}
	\end{array}\,.
\end{equation}
The respective ansatz for \eqref{fp_eq} reads:
\begin{equation}
	\begin{aligned}
		&\check v_1:\;
		C_1=\left(
		\begin{array}{c}
			0 \\
		\end{array}
		\right),\;C_2=\left(
		\begin{array}{cc}
			0 & 0 \\
			0 & 0 \\
		\end{array}
		\right),\;C_3=\left(
		\begin{array}{c}
			0 \\
		\end{array}
		\right),\; A_1=\left(
		\begin{array}{c}
			0 \\
			0 \\
		\end{array}
		\right),\; B_1=\left(
		\begin{array}{cc}
			x_1 & 0 \\
		\end{array}
		\right)\,,\\
		& A_2=\left(
		\begin{array}{cc}
			x_2 & 0 \\
		\end{array}
		\right),\;B_2=\left(
		\begin{array}{c}
			0 \\
			0\\
		\end{array}
		\right),\;
		R=\left(
		\begin{array}{c}
			x_3 \\
			0 \\
		\end{array}
		\right),\;S=\left(
		\begin{array}{cc}
			0 & x_4 \\
		\end{array}
		\right)\,,\\
		& \Phi_1=\left(\begin{array}{c}
			\mu+\epsilon_2
		\end{array}\right),\quad \Phi_2=\left(\begin{array}{cc}
			\mu & 0\\ 0 & \mu+h\left(\epsilon_1+\epsilon_2\right)
		\end{array}\right),\quad \Phi_3=\left(\begin{array}{c}
			\mu+\epsilon_1
		\end{array}\right)\,.
	\end{aligned}
\end{equation} 
And a specification to $G_{\IR}(\fQ)$ orbit reads:
\begin{equation}
	x_1=\sqrt{ \zeta _1},\quad x_2=\sqrt{ \zeta _3},\quad x_3=\sqrt{ \zeta _1+\zeta _2+\zeta _3},\quad x_4=\sqrt{ -\zeta _2}\,.
\end{equation}

For $v_2$ we find:
\begin{equation}\label{alpha}
	\beta(v_2)=\left(
	\begin{array}{cccc}
		0 & 1 & 0 & 0 \\
		1 & 0 & 0 & 0 \\
	\end{array}
	\right)\;\Longrightarrow\;\alpha=\left(
	\begin{array}{cc}
		0 & 0 \\
		0 & 0 \\
		{\bf\color{black!40!green}1} & {\bf\color{black!40!red}1} \\
		{\bf\color{black!40!green}1} & {\bf\color{black!40!red}-1} \\
	\end{array}
	\right)\,.
\end{equation}
We were searching for $\alpha$ as a matrix constructed of null vectors of $\beta$ and this particular choice of $\alpha$ is not a minimal choice in the null vector space, however we should note that the terms marked by green color should coincide with those in \eqref{abv_1} as they correspond to paths $B_1R$ and $A_2R$ preserved by the mutation.
Looking at the remaining terms marked by red color we note that the mutated green atom should be connected to red and blue atom by $B_1$ and $A_2$ arrows:
\begin{equation}
	\check v_2=s_{\bf\color{black!40!green}2}(v_2)=s_{\bf\color{black!40!green}2}\left(\begin{array}{c}
		\begin{tikzpicture}
			\tikzset{col1/.style={fill=black!40!red}}
			\tikzset{col2/.style={fill=black!40!green}}
			\tikzset{col3/.style={fill=\myblue}}
			\draw[postaction=decorate, decoration={markings, mark= at position 0.7 with {\arrow{stealth}}}] (-1,0.5) -- (0,0) node[pos=0.3,above right] {$\scriptstyle R$};
			\draw[postaction=decorate, decoration={markings, mark= at position 0.7 with {\arrow{stealth}}}] (0.,0.) -- (-0.45399,-0.52372) node[pos=0.5,above left] {$\scriptstyle B_1$};
			\draw[postaction=decorate, decoration={markings, mark= at position 0.7 with {\arrow{stealth}}}] (0.,0.) -- (0.891007,-0.266849) node[pos=0.8,below left] {$\scriptstyle A_1$};
			\draw[postaction=decorate, decoration={markings, mark= at position 0.7 with {\arrow{stealth}}}] (0.,0.) -- (0.,0.809017) node[pos=0.5,right] {$\scriptstyle C_1$};
			\foreach \a/\b/\c in {0./0./col2,0./0.809017/col2,-0.45399/-0.52372/col1,0.891007/-0.266849/col3}
			{
				\draw[\c] (\a,\b) circle (0.1);
			}
			\begin{scope}[shift={(-1,0.5)}]
				\draw[fill=gray] (-0.1,-0.1) -- (-0.1,0.1) -- (0.1,0.1) -- (0.1,-0.1) -- cycle;
			\end{scope}
		\end{tikzpicture}
	\end{array}\right)=\begin{array}{c}
		\begin{tikzpicture}
			\tikzset{col1/.style={fill=black!40!red}}
			\tikzset{col2/.style={fill=black!40!green}}
			\tikzset{col3/.style={fill=\myblue}}
			\draw[postaction=decorate, decoration={markings, mark= at position 0.7 with {\arrow{stealth}}}] (-1,0.5) -- (0,0) node[pos=0.3,above right] {$\scriptstyle R$};
			\draw[postaction=decorate, decoration={markings, mark= at position 0.7 with {\arrow{stealth}}}] (0.,0.) -- (-0.45399,-0.52372) node[pos=0.5,above left] {$\scriptstyle B_1$};
			\draw[thick, white!50!violet, postaction=decorate, decoration={markings, mark= at position 0.3 with {\arrowreversed{stealth}}}] (0.891007,-0.266849) -- (0.437016,-0.790569)  node[pos=0.5,below right, black] {$\scriptstyle A_2$};
			\draw[postaction=decorate, decoration={markings, mark= at position 0.7 with {\arrow{stealth}}}] (0.,0.) -- (0.891007,-0.266849) node[pos=0.5,above right] {$\scriptstyle A_2$};
			\draw[thick, white!50!violet, postaction=decorate, decoration={markings, mark= at position 0.3 with {\arrowreversed{stealth}}}] (-0.45399,-0.52372) -- (0.437016,-0.790569) node[pos=0.5,below left, black] {$\scriptstyle B_1$};
			\foreach \a/\b/\c in {0./0./col2,-0.45399/-0.52372/col1,0.891007/-0.266849/col3,0.437016/-0.790569/col2}
			{
				\draw[\c] (\a,\b) circle (0.1);
			}
			\begin{scope}[shift={(-1,0.5)}]
				\draw[fill=gray] (-0.1,-0.1) -- (-0.1,0.1) -- (0.1,0.1) -- (0.1,-0.1) -- cycle;
			\end{scope}
		\end{tikzpicture}
	\end{array}\,,
\end{equation}
where we marked edges directed not along respective coordinates as in Fig.~\ref{fig:boxedYoung} by the violet color.
From this atomic picture plot we find the following ansatz:
\begin{equation}
	\begin{aligned}
		&\check v_2:\;
		C_1=\left(
		\begin{array}{c}
			0 \\
		\end{array}
		\right),\;C_2=\left(
		\begin{array}{cc}
			0 & 0 \\
			0 & 0 \\
		\end{array}
		\right),\;C_3=\left(
		\begin{array}{c}
			0 \\
		\end{array}
		\right),\; A_1=\left(
		\begin{array}{c}
			0 \\
			0 \\
		\end{array}
		\right),\; B_1=\left(
		\begin{array}{cc}
			x_1 & x_2 \\
		\end{array}
		\right)\,,\\
		& A_2=\left(
		\begin{array}{cc}
			x_3 & x_4 \\
		\end{array}
		\right),\;B_2=\left(
		\begin{array}{c}
			0 \\
			0 \\
		\end{array}
		\right),\;
		R=\left(
		\begin{array}{c}
			x_5 \\
			0 \\
		\end{array}
		\right),\;S=\left(
		\begin{array}{cc}
			0 & 0 \\
		\end{array}
		\right)\,,\\
		& \Phi_1=\left(\begin{array}{c}
			\mu+\epsilon_2
		\end{array}\right),\quad \Phi_2=\left(\begin{array}{cc}
			\mu & 0\\ 0 & \mu
		\end{array}\right),\quad \Phi_3=\left(\begin{array}{c}
			\mu+\epsilon_1
		\end{array}\right)\,.
	\end{aligned}
\end{equation} 
In this case we arrive to the following set of equations:
\begin{equation}
	x_1 x_2+x_3 x_4=0,\;x_1^2+x_2^2=\zeta _1,\;x_2^2-x_4^2=\zeta _2,\;-x_1^2-x_3^2+x_5^2=\zeta _2,\;x_3^2+x_4^2=\zeta _3\,.
\end{equation}
Let us stress that the \emph{first} equation in the row has no non-zero solution when all $x_i>0$, so to acquire the $G_{\IR}(\fQ)$-orbit representative one should switch a sign of one of $x_i$'s, say, $x_4$ (this sign switch corresponds to the minus sign of the matrix $\alpha$ element in \eqref{alpha}):
\begin{equation}
	\begin{aligned}
	& x_1=\sqrt{ \frac{\left(\zeta _1+\zeta _2\right) \left(\zeta _1+\zeta _2+\zeta _3\right)}{\zeta _1+2 \zeta _2+\zeta _3}},\; x_2=\sqrt{ -\frac{\zeta _2 \left(\zeta _2+\zeta _3\right)}{\zeta _1+2 \zeta _2+\zeta _3}},\; x_3=\sqrt{ \frac{\left(\zeta _2+\zeta _3\right) \left(\zeta _1+\zeta _2+\zeta _3\right)}{\zeta _1+2 \zeta _2+\zeta _3}}\,,\\
	& x_4=-\sqrt{ -\frac{\zeta _2 \left(\zeta _1+\zeta _2\right)}{\zeta _1+2 \zeta _2+\zeta _3}},\; x_5=\sqrt{ \zeta _1+2 \zeta _2+\zeta _3}\,.
	\end{aligned}
\end{equation}

{\bf Remark:} we should note that solution $\check v_2$ is reminiscent of a ``crooked glass'' solution introduced in \cite[Sec.~6.3]{Galakhov:2024foa}.
Canonically atoms in atomic picture plots are treated as operators acquiring expectation values in the respective vacuum, they are positioned in a 3d space lattice according to 3 charges: $\epsilon_{1,2}$-components of the flavor charge and the R-charge.
All these charges could be calculated from the atomic picture plot by \eqref{flavor}, where $\epsilon$'s are substituted by charges of elementary chiral fields associated to arrows.
Applying this formula we find that all the charges for two operators corresponding to two green atoms in $\check v_2$ are identical.

%%%%%%%%%%%%%%%%%%%%%%%%%%%%%%%%%%%%%%%%%%%%%%%%%%%%%%%%%%%%%%%%%%%%%%%%%%%%%%%%%
%%%%%%%%%%%%%%%%%%%%%%%%%%%%%%%%%%%%%%%%%%%%%%%%%%%%%%%%%%%%%%%%%%%%%%%%%%%%%%%%%
%%%%%%%%%%%%%%%%%%%%%%%%%%%%%%%%%%%%%%%%%%%%%%%%%%%%%%%%%%%%%%%%%%%%%%%%%%%%%%%%%

\section{Quiver Yangians \& Weyl mutations}\label{sec:MutYang}

\subsection{Yangian algebra \texorpdfstring{$Y(\fs\fl_{n+1})$}{Y(sl(n+1))}}

For Yangian algebras Drinfeld introduced two realizations \cite{drinfel1985hopf,drinfel1988new}.
Each realization has its own advantages.
The first realization (also known as a $J$-realization) is written in terms of generators of the original algebra and its $J$-deformations \cite{GUAY2018865,Bao:2023ece}.
It is easier to formulate the co-product structure and the R-matrix in terms of the $J$-realization. 
On the other hand the second realization has a simpler, human readable form, even despite it requires now an infinite set of generator modes and relations.
As well a field theoretic construction of quiver Yangians reveals a form closer to the second realization.
So in what follows we exploit the second realization option.

The Yangian algebra $Y(\fs\fl_{n+1})$ is formulated in terms of an infinite family of generators:
\begin{equation}
	\left(e_a^{(k)},f_a^{(k)},h_a^{(k)}\right),\quad a=1,\ldots,n,\quad k\in\IZ_{\geq 0}\,.
\end{equation}

Whereas zero-mode generators $e_a^{(0)}$, $f_a^{(0)}$, $h_a^{(0)}$ form the ordinary $\fs\fl_{n+1}$ subalgebra \eqref{Lie_algebra_relations} of $Y(\fs\fl_{n+1})$, higher modes satisfy the following set of relations \cite{2019SIGMA..15..020K}:
\begin{equation}
	\begin{aligned}
		& \left[h_i^{(k)},h_j^{(m)}\right]=0,\quad \left[h_i^{(k)},e_j^{(m)}\right]=\CA_{ij}e_j^{(m)},\quad \left[h_i^{(k)},e_j^{(m)}\right]=-\CA_{ij}f_j^{(m)},\quad \left[e_i^{(k)},f_j^{(m)}\right]=\delta_{ij}h_i^{(k+m)}\,,\\
		& \left[e_i^{(k+1)},e_j^{(m)}\right]-\left[e_i^{(k)},e_j^{(m+1)}\right]=-\frac{\epsilon_1+\epsilon_2}{2}\CA_{ij}\left\{e_i^{(k)},e_j^{(m)}\right\}+\frac{\epsilon_1-\epsilon_2}{2}\CM_{ij}\left[e_i^{(k)},e_j^{(m)}\right]\,,\\
		& \left[f_i^{(k+1)},f_j^{(m)}\right]-\left[f_i^{(k)},f_j^{(m+1)}\right]=\frac{\epsilon_1+\epsilon_2}{2}\CA_{ij}\left\{f_i^{(k)},f_j^{(m)}\right\}+\frac{\epsilon_1-\epsilon_2}{2}\CM_{ij}\left[f_i^{(k)},f_j^{(m)}\right]\,,\\
		& \left[h_i^{(k+1)},e_j^{(m)}\right]-\left[h_i^{(k)},e_j^{(m+1)}\right]=-\frac{\epsilon_1+\epsilon_2}{2}\CA_{ij}\left\{h_i^{(k)},e_j^{(m)}\right\}+\frac{\epsilon_1-\epsilon_2}{2}\CM_{ij}\left[h_i^{(k)},e_j^{(m)}\right]\,,\\
		& \left[h_i^{(k+1)},f_j^{(m)}\right]-\left[h_i^{(k)},f_j^{(m+1)}\right]=\frac{\epsilon_1+\epsilon_2}{2}\CA_{ij}\left\{h_i^{(k)},f_j^{(m)}\right\}+\frac{\epsilon_1-\epsilon_2}{2}\CM_{ij}\left[h_i^{(k)},f_j^{(m)}\right]\,,
	\end{aligned}
\end{equation}
where auxiliary matrix $\CM_{ij}=\delta_{i,j+1}-\delta_{i+1,j}$.
In addition to these quadratic relations one imposes cubic Serre relations.

We should stress that to define a homomorphic representation of this algebra it is sufficient to consider only the $\fs\fl_{n+1}$-subalgebra and either of $e_k^{(1)}$, $f_k^{(1)}$ or $h_k^{(1)}$, the rest could be re-constructed from the relations.
For example, $h_a^{(1)}$ play the role of mode shifting operators:
\begin{equation}
	e_a^{(k+1)}=\frac{1}{2}\left[h_a^{(1)},e_a^{(k)}\right]+\frac{\epsilon_1+\epsilon_2}{2}\left\{h_a^{(0)},e_a^{(k)}\right\},\quad
	f_a^{(k+1)}=-\frac{1}{2}\left[h_a^{(1)},f_a^{(k)}\right]+\frac{\epsilon_1+\epsilon_2}{2}\left\{h_a^{(0)},f_a^{(k)}\right\}\,.
\end{equation}

The action of the Weyl group $\CW_{\fs\fl_{n+1}}$ is naturally extended to the Yangian algebra by applying exponential operators \cite{2019SIGMA..15..020K} as defined by reflections $\tau_i$ in \eqref{Weyl_Tits}.
As we mentioned it is sufficient to define this action for, say, $e_a^{(1)}$, $f_a^{(1)}$:
\begin{equation}\label{Tits_refl}
	\begin{aligned}
	&\tau_i\left(e_i^{(1)}\right)=-f_i^{(1)}-\frac{\epsilon_1+\epsilon_2}{2}\left\{h_i^{(0)},f_i^{(0)}\right\},\quad \tau_i\left(e_{i\pm 1}^{(1)}\right)=\left[e_i^{(0)},e_{i\pm 1}^{(1)}\right],\quad \tau_i\left(e_{j}^{(1)}\right)=e_{j}^{(1)},\mbox{ for }|i-j|>1\,,\\
	&\tau_i\left(f_i^{(1)}\right)=-e_i^{(1)}-\frac{\epsilon_1+\epsilon_2}{2}\left\{h_i^{(0)},e_i^{(0)}\right\},\quad \tau_i\left(f_{i\pm 1}^{(1)}\right)=-\left[f_i^{(0)},f_{i\pm 1}^{(1)}\right],\quad \tau_i\left(f_{j}^{(1)}\right)=f_{j}^{(1)},\mbox{ for }|i-j|>1\,.
	\end{aligned}
\end{equation}

%%%%%%%%%%%%%%%%%%%%%%%%%%%%%%%%%%%%%%%%%%%%%%%%%%%%%%%%%%%%%%%%%%%%%%%%%%%%%%%%%
%%%%%%%%%%%%%%%%%%%%%%%%%%%%%%%%%%%%%%%%%%%%%%%%%%%%%%%%%%%%%%%%%%%%%%%%%%%%%%%%%
%%%%%%%%%%%%%%%%%%%%%%%%%%%%%%%%%%%%%%%%%%%%%%%%%%%%%%%%%%%%%%%%%%%%%%%%%%%%%%%%%

\subsection{BPS algebra on 3d Young diagrams}

A quiver BPS algebra that turns out to be the quiver Yangian could be defined physically on molten crystals as fixed point vacua in SQM.
In our case as we explained in Sec.~\ref{sec:BoxYoung} earlier generic molten crystals reduce to a special case of boxed 3d Young diagrams.
Tunneling processes between ``neighboring'' diagrams differing by adding/subtracting a cell give rise to a representation of the respective quiver Yangian, that is $Y(\fs\fl_{n+1})$ in our case.
Here let us give a brief review of this construction, details could be found, for example, in \cite{Galakhov:2020vyb,Galakhov:2024bzs,Li:2023zub}.

Here we denote by $\lambda$ boxed 3d Young diagrams as it is depicted in Fig.~\ref{fig:boxedYoung}, by $\lambda^+$ and by $\lambda^-$ we denote sets of cells that could be added or subtracted respectively, so that a resulting diagram is again a valid boxed 3d Young diagram.
For brevity we denote cells of specific color $a$ as $\sqbox{$a$}$, and a diagram with added/subtracted cell as $\lambda \pm \sqbox{$a$}$.
Eventually, by $\phi_{\ssqbox{$s$}}$ we denote value \eqref{flavor} corresponding to an atom obtained from cell $\sqbox{$s$}$.
If a cell is located at coordinates $(b,a,c)$ in the frame $(\vec B,\vec A,\vec C)$ depicted in Fig.~\ref{fig:boxedYoung} then
\begin{equation}
	\phi_{\ssqbox{$s$}}=\mu+\epsilon_1 a+ \epsilon_2 b-(\epsilon_1+\epsilon_2)c\,.
\end{equation}

For given $\fQ_{n,u,h}$ we consider a representation spanned by vectors $|\lambda\rangle$.
The generators are represented by the following matrix elements:
\begin{equation}\label{YrepYoung}
\begin{aligned}
	&e_a^{(k)}|\lambda\rangle=\sum\lm_{\ssqbox{$a$}\in\lambda^+}{\bf E}\left[\lambda,\lambda+\sqbox{$a$}\right]\,\phi_{\ssqbox{$a$}}^k\,|\lambda+\sqbox{$a$}\rangle\,,\\
	&f_a^{(k)}|\lambda\rangle=\sum\lm_{\ssqbox{$a$}\in\lambda^-}{\bf F}\left[\lambda,\lambda-\sqbox{$a$}\right]\,\phi_{\ssqbox{$a$}}^k\,|\lambda-\sqbox{$a$}\rangle\,,\\
	&\left(1+\sum\lm_{k=0}^{\infty}\frac{h_a^{(k)}}{z^{k+1}}\right)|\lambda\rangle=\underbrace{\psi_a\times\frac{z-\left(\mu-h\left(\epsilon_1+\epsilon_2\right)\right)}{z-\mu}\times\prod\lm_{b=1}^n\prod\lm_{\ssqbox{$b$}\in \lambda}\varphi_{a,b}(z-\phi_{\ssqbox{$b$}})}_{\Psi_{a,\lambda}(z)}|\lambda\rangle\,,
\end{aligned}
\end{equation}
where we introduced the following notations:
\begin{equation}
	\varphi_{i,i}(z)=\frac{z-\epsilon_1-\epsilon_2}{z+\epsilon_1+\epsilon_2},\quad \varphi_{i,i+1}(z)=\frac{z+\epsilon_1}{z-\epsilon_2},\quad \varphi_{i+1,i}(z)=\frac{z+\epsilon_2}{z-\epsilon_1},\quad \varphi_{i,j}(z)=1,\mbox{ for }|i-j|>1\,.
\end{equation}
Numeric values $\psi_a^{(0)}$ are defined in the following way:
\begin{equation}
	\psi_a=\left(-\frac{1}{\epsilon_1+\epsilon_2}\right)^{\delta_{a,u}}\,.
\end{equation}

For \eqref{YrepYoung} to form a representation of $Y(\fs\fl_{n+1})$ it suffices to subject matrix elements to so called \emph{hysteresis} relations:
\begin{equation}\label{hyster}
	\begin{aligned}
		&{\bf E}\left[\lambda+\sqbox{$a$},\lambda+\sqbox{$a$}+\sqbox{$b$}\right]{\bf F}\left[\lambda+\sqbox{$a$}+\sqbox{$b$},\lambda+\sqbox{$b$}\right]={\bf F}\left[\lambda+\sqbox{$a$},\lambda\right]{\bf E}\left[\lambda,\lambda+\sqbox{$b$}\right]\,,\\
		&\frac{{\bf E}\left[\lambda,\lambda+\sqbox{$a$}\right]{\bf E}\left[\lambda+\sqbox{$a$},\lambda+\sqbox{$a$}+\sqbox{$b$}\right]}{{\bf E}\left[\lambda,\lambda+\sqbox{$b$}\right]{\bf E}\left[\lambda+\sqbox{$b$},\lambda+\sqbox{$a$}+\sqbox{$b$}\right]}\,\varphi_{a,b}\left(\phi_{\ssqbox{$a$}}-\phi_{\ssqbox{$b$}}\right)=1\,,\\
		&\frac{{\bf F}\left[\lambda+\sqbox{$a$}+\sqbox{$b$},\lambda+\sqbox{$a$}\right]{\bf F}\left[\lambda+\sqbox{$a$},\lambda\right]}{{\bf F}\left[\lambda+\sqbox{$a$}+\sqbox{$b$},\lambda+\sqbox{$b$}\right]{\bf F}\left[\lambda+\sqbox{$b$},\lambda\right]}\,\varphi_{a,b}\left(\phi_{\ssqbox{$a$}}-\phi_{\ssqbox{$b$}}\right)=1\,,\\
		& {\bf E}\left[\lambda,\lambda+\sqbox{$a$}\right]={\bf F}\left[\lambda+\sqbox{$a$},\lambda\right]=\mathop{\rm res}\lm_{z=\phi_{\ssqbox{$a$}}}\Psi_{a,\lambda}(z)=-\mathop{\rm res}\lm_{z=\phi_{\ssqbox{$a$}}}\Psi_{a,\lambda+\ssqbox{$a$}}(z)\,.
	\end{aligned}
\end{equation}

A combinatorial solution to \eqref{hyster} could be found \cite{Prochazka:2015deb,Li:2020rij,Li:2023zub}.
This is a so called \emph{bootstrap} or a \emph{square-root} representation.

Also one could interpret matrix elements $\bf E$ and $\bf F$ as tunneling amplitudes \cite{Galakhov:2025fhd} in SQM between vacua labeled by corresponding diagrams.
In this interpretation matrix elements acquire geometric treatment as Hecke modifications.
For a pair of fixed points $v_1$, $v_2$ these matrix element are defined in the following way:
\begin{equation}\label{EFgeom}
	{\bf E}\left[v_1,v_2\right]=\frac{{\bf e}\left(T_{v_1}^G\mathscr{M}\right)}{{\bf e}\left(T_{v_2\to v_1}^G\Sigma\right)},\quad {\bf F}\left[v_2,v_1\right]=\frac{{\bf e}\left(T_{v_2}^G\mathscr{M}\right)}{{\bf e}\left(T_{v_2\to v_1}^G\Sigma\right)}\,,
\end{equation}
where $T_{v}^G\mathscr{M}$ is a tangent space to fixed point $v$ in the space of quiver morphism modulo the linearized action of the complexified group $G_{\IC}(\fQ)$; $T_{v_2\to v_1}^G\Sigma$ is a tangent space to a surface $\Sigma$ in the Cartesian product of quiver morphism spaces, where there is a quiver homomorphism of the second representation into the first one, passing through a pair $v_1\times v_2$; and ${\bf e}(X)$ denotes the Euler class of space $X$.
There are a few comments required by this definition.

A quiver homomorphism $\{M\}\to \{M'\}$ is defined as a ``\emph{singular}'' gauge transform (Hecke correspondence) $H_a$ in all quiver nodes $a$ (including the framing node where $H_{\rm framing}=\bbone$) between quiver morphism spaces $M$ and $M'$.
Sets of morphisms $\{M\}$ and $\{M'\}$ are allowed to correspond to different quiver dimensions.
A singular version of the gauge transform takes the following form:
\begin{equation}\label{homomorphism}
	H_b\cdot M=M\cdot H_a,\quad \forall\,M\in{\rm Hom}(V_a,V_b),\quad \forall\,a,b=0,\ldots,n\,,
\end{equation}
where by index 0 we imply the framing node.

Since spaces discussed above are singular in general we apply here a version of a regularized  Euler class suggested in \cite{Galakhov:2020vyb}.
For quiver morphisms the vector field induced by flavor symmetry could be transformed to local coordinates where it becomes diagonal:
\begin{equation}
	V=\sum\lm_{a,b}\sum\lm_{M\in{\rm Hom}(V_a,V_b)}\Tr\left(\Phi_bM-M\Phi_a-\epsilon_M M\right)\frac{\p}{\p M}\;\longrightarrow\; V=\sum\lm_{\nu}\Omega_{\nu}m_{\nu}\frac{\p}{\p m_{\nu}}\,.
\end{equation}
In this terms the Euler class is defined:
\begin{equation}
	{\bf e}\left(X\right)=(-1)^{\left\lfloor \sum\lm_{\nu:\;\Omega_{\nu}=0}\frac{1}{2}\right\rfloor}\prod\lm_{\nu:\;\Omega_{\nu}\neq 0}\Omega_{\nu}\,.
\end{equation}

%%%%%%%%%%%%%%%%%%%%%%%%%%%%%%%%%%%%%%%%%%%%%%%%%%%%%%%%%%%%%%%%%%%%%%%%%%%%%%%%%
%%%%%%%%%%%%%%%%%%%%%%%%%%%%%%%%%%%%%%%%%%%%%%%%%%%%%%%%%%%%%%%%%%%%%%%%%%%%%%%%%
%%%%%%%%%%%%%%%%%%%%%%%%%%%%%%%%%%%%%%%%%%%%%%%%%%%%%%%%%%%%%%%%%%%%%%%%%%%%%%%%%

\subsection{Weyl mutations on quiver Yangians}

A physically motivated prescription for a quiver Yangian representation discussed in the previous section depends \emph{explicitly} on the structure of molten crystals in general and of boxed 3d Young diagrams in our case.
As we could see in many examples illustrated in App.~\ref{app:examples} this structure is lost outside the cyclic chamber.
If we would follow these examples we would notice that many aspects of the construction were broken.

For instance, let us take a look at fixed points in phase III of $\fQ_{\fs\fl_3,[3,3]}$ \eqref{PhIII}.
Despite phase III is not cyclic these atomic structure plots look being shaped again as two row Young diagrams.
And we might have tried to define raising and lowering operators as adding/subtracting processes for atoms in atomic structure plots, like in the following hypothetical transition the number of red atoms is increased from 2 to 3:
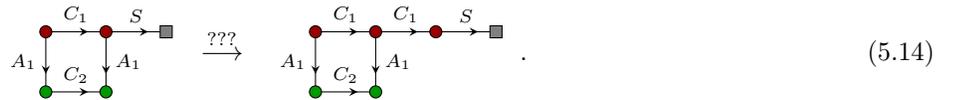
\begin{equation}
	\begin{array}{c}
		\begin{tikzpicture}[scale=0.8]
			\draw[postaction=decorate, decoration={markings, mark= at position 0.7 with {\arrow{stealth}}}] (-1,0) -- (0,0) node[pos=0.5,above] {$\scriptstyle S$};
			\draw[postaction=decorate, decoration={markings, mark= at position 0.7 with {\arrow{stealth}}}] (-2,0) -- (-1,0) node[pos=0.5,above] {$\scriptstyle C_1$};
			\draw[postaction=decorate, decoration={markings, mark= at position 0.7 with {\arrow{stealth}}}] (-2,0) -- (-2,-1) node[pos=0.5,left] {$\scriptstyle A_1$};
			\draw[postaction=decorate, decoration={markings, mark= at position 0.7 with {\arrow{stealth}}}] (-1,0) -- (-1,-1) node[pos=0.5,right] {$\scriptstyle A_1$};
			\draw[postaction=decorate, decoration={markings, mark= at position 0.7 with {\arrow{stealth}}}] (-2,-1) -- (-1,-1) node[pos=0.5,above] {$\scriptstyle C_2$};
			\draw[fill=black!40!red] (-1,0) circle (0.1) (-2,0) circle (0.1);
			\draw[fill=black!40!green] (-2,-1) circle (0.1) (-1,-1) circle (0.1);
			\begin{scope}[shift={(0,0)}]
				\draw[fill=gray] (-0.1,-0.1) -- (-0.1,0.1) -- (0.1,0.1) -- (0.1,-0.1) -- cycle;
			\end{scope}
		\end{tikzpicture}
	\end{array}\;\overset{???}{\longrightarrow}\;\begin{array}{c}
		\begin{tikzpicture}[scale=0.8]
			\draw[postaction=decorate, decoration={markings, mark= at position 0.7 with {\arrow{stealth}}}] (-1,0) -- (0,0) node[pos=0.5,above] {$\scriptstyle S$};
			\draw[postaction=decorate, decoration={markings, mark= at position 0.7 with {\arrow{stealth}}}] (-2,0) -- (-1,0) node[pos=0.5,above] {$\scriptstyle C_1$};
			\draw[postaction=decorate, decoration={markings, mark= at position 0.7 with {\arrow{stealth}}}] (-3,0) -- (-2,0) node[pos=0.5,above] {$\scriptstyle C_1$};
			\draw[postaction=decorate, decoration={markings, mark= at position 0.7 with {\arrow{stealth}}}] (-3,0) -- (-3,-1) node[pos=0.5,left] {$\scriptstyle A_1$};
			\draw[postaction=decorate, decoration={markings, mark= at position 0.7 with {\arrow{stealth}}}] (-2,0) -- (-2,-1) node[pos=0.5,right] {$\scriptstyle A_1$};
			\draw[postaction=decorate, decoration={markings, mark= at position 0.7 with {\arrow{stealth}}}] (-3,-1) -- (-2,-1) node[pos=0.5,above] {$\scriptstyle C_2$};
			\draw[fill=black!40!red] (-1,0) circle (0.1) (-2,0) circle (0.1) (-3,0) circle (0.1);
			\draw[fill=black!40!green] (-3,-1) circle (0.1) (-2,-1) circle (0.1);
			\begin{scope}[shift={(0,0)}]
				\draw[fill=gray] (-0.1,-0.1) -- (-0.1,0.1) -- (0.1,0.1) -- (0.1,-0.1) -- cycle;
			\end{scope}
		\end{tikzpicture}
	\end{array}\,.
\end{equation}
However we see in this pictures that before putting an extra red atom in a bridge connecting the framing atom with the remaining square, the positions of atoms in the square are all shifted towards the left.
So this process is more complicated than adding a single extra atom to a vacant position.
Also if we would try to construct a singular gauge map $H_a$ there is no such map satisfying \eqref{homomorphism}.

In this situation we would suggest an approach to quiver Yangians proposed in \cite{Galakhov:2024foa} for non-cyclic phases.
We \emph{checked explicitly} an applicability of this algorithm for the simplest non-trivial case of $\fs\fl_3$.

Consider some element ${\bf w}\in \CW_{\fs\fl_{n+1}}$ of the Weyl group.
We would say that $\bf w$ characterizes a phase locus (let us call it a $\bf w$-phase) in the moduli space if this locus could be described as a set of inequalities ${\bf w}^{-1}\cdot\vec \zeta>0$ implying this inequality for each component of the vector: $\left({\bf w}^{-1}\cdot\vec \zeta\right)_a>0$, $a=1,\ldots,n$.
This inequality implies that to arrive to this phase from the cyclic chamber one should decompose element ${\bf w}=\ldots s_c s_b s_a$ over elementary reflections and then apply Weyl mutations in the respective order: first $s_a$, then $s_b$, then $s_c$, and so on.

Element $\bf w$ of the Weyl group delivers an element of the Weyl mutation group acting on fixed points.
In particular it establishes a 1-to-1 map of fixed points in the $\bf w$-phase to ones in the cyclic chamber.
So that to a fixed point $v$ in the $\bf w$-phase we are able to associate a fixed point ${\bf w}^{-1}(v)$ in the cyclic chamber, moreover ${\bf w}^{-1}(v)$ repeats all the structures of a boxed 3d Young diagram according to Sec.~\ref{sec:BoxYoung}.
Then we define generators for quiver Yangian in the $\bf w$-phase (we would like to denote it as ${}^{\bf w}Y(\fs\fl_{n+1})$) in the following way:
\begin{equation}\label{w_quivYang}
	\begin{aligned}
		&{}^{\bf w}e_a^{(k)}|v\rangle=\sum\lm_{\ssqbox{$a$}\in\left({\bf w}^{-1}(v)\right)^+}{\bf E}\left[v,{\bf w}\left({\bf w}^{-1}(v)+\sqbox{$a$}\right)\right]\,\phi_{\ssqbox{$a$}}^k\,\left|{\bf w}\left({\bf w}^{-1}(v)+\sqbox{$a$}\right)\right\rangle\,,\\
		&{}^{\bf w}f_a^{(k)}|v\rangle=\sum\lm_{\ssqbox{$a$}\in\left({\bf w}^{-1}(v)\right)^-}{\bf F}\left[v,{\bf w}\left({\bf w}^{-1}(v)-\sqbox{$a$}\right)\right]\,\phi_{\ssqbox{$a$}}^k\,\left|{\bf w}\left({\bf w}^{-1}(v)-\sqbox{$a$}\right)\right\rangle\,,\\
		&\left(1+\sum\lm_{k=0}^{\infty}\frac{{}^{\bf w}h_a^{(k)}}{z^{k+1}}\right)|v\rangle=\Psi_{a,{\bf w}^{-1}(v)}(z)|v\rangle\,.
	\end{aligned}
\end{equation}
Let us \emph{stress} that if the 3d Young diagram structure is read off from $\bf w$ pre-images of fixed points in the cyclic chamber, matrix elements are calculated according to \eqref{EFgeom} for actual fixed points in the $\bf w$-phase that \emph{are not} Young diagrams.

We observe that the quiver Yangian algebra and the respective representation on the fixed points are isomorphic to the ordinary $Y(\fs\fl_{n+1})$ constructed in the cyclic chamber:
\begin{tcolorbox}
\begin{equation}\label{mainiso}
	{}^{\bf w}Y(\fs\fl_{n+1})\cong Y(\fs\fl_{n+1})\,.
\end{equation}
\end{tcolorbox}
We confirm this isomorphism by checking explicitly that $\bf E$- and $\bf F$-matrix elements satisfy hysteresis relations \eqref{hyster} for the case of $\fs\fl_3$.
For example, for phase $\bf V$ \eqref{PhV} we observe the following equality:
\begin{equation}
\begin{aligned}
	&\frac{{\bf E}\left[\begin{array}{c}
			\begin{tikzpicture}[scale=0.5]
				\foreach \a/\b/\c/\d in {-3/1/-3/0,-2/1/-2/0,-1/1/-1/0,-3/0/-2/0,-2/0/-1/0,-1/0/0/0,-3/1/-2/1,-2/1/-1/1}
				{\draw[postaction=decorate, decoration={markings, mark= at position 0.7 with {\arrow{stealth}}}] (\a,\b) -- (\c,\d);}
				\foreach \a/\b in {-3/0,-2/0,-1/0}
				{\draw[fill=black!40!red] (\a,\b) circle (0.15);}
				\foreach \a/\b in {-3/1,-2/1,-1/1}
				{\draw[fill=black!40!green] (\a,\b) circle (0.15);}
				\draw[fill=gray] (-0.15,-0.15) -- (-0.15,0.15) -- (0.15,0.15) -- (0.15,-0.15) -- cycle;
			\end{tikzpicture}
		\end{array},\begin{array}{c}
		\begin{tikzpicture}[scale=0.5]
			\foreach \a/\b/\c/\d in {-2/1/-2/0,-1/1/-1/0,-2/0/-1/0,-1/0/0/0,-2/1/-1/1}
			{\draw[postaction=decorate, decoration={markings, mark= at position 0.7 with {\arrow{stealth}}}] (\a,\b) -- (\c,\d);}
			\foreach \a/\b in {-2/0,-1/0}
			{\draw[fill=black!40!red] (\a,\b) circle (0.15);}
			\foreach \a/\b in {-2/1,-1/1}
			{\draw[fill=black!40!green] (\a,\b) circle (0.15);}
			\draw[fill=gray] (-0.15,-0.15) -- (-0.15,0.15) -- (0.15,0.15) -- (0.15,-0.15) -- cycle;
		\end{tikzpicture}
		\end{array}\right]{\bf E}\left[\begin{array}{c}
		\begin{tikzpicture}[scale=0.5]
			\foreach \a/\b/\c/\d in {-2/1/-2/0,-1/1/-1/0,-2/0/-1/0,-1/0/0/0,-2/1/-1/1}
			{\draw[postaction=decorate, decoration={markings, mark= at position 0.7 with {\arrow{stealth}}}] (\a,\b) -- (\c,\d);}
			\foreach \a/\b in {-2/0,-1/0}
			{\draw[fill=black!40!red] (\a,\b) circle (0.15);}
			\foreach \a/\b in {-2/1,-1/1}
			{\draw[fill=black!40!green] (\a,\b) circle (0.15);}
			\draw[fill=gray] (-0.15,-0.15) -- (-0.15,0.15) -- (0.15,0.15) -- (0.15,-0.15) -- cycle;
		\end{tikzpicture}
		\end{array},\begin{array}{c}
		\begin{tikzpicture}[scale=0.5]
			\foreach \a/\b/\c/\d in {-2/1/-2/0,-1/1/-1/0,-2/0/-1/0,-1/0/0/0,-2/1/-1/1,0/0/1/0}
			{\draw[postaction=decorate, decoration={markings, mark= at position 0.7 with {\arrow{stealth}}}] (\a,\b) -- (\c,\d);}
			\foreach \a/\b in {-2/0,-1/0,1/0}
			{\draw[fill=black!40!red] (\a,\b) circle (0.15);}
			\foreach \a/\b in {-2/1,-1/1}
			{\draw[fill=black!40!green] (\a,\b) circle (0.15);}
			\draw[fill=gray] (-0.15,-0.15) -- (-0.15,0.15) -- (0.15,0.15) -- (0.15,-0.15) -- cycle;
		\end{tikzpicture}
		\end{array}\right]}{{\bf E}\left[\begin{array}{c}
		\begin{tikzpicture}[scale=0.5]
			\foreach \a/\b/\c/\d in {-3/1/-3/0,-2/1/-2/0,-1/1/-1/0,-3/0/-2/0,-2/0/-1/0,-1/0/0/0,-3/1/-2/1,-2/1/-1/1}
			{\draw[postaction=decorate, decoration={markings, mark= at position 0.7 with {\arrow{stealth}}}] (\a,\b) -- (\c,\d);}
			\foreach \a/\b in {-3/0,-2/0,-1/0}
			{\draw[fill=black!40!red] (\a,\b) circle (0.15);}
			\foreach \a/\b in {-3/1,-2/1,-1/1}
			{\draw[fill=black!40!green] (\a,\b) circle (0.15);}
			\draw[fill=gray] (-0.15,-0.15) -- (-0.15,0.15) -- (0.15,0.15) -- (0.15,-0.15) -- cycle;
		\end{tikzpicture}
		\end{array},\begin{array}{c}
		\begin{tikzpicture}[scale=0.5]
			\foreach \a/\b/\c/\d in {-3/1/-3/0,-2/1/-2/0,-1/1/-1/0,-3/0/-2/0,-2/0/-1/0,-1/0/0/0,-3/1/-2/1,-2/1/-1/1,0/0/1/0}
			{\draw[postaction=decorate, decoration={markings, mark= at position 0.7 with {\arrow{stealth}}}] (\a,\b) -- (\c,\d);}
			\foreach \a/\b in {-3/0,-2/0,-1/0,1/0}
			{\draw[fill=black!40!red] (\a,\b) circle (0.15);}
			\foreach \a/\b in {-3/1,-2/1,-1/1}
			{\draw[fill=black!40!green] (\a,\b) circle (0.15);}
			\draw[fill=gray] (-0.15,-0.15) -- (-0.15,0.15) -- (0.15,0.15) -- (0.15,-0.15) -- cycle;
		\end{tikzpicture}
		\end{array}\right]{\bf E}\left[\begin{array}{c}
		\begin{tikzpicture}[scale=0.5]
			\foreach \a/\b/\c/\d in {-3/1/-3/0,-2/1/-2/0,-1/1/-1/0,-3/0/-2/0,-2/0/-1/0,-1/0/0/0,-3/1/-2/1,-2/1/-1/1,0/0/1/0}
			{\draw[postaction=decorate, decoration={markings, mark= at position 0.7 with {\arrow{stealth}}}] (\a,\b) -- (\c,\d);}
			\foreach \a/\b in {-3/0,-2/0,-1/0,1/0}
			{\draw[fill=black!40!red] (\a,\b) circle (0.15);}
			\foreach \a/\b in {-3/1,-2/1,-1/1}
			{\draw[fill=black!40!green] (\a,\b) circle (0.15);}
			\draw[fill=gray] (-0.15,-0.15) -- (-0.15,0.15) -- (0.15,0.15) -- (0.15,-0.15) -- cycle;
		\end{tikzpicture}
		\end{array},\begin{array}{c}
		\begin{tikzpicture}[scale=0.5]
			\foreach \a/\b/\c/\d in {-2/1/-2/0,-1/1/-1/0,-2/0/-1/0,-1/0/0/0,-2/1/-1/1,0/0/1/0}
			{\draw[postaction=decorate, decoration={markings, mark= at position 0.7 with {\arrow{stealth}}}] (\a,\b) -- (\c,\d);}
			\foreach \a/\b in {-2/0,-1/0,1/0}
			{\draw[fill=black!40!red] (\a,\b) circle (0.15);}
			\foreach \a/\b in {-2/1,-1/1}
			{\draw[fill=black!40!green] (\a,\b) circle (0.15);}
			\draw[fill=gray] (-0.15,-0.15) -- (-0.15,0.15) -- (0.15,0.15) -- (0.15,-0.15) -- cycle;
		\end{tikzpicture}
		\end{array}\right]}=\\
	&=\frac{{\bf E}\left[\begin{array}{c}
			\begin{tikzpicture}[scale=0.5]
				\foreach \a/\b/\c/\d in {0/0/1/0, 1/0/2/0, 2/0/3/0}
				{\draw[postaction=decorate, decoration={markings, mark= at position 0.7 with {\arrow{stealth}}}] (\a,\b) -- (\c,\d);}
				\foreach \a/\b in {1/0,2/0,3/0}
				{\draw[fill=black!40!red] (\a,\b) circle (0.15);}
				\draw[fill=gray] (-0.15,-0.15) -- (-0.15,0.15) -- (0.15,0.15) -- (0.15,-0.15) -- cycle;
			\end{tikzpicture}
		\end{array},\begin{array}{c}
		\begin{tikzpicture}[scale=0.5]
			\foreach \a/\b/\c/\d in {0/0/1/0, 1/0/2/0, 2/0/3/0, 3/0/4/0}
			{\draw[postaction=decorate, decoration={markings, mark= at position 0.7 with {\arrow{stealth}}}] (\a,\b) -- (\c,\d);}
			\foreach \a/\b in {1/0,2/0,3/0,4/0}
			{\draw[fill=black!40!red] (\a,\b) circle (0.15);}
			\draw[fill=gray] (-0.15,-0.15) -- (-0.15,0.15) -- (0.15,0.15) -- (0.15,-0.15) -- cycle;
		\end{tikzpicture}
		\end{array}\right]{\bf E}\left[\begin{array}{c}
		\begin{tikzpicture}[scale=0.5]
			\foreach \a/\b/\c/\d in {0/0/1/0, 1/0/2/0, 2/0/3/0, 3/0/4/0}
			{\draw[postaction=decorate, decoration={markings, mark= at position 0.7 with {\arrow{stealth}}}] (\a,\b) -- (\c,\d);}
			\foreach \a/\b in {1/0,2/0,3/0,4/0}
			{\draw[fill=black!40!red] (\a,\b) circle (0.15);}
			\draw[fill=gray] (-0.15,-0.15) -- (-0.15,0.15) -- (0.15,0.15) -- (0.15,-0.15) -- cycle;
		\end{tikzpicture}
		\end{array},\begin{array}{c}
		\begin{tikzpicture}[scale=0.5]
			\foreach \a/\b/\c/\d in {0/0/1/0, 1/0/2/0, 2/0/3/0, 3/0/4/0, 1/0/1/-1}
			{\draw[postaction=decorate, decoration={markings, mark= at position 0.7 with {\arrow{stealth}}}] (\a,\b) -- (\c,\d);}
			\foreach \a/\b in {1/0,2/0,3/0,4/0}
			{\draw[fill=black!40!red] (\a,\b) circle (0.15);}
			\foreach \a/\b in {1/-1}
			{\draw[fill=black!40!green] (\a,\b) circle (0.15);}
			\draw[fill=gray] (-0.15,-0.15) -- (-0.15,0.15) -- (0.15,0.15) -- (0.15,-0.15) -- cycle;
		\end{tikzpicture}
		\end{array}\right]}{{\bf E}\left[\begin{array}{c}
		\begin{tikzpicture}[scale=0.5]
			\foreach \a/\b/\c/\d in {0/0/1/0, 1/0/2/0, 2/0/3/0}
			{\draw[postaction=decorate, decoration={markings, mark= at position 0.7 with {\arrow{stealth}}}] (\a,\b) -- (\c,\d);}
			\foreach \a/\b in {1/0,2/0,3/0}
			{\draw[fill=black!40!red] (\a,\b) circle (0.15);}
			\draw[fill=gray] (-0.15,-0.15) -- (-0.15,0.15) -- (0.15,0.15) -- (0.15,-0.15) -- cycle;
		\end{tikzpicture}
		\end{array},\begin{array}{c}
		\begin{tikzpicture}[scale=0.5]
			\foreach \a/\b/\c/\d in {0/0/1/0, 1/0/2/0, 2/0/3/0, 1/0/1/-1}
			{\draw[postaction=decorate, decoration={markings, mark= at position 0.7 with {\arrow{stealth}}}] (\a,\b) -- (\c,\d);}
			\foreach \a/\b in {1/0,2/0,3/0}
			{\draw[fill=black!40!red] (\a,\b) circle (0.15);}
			\foreach \a/\b in {1/-1}
			{\draw[fill=black!40!green] (\a,\b) circle (0.15);}
			\draw[fill=gray] (-0.15,-0.15) -- (-0.15,0.15) -- (0.15,0.15) -- (0.15,-0.15) -- cycle;
		\end{tikzpicture}
		\end{array}\right]{\bf E}\left[\begin{array}{c}
		\begin{tikzpicture}[scale=0.5]
			\foreach \a/\b/\c/\d in {0/0/1/0, 1/0/2/0, 2/0/3/0, 1/0/1/-1}
			{\draw[postaction=decorate, decoration={markings, mark= at position 0.7 with {\arrow{stealth}}}] (\a,\b) -- (\c,\d);}
			\foreach \a/\b in {1/0,2/0,3/0}
			{\draw[fill=black!40!red] (\a,\b) circle (0.15);}
			\foreach \a/\b in {1/-1}
			{\draw[fill=black!40!green] (\a,\b) circle (0.15);}
			\draw[fill=gray] (-0.15,-0.15) -- (-0.15,0.15) -- (0.15,0.15) -- (0.15,-0.15) -- cycle;
		\end{tikzpicture}
		\end{array},\begin{array}{c}
		\begin{tikzpicture}[scale=0.5]
			\foreach \a/\b/\c/\d in {0/0/1/0, 1/0/2/0, 2/0/3/0, 3/0/4/0, 1/0/1/-1}
			{\draw[postaction=decorate, decoration={markings, mark= at position 0.7 with {\arrow{stealth}}}] (\a,\b) -- (\c,\d);}
			\foreach \a/\b in {1/0,2/0,3/0,4/0}
			{\draw[fill=black!40!red] (\a,\b) circle (0.15);}
			\foreach \a/\b in {1/-1}
			{\draw[fill=black!40!green] (\a,\b) circle (0.15);}
			\draw[fill=gray] (-0.15,-0.15) -- (-0.15,0.15) -- (0.15,0.15) -- (0.15,-0.15) -- cycle;
		\end{tikzpicture}
		\end{array}\right]}=\frac{4}{3}\,.
\end{aligned}
\end{equation}

However we should stress that the value of $\bf w$ affects some parameters of Yangians, in particular from phase to phase a set of scaling numbers $\psi_a$ varies.
For the $\fs\fl_3$ quiver family with the framing node attached to the first gauge node we find:
\begin{equation}
\begin{aligned}
	&{\bf I}:\;\psi_1=-\frac{1}{\epsilon_1+\epsilon_2},\;\psi_2=1,\quad {\bf II}:\;\psi_1=-\frac{1}{\epsilon_1+\epsilon_2},\;\psi_2=\frac{1}{\left(\epsilon_1+\epsilon_2\right)^2},\quad {\bf III}:\;\psi_1=\frac{1}{\epsilon_1+\epsilon_2},\;\psi_2=\frac{1}{\left(\epsilon_1+\epsilon_2\right)^2}\,,\\
	&{\bf IV}:\;\psi_1=\frac{1}{\epsilon_1+\epsilon_2},\;\psi_2=-1,\quad {\bf V}:\;\psi_1=\frac{1}{\left(\epsilon_1+\epsilon_2\right)^3},\;\psi_2=-1,\quad {\bf VI}:\;\psi_1=\frac{1}{\left(\epsilon_1+\epsilon_2\right)^3},\;\psi_2=1\,.
\end{aligned}
\end{equation}

We should note that unlike \eqref{YrepYoung} generators \eqref{w_quivYang} do not shift quiver dimensions by unit vectors, rather $\vec d\to \vec d\pm {\bf w}(\bbone_a)$ for ${}^{\bf w}e_a^{(k)}$/${}^{\bf w}f_a^{(k)}$.
However due to isomorphism \eqref{mainiso} we conclude that reflections \eqref{Tits_refl} are isomorphisms of ${}^{\bf w}Y(\fs\fl_{n+1})$ as well.
Therefore if one would like to make a canonical definition of the raising/lowering operators corresponding to Dynkin/quiver node $a$ such that they necessarily shift the dimension as $d_a\to d_a\pm 1$ it is sufficient to consider an image of ${}^{\bf w}Y(\fs\fl_{n+1})$ under map $\tau_{{\bf w}^{-1}}$ corresponding to Weyl group element ${\bf w}^{-1}$.

%%%%%%%%%%%%%%%%%%%%%%%%%%%%%%%%%%%%%%%%%%%%%%%%%%%%%%%%%%%%%%%%%%%%%%%%%%%%%%%%%
%%%%%%%%%%%%%%%%%%%%%%%%%%%%%%%%%%%%%%%%%%%%%%%%%%%%%%%%%%%%%%%%%%%%%%%%%%%%%%%%%
%%%%%%%%%%%%%%%%%%%%%%%%%%%%%%%%%%%%%%%%%%%%%%%%%%%%%%%%%%%%%%%%%%%%%%%%%%%%%%%%%

\section{Towards Yangians of generic Lie algebras}\label{sec:Towards_Lie}

Surely, we would like to extend the machinery presented in this paper to all simple Lie algebras and respective Yangians, yet much more practical calculations of BPS spectra structures in various chambers need to be done.
So in this section we make a step in this direction by proposing a \emph{hypothesis} of how duality rules of Sec.~\ref{sec:Weyl_mut} should be generalized to the case of non-simply laced Dynkin diagrams $\myB_n$, $\myC_n$, $\myF_4$ and $\myG_2$ since those seem the most intriguing ones from the point of view of generalizing rule \ref{majoritem} in Sec.~\ref{sec:Weyl_mut}.

Following \cite{Bao:2023ece,Bershadsky:1996nh,Cecotti:2012gh,Bao:2025hfu,Chen:2025xoe} to obtain D-brains probing singularities associated with non-simply laced Dynkin diagrams one should apply a so called Slodowy correspondence \cite{slodowy2006simple,2005math.....10216S} generalizing the McKay correspondence.
Surprisingly, the fact that the root lattice is non-simply laced does not affect quiver arrow dispositions, rather it modifies superpotentials.
If one denotes the adjoint matter field corresponding to node $a$ as $Y_a$, then a Dynkin edge connecting nodes $a$ and $b$ contributes with oppositely flowing arrow fields $X_{ab}$ and $X_{ba}$ and the following superpotnetial term:
\begin{equation}
	W_{ab}=\Tr\left(X_{ba}Y_a^{-\CA_{ab}}X_{ab}-X_{ab}Y_b^{-\CA_{ba}}X_{ba}\right)\,,
\end{equation}
where $\CA_{ab}$ are corresponding elements of the Cartan matrix.
In other words, using Dynkin notations an oriented $\ell$-edge induces the following superpotential:
\begin{equation}\label{DynkinSuper}
	\raisebox{0.1cm}{$\begin{array}{c}
		\begin{tikzpicture}[xscale=-1]
			\draw (0,0.05) -- (1,0.05) (0,-0.05) -- (1,-0.05) (0,0) -- (1,0);
			\begin{scope}[shift={(0.5,0)}]
				\draw (-0.1,0) -- (0,0.1) (-0.1,0) -- (0,-0.1); 
			\end{scope}
			\draw[fill=white] (0,0) circle (0.08) (1,0) circle (0.08);
			\node[right] at (-0.1,0) {$\scriptstyle b$};
			\node[left] at (1.1,0) {$\scriptstyle a$};
			\node[above, \mygreen] at (0.5,0.05) {$\scriptstyle \ell$};
		\end{tikzpicture}
	\end{array}$}\hspace{-0.3cm}\to W_{ab}=\Tr\left(X_{ba}Y_aX_{ab}-X_{ab}Y_b^{{\color{\mygreen}\ell}}X_{ba}\right),\; \raisebox{0.1cm}{$\begin{array}{c}
	\begin{tikzpicture}
		\draw (0,0.05) -- (1,0.05) (0,-0.05) -- (1,-0.05) (0,0) -- (1,0);
		\begin{scope}[shift={(0.5,0)}]
			\draw (-0.1,0) -- (0,0.1) (-0.1,0) -- (0,-0.1); 
		\end{scope}
		\draw[fill=white] (0,0) circle (0.08) (1,0) circle (0.08);
		\node[left] at (-0.1,0) {$\scriptstyle a$};
		\node[right] at (1.1,0) {$\scriptstyle b$};
		\node[above, \mygreen] at (0.5,0.05) {$\scriptstyle \ell$};
	\end{tikzpicture}
	\end{array}$}\hspace{-0.3cm}\to W_{ab}=\Tr\left(X_{ba}Y_a^{{\color{\mygreen}\ell}}X_{ab}-X_{ab}Y_bX_{ba}\right)\,.
\end{equation}

So here we continue to work with \emph{framed} quivers of form \eqref{quiver}, yet the superpotential specification depends on the Dynkin diagram choice (see Table \ref{fig:nslD}).

\begin{table}[ht!]
	\centering
	\begingroup
	\renewcommand*{\arraystretch}{1.5}
	$\begin{array}{|c|c|c|}
		\hline
		&\mbox{Dynkin diagram} & \mbox{Superpotential}\\
		\hline
		\myB_n & \begin{array}{c}
			\begin{tikzpicture}
				\draw (0,0) -- (1,0) (2,0.03) -- (3,0.03) (2,-0.03) -- (3,-0.03);
				\draw[dashed] (1,0) -- (2,0);
				\begin{scope}[shift={(2.5,0)}]
					\draw (0.1,0) -- (0,0.1) (0.1,0) -- (0,-0.1); 
				\end{scope}
				\draw[fill=white] (0,0) circle (0.08) (1,0) circle (0.08) (2,0) circle (0.08) (3,0) circle (0.08); 
			\end{tikzpicture}
		\end{array} & W=\Tr\left[\sum\lm_{a=1}^{n-2}\left(A_aC_aB_a-B_{a}C_{a+1}A_{a}\right)+{\color{\mygreen}\left(C_{n-1}B_{n-1}A_{n-1}-C_n^{2}A_{n-1}B_{n-1}\right)}+{\color{burgundy}S C_u^{h}R}\right]\\
		\hline
		\myC_n & \begin{array}{c}
			\begin{tikzpicture}
				\draw (0,0) -- (1,0) (2,0.03) -- (3,0.03) (2,-0.03) -- (3,-0.03);
				\draw[dashed] (1,0) -- (2,0);
				\begin{scope}[shift={(2.5,0)}]
					\draw (-0.1,0) -- (0,0.1) (-0.1,0) -- (0,-0.1); 
				\end{scope}
				\draw[fill=white] (0,0) circle (0.08) (1,0) circle (0.08) (2,0) circle (0.08) (3,0) circle (0.08); 
			\end{tikzpicture}
		\end{array} & W=\Tr\left[\sum\lm_{a=1}^{n-2}\left(A_aC_aB_a-B_{a}C_{a+1}A_{a}\right)+{\color{\mygreen}\left(C_{n-1}^{2}B_{n-1}A_{n-1}-C_nA_{n-1}B_{n-1}\right)}+{\color{burgundy}S C_u^{h}R}\right]\\
		\hline
		\myF_4 & \begin{array}{c}
			\begin{tikzpicture}
				\draw (0,0) -- (1,0) (1,0.03) -- (2,0.03) (1,-0.03) -- (2,-0.03) (2,0) -- (3,0);
				\begin{scope}[shift={(1.5,0)}]
					\draw (0.1,0) -- (0,0.1) (0.1,0) -- (0,-0.1); 
				\end{scope}
				\draw[fill=white] (0,0) circle (0.08) (1,0) circle (0.08) (2,0) circle (0.08) (3,0) circle (0.08); 
			\end{tikzpicture}
		\end{array} &
		\begin{array}{c}
			W=\Tr\bigg[\left(C_1B_1A_1-C_2A_1B_1\right)+{\color{\mygreen}\left(C_2B_2A_2-C_3^2A_2B_2\right)}+\\
			+\left(C_3B_3A_3-C_4A_3B_3\right)+{\color{burgundy}S C_u^{h}R}\bigg]
		\end{array}\\
		\hline
		\myG_2 & \begin{array}{c}
			\begin{tikzpicture}
				\draw (0,0.05) -- (1,0.05) (0,-0.05) -- (1,-0.05) (0,0) -- (1,0);
				\begin{scope}[shift={(0.5,0)}]
					\draw (-0.1,0) -- (0,0.1) (-0.1,0) -- (0,-0.1); 
				\end{scope}
				\draw[fill=white] (0,0) circle (0.08) (1,0) circle (0.08); 
			\end{tikzpicture}
		\end{array} & W=\Tr\left[{\color{\mygreen}\left(C_1^3B_1A_1-C_2A_1B_1\right)}+{\color{burgundy}S C_u^{h}R}\right]\\
		\hline
	\end{array}$
	\endgroup
	\caption{Superpotentials for non-simply laced Dynkin diagrams}\label{fig:nslD}
\end{table}

We expect that the choice of the quiver framing in this form leads again to a correspondence of fixed points to vectors of an irreducible lowest weight representation of weight $-h \omega_u$.
So that the weight-dimension formula \eqref{weight} holds and allows one to identify weight $w$ of the weight subspace $U_{w}$ with quiver dimensions $d_a(w)$.
Also we expect that relation \eqref{DTSchur} transforms into the following relation:
\begin{equation}
	{\bf DT}_{\fQ_{n,u,h}}\left(\{q_a\}\right)=\sum\lm_{d_a\geq 0} \left({\rm dim}\,U_{-h\omega_u+\sum\lm_a d_a\alpha_a}\right)\,\prod\lm_a q_a^{d_a}\,,
\end{equation}
where the weight subspace dimensions are read off from the standard Lie algebra character expression (see e.g. \cite[eq. (13.161)]{DiFrancesco:1997nk}):
\begin{equation}
	\chi_{\lambda}=\sum\lm_w {\rm dim}\,U_{w}\; e^w\,.
\end{equation}

The action of the Weyl mutations in this case is expected to comply with analogies established in Table \ref{thetable}.
So we expect the following transforms for stability parameters and for quiver dimensions:
\begin{equation}
	s_{a}(\zeta_{b}) = \zeta_{b} - \CA_{ba}\zeta_{a}\,,\qquad s_{a}(d_{b}) = \delta_{ab}\delta_{au}h + d_{b} - \delta_{ab}\sum\lm_{c = 1}^{n}\CA_{cb}d_{c}\,.
\end{equation}

Clearly the local structure of the superpotential for node $a$ is the same as in the simply laced case unless $a$ is a short root connected to a long root (see \eqref{DynkinSuper}).
In the first case we assume the rules of Sec.~\ref{sec:Weyl_mut} are unchanged.
Let au suppose that a thick Dynkin $\ell$-edge connects nodes in the following way:
\begin{equation}
	\begin{array}{c}
		\begin{tikzpicture}[xscale=-1]
			\draw (0,0.05) -- (1,0.05) (0,-0.05) -- (1,-0.05) (0,0) -- (1,0) (0,0) -- (-1,0);
			\begin{scope}[shift={(0.5,0)}]
				\draw (-0.1,0) -- (0,0.1) (-0.1,0) -- (0,-0.1); 
			\end{scope}
			\draw[fill=white] (0,0) circle (0.08) (1,0) circle (0.08) (-1,0) circle (0.08);
			\node[above, \mygreen] at (0.5,0.05) {$\scriptstyle \ell$};
			\node[above] at (-1.1,0.1) {$\scriptstyle a+1$};
			\node[above] at (0,0.1) {$\scriptstyle a$};
			\node[above] at (1.1,0.1) {$\scriptstyle a-1$};
		\end{tikzpicture}
	\end{array}\,.
\end{equation}
In this case dimension of $V_{a}$ is transformed under mutation in the following way:
\begin{equation}
	\check{d}_{a} = \delta_{a,u}h_{u} + d_{a + 1} + {\color{\mygreen}\ell} \, d_{a - 1} - d_{a}\,.
\end{equation}
Motivated by this, we suggest the following modification of the maps $\alpha$ and $\beta$ in rule \eqref{majoritem}:
\begin{equation}\label{non-simp1}
	\begin{aligned}
		&\alpha=\left(\begin{array}{ccccc}
			\check B_{a-1} & \color{\mygreen}{\check{B}_{a - 1}\check{C}_{a}} & \dots & \color{\mygreen}{\check{B}_{a - 1}\check{C}_{a}^{\ell - 1}} & \check A_{a}
		\end{array}\right)^T,\; \beta=\left(\begin{array}{ccccc}
			\color{\mygreen}{-C_{a}^{\ell - 1}A_{a - 1}} & \dots & \color{\mygreen}{-C_{a}A_{a - 1}} & -A_{a-1} & B_{a}
		\end{array}\right)\,,
	\end{aligned}
\end{equation}
in the case $a\neq u$, otherwise
\begin{equation}\label{non-simp2}
	\begin{aligned}	
		&\alpha=\left(\begin{array}{ccccccccc}
			\check S & \check S\check C_u &\ldots & \check S\check C_u^{h-1} & \check B_{u-1} & \color{\mygreen}{\check{B}_{u - 1}\check{C}_{u}} & \dots & \color{\mygreen}{\check{B}_{u - 1}\check{C}_{u}^{\ell - 1}} & \check A_{u}
		\end{array}\right)^T\,,\\
		&\beta=\left(\begin{array}{ccccccccc}
			C_u^{h-1}R & C_u^{h-2}R &\ldots & R& \color{\mygreen}{-C_{u}^{\ell - 1}A_{u - 1}} & \dots & \color{\mygreen}{-C_{u}A_{u - 1}} &-A_{u-1} & B_{u}
		\end{array}\right)\,.
	\end{aligned}
\end{equation}
These maps induce a short exact sequence:
\begin{equation}
	\begin{array}{c}
		\begin{tikzpicture}[baseline=(C.base)]
			\node (A) at (-6,0) {$0$};
			\node (B) at (-4.5,0) {$\check V_a$};
			\node (C) at (0,0) {$\IC^{h\delta_{u,a}}\oplus {\color{\mygreen}\smash{\overset{\ell}{\overbrace{V_{a-1}\oplus \dots \oplus V_{a - 1}}}}} \oplus V_{a+1}$};
			\node (D) at (4.5,0) {$V_a$};
			\node (E) at (6,0) {$0$};
			\path (A) edge[->] (B) (B) edge[->] node[above] {$\scriptstyle \alpha$} (C) (C) edge[->] node[above] {$\scriptstyle \beta$} (D) (D) edge[->] (E);
		\end{tikzpicture}
	\end{array}\,.
\end{equation}
We tested our hypothesis on the simplest examples in App.~\ref{app:B2} and App.~\ref{app:G2}.

%%%%%%%%%%%%%%%%%%%%%%%%%%%%%%%%%%%%%%%%%%%%%%%%%%%%%%%%%%%%%%%%%%%%%%%%%%%%%%%%%
%%%%%%%%%%%%%%%%%%%%%%%%%%%%%%%%%%%%%%%%%%%%%%%%%%%%%%%%%%%%%%%%%%%%%%%%%%%%%%%%%
%%%%%%%%%%%%%%%%%%%%%%%%%%%%%%%%%%%%%%%%%%%%%%%%%%%%%%%%%%%%%%%%%%%%%%%%%%%%%%%%%

\section{Conclusion}

In this paper we exposed the action of Weyl mutations on various elements of quiver Yangians associated to $\myA_n$-type quiver diagrams.
Weyl mutations are a form of Seiberg-like electro-magnetic duality mapping quiver theories into each other.
Under this action the very quiver remains self-dual, yet various parameters vary covariantly analogously to the Weyl group action on Lie algebras (see Table \ref{thetable}).
The moduli space of the quiver gauge theory is cut into stability chambers with constant spectra of BPS particles.
We showed that for quiver theories of $\myA_n$-type all the stability chambers form an orbit of the Weyl mutations starting with the cyclic chamber.

We constructed the action of Weyl mutations on fixed points points on quiver varieties corresponding to representation vectors and on the very generators of the quiver Yangian algebra.
We showed that the quiver Yangian algebra is a wall-crossing invariant and is isomorphic to the Yangian algebra $Y(\fs\fl_{n+1})$ constructed from the Lie algebra $\fs\fl_{n+1}$.

To conclude let us list some open problems for further investigation:
\begin{itemize}
	\item Rules for Weyl mutations on Nakajima quiver varieties \cite{lusztig2000quiver} and their extension presented in Sec.~\ref{sec:Weyl_mut}, as we mentioned, have many similarities with Seiberg dualities.
	Yet as well there are certainly \emph{unusual} rules for Seiberg duality and for generalized Seiberg-Kutasov duality.
	It would be illuminating to give an explanation for these unusual rules purely within the physical context of dualities.
	%%%%%%%%%%%%%%%%%%%%%%%%%%%%%%%%%%%%%%%%%%%%%%%%%%%%%%%%%%%
	\item A mutation maps a solution to highly non-linear equations \eqref{fp_eq} to another solution.
	We can not help noticing this job is analogous to an action of the B\"acklund transformation in integrable hierarchies \cite{chern1980analogue}.
	Surely, it would be interesting to extend this analogy.
	On the other hand, the B\"acklund transformation becomes a linear transform on an auxiliary space to the soliton equation, whereas for Nakajima quiver varieties describing instantons on ALE spaces \cite{kronheimer1989construction,kronheimer1990yang} the Weyl group action looks more natural on instanton solutions rather than on Nahm-dual ADHM equations describing a quiver variety.
	%%%%%%%%%%%%%%%%%%%%%%%%%%%%%%%%%%%%%%%%%%%%%%%%%%%%%%%%%%%
	\item It would be interesting to extend the action of the Weyl mutations form quiver Yangians to quantum toroidal algebras and compare this action to the dynamical Weyl group on the K-theory for  Nakajima quiver varieties \cite{Okounkov:2016sya} and beyond.
	%%%%%%%%%%%%%%%%%%%%%%%%%%%%%%%%%%%%%%%%%%%%%%%%%%%%%%%%%%%
	\item Exponential map \eqref{Weyl_Tits} performing Weyl reflections on Lie algebras and Weyl mutations on the quiver Yangian suggests that other transforms of similar type like transition between commuting rays in the DIM algebra \cite{Mironov:2020pcd,Mironov:2023wga} could be translated to some duality of physical systems.
	Another argument in favor of this strategy might be a collection of wall-crossing phenomena for elliptic stable envelopes \cite{Smirnov:2018drm,Crew:2020psc}.
	%%%%%%%%%%%%%%%%%%%%%%%%%%%%%%%%%%%%%%%%%%%%%%%%%%%%%%%%%%%
	\item In Sec.~\ref{sec:MutYang} we considered the action of Weyl mutations only in the next to trivial case of $Y(\fs\fl_3)$.
	Surely, it would be interesting to check the validity of the proposed Weyl action in more general cases of $Y(\fs\fl_{n+1})$.
	%%%%%%%%%%%%%%%%%%%%%%%%%%%%%%%%%%%%%%%%%%%%%%%%%%%%%%%%%%%
	\item To construct generators of quiver Yangians outside the cyclic chamber we exploited for fixed points Weyl mutation pre-imaged Young diagrams in the cyclic chamber.
	This definition does not seem very optimal.
	At least it depends explicitly on an existence of a Weyl group orbit connecting the stability chamber in question with the cyclic chamber.
	Apparently, it is not an obstruction in our case of $\myA_n$ quivers.
	However it would be quite useful to present a universal, invariant definition of quiver Yangians independent of the stability chamber choice.
	%%%%%%%%%%%%%%%%%%%%%%%%%%%%%%%%%%%%%%%%%%%%%%%%%%%%%%%%%%%
	 \item A next big question to be addressed in this field is \emph{the possibility that} and \emph{the way how}
	the Tannaka-Krein approach (for a review see \cite{joyal2006introduction} and references therein)
	can be extended to Yangians and further -- to quantum toroidal and elliptic algebras.
	For ordinary Lie algebras it appears possible to change the commutation relations for multiplication
	and decomposition of representations.
	In a sense this is a possibility to switch between multiplication and comultiplication structures.
	An immediate problem is that at the simplest level the algebra of representations is commutative,
	thus to get chances for equivalence with the underlying non-commutative algebra
	one needs to consider a more sophisticated notion of 
	category of representations, where it is important that the product of irreducibles 
	can be a sophisticated composition of those, with non-trivial multiplicities and braiding.
	For Yangians we saw that the structure of representations looks very rich and involved,
	with non-trivial action of mutations -- which at the same time have a clear meaning and interpretation.
	This gives chances to describe the algebra-representation duality in a more natural way,
	than provided by its reduction to just quantum groups, rational conformal theories and RT theory of knots
	\cite{reshetikhin1991invariants,Mironov:2011aa}.
	%perhaps, even more straightforward than for ordinary Lie algebras. 
	Tannaka-Krein approach is a kind of crucial for physics \cite{Morozov:1992ea}, 
	because ordinary algebraic structures 
	are hidden in Lagrangian/Hamiltonian \emph{formalism},
	sometimes rather deeply \cite{Mironov:2025yak}, 
	while representations, their characters and braiding relations
	are directly observable quantities, independent of the underground formalism.
	Understanding Tannaka-Krein duality is thus a question of principal importance,
	at least conceptually, since it relates \emph{hidden} symmetries with their \emph{actual} realizations.
	\item At this point it seems natural to draw some attention to a problem of constructing 3-Schur functions \cite{Morozov:2018fjb,Wang:2020zlv,Cui:2022dqp,10.1063/5.0033847,Morozov:2022ndt,Morozov:2023vra} (an analog of Schur functions for plane partitions).
	It is well-known that Schur, Jack and Macdonald polynomials are D-brane BPS wave functions corresponding to fixed points associated with 2d Young diagrams (ordinary partitions).
	A success in this description is achieved since in these so called Fock representations one is able to determine families of mutually commuting ``time operators'' $p_k$ adding $k$ cells, and their Heisenberg counterparts $\p/\p p_k$ subtracting $k$ cells.
	So a Schur/Jack/Macdonald polynomial corresponding to a 2d diagram $\lambda$ ``explains'' how $\lambda$ is constructed by applying canonical operators $p_k$.
	It turns out to be notoriously difficult to present a similar construction for 3d Young diagrams, a.k.a. plane partitions, a.k.a. Macmahon modules.
	It looks even more surprising in the current context where vectors of considered representations for Yangians $Y(\fs\fl_{n+1})$ are labeled by box-bounded, yet honest 3d Young diagrams.
	On the other hand as we have seen by construction $Y(\fs\fl_{n+1})$ captures many features of ordinary Lie algebras having nice free field representations \cite{Gerasimov:1990fi,Morozov:2021hwr,Morozov:2022ocp}.
	\item Preliminary estimates indicate that all the structures could be extended naturally beyond the Dynkin $\myA$-series, probably even to Kac-Moody algebras.
	However various intriguing nuances start to appear in these considerations: for example, fixed points even in the cyclic chamber are \emph{not} 3d Young diagrams anymore.
	And one has to develop more rigorous approaches to new emerging objects.
	We will provide details separately elsewhere.
\end{itemize}

%%%%%%%%%%%%%%%%%%%%%%%%%%%%%%%%%%%%%%%%%%%%%%%%%%%%%%%%%%%%%%%%%%%%%%%%%%%%%%%%%%%%%%%%%%%%%%%%%%%
%%%%%%%%%%%%%%%%%%%%%%%%%%%%%%%%%%%%%%%%%%%%%%%%%%%%%%%%%%%%%%%%%%%%%%%%%%%%%%%%%%%%%%%%%%%%%%%%%%%
%%%%%%%%%%%%%%%%%%%%%%%%%%%%%%%%%%%%%%%%%%%%%%%%%%%%%%%%%%%%%%%%%%%%%%%%%%%%%%%%%%%%%%%%%%%%%%%%%%%

\section*{Acknowledgments}
We would like to thank Andrei Mironov and Nikita Tselousov for illuminating discussions.
The work
was supported by the state assignment of the Institute for Information Transmission Problems of RAS.

%%%%%%%%%%%%%%%%%%%%%%%%%%%%%%%%%%%%%%%%%%%%%%%%%%%%%%%%%%%%%%%%%%%%%%%%%%%%%%%%%%%%%%%%%%%%%%%%%%%
%%%%%%%%%%%%%%%%%%%%%%%%%%%%%%%%%%%%%%%%%%%%%%%%%%%%%%%%%%%%%%%%%%%%%%%%%%%%%%%%%%%%%%%%%%%%%%%%%%%
%%%%%%%%%%%%%%%%%%%%%%%%%%%%%%%%%%%%%%%%%%%%%%%%%%%%%%%%%%%%%%%%%%%%%%%%%%%%%%%%%%%%%%%%%%%%%%%%%%%

\appendix

\section{Weyl Automorphisms of Lie algebras}\label{App: Weyl Automorphisms}

In this section, we describe how the Weyl group of a chosen Lie algebra $\fg$ induce the algebra automorphisms. This mathematical exercise is fairly simple and can be found, for example, in \cite{humphreys2012introduction, fulton2013representation}.

To be more precise, we choose the algebra automorphism
\begin{equation}\label{Weyl_Tits_2}
	\tau_{i} = e^{{\rm ad}(e_{i})}e^{-{\rm ad}(f_{i})}e^{{\rm ad}(e_{i})}\,,
\end{equation}
and show that it induces simple reflection $s_{i}$, meaning:
\begin{equation}
	\tau_{i}(\fg_{\alpha_{j}}) = \fg_{s_{i}(\alpha_{j})}\,, \quad \tau_{i}(\fh) = \fh\,,
\end{equation}
where we use the root decomposition $\fg = \fh \bigoplus_{\alpha \in \Delta}\fg_{\alpha}$.

First, we focus on an arbitrary $\fs\fl_{2}^{(\alpha_{i})}$ subalgebra $\{e_{i}, f_{i}, h_{i}\}$. 
We observe that the exponential series of the adjoint action are truncated:
\begin{equation}
	\begin{aligned}
		e^{{\rm ad}(e_{i})}(e_{i}) = e_{i},\quad e^{{\rm ad}(e_{i})}(h_{i}) = h_{i} - 2e_{i},\quad e^{{\rm ad}(e_{i})}(f_{i}) = f_{i} + h_{i} - e_{i}\,.
	\end{aligned}
\end{equation}
We also find rather useful to calculate the action of the lowering operator:
\begin{equation}
	\begin{aligned}
		e^{-{\rm ad}(f_{i})}(e_{i}) = e_{i} + h_{i} - f_{i},\quad e^{-{\rm ad}(f_{i})}(h_{i}) = h_{i} - 2f_{i},\quad e^{-{\rm ad}(f_{i})}(f_{i}) = f_{i}\,.
	\end{aligned}
\end{equation}
Combining these actions we derive:
\begin{equation}
	\begin{aligned}
		\tau_{i}(e_{i}) = -f_{i},\quad \tau_{i}(h_{i}) = -h_{i},\quad  \tau_{i}(f_{i}) = -e_{i}\,.
	\end{aligned}
\end{equation}
While the calculations between different $\fs\fl_{2}$ subalgebras are more involved, the Serre relations truncate the operators again.

We start with the Cartan subalgebra:
\begin{equation}
	\begin{aligned}
		&\tau_{i}(h_{j}) = e^{{\rm ad}(e_{i})}e^{-{\rm ad}(f_{i})}e^{{\rm ad}(e_{i})}(h_{j})\,,\\
		&e^{{\rm ad}(e_{i})}(h_{j}) = h_{j} - \CA_{ji}e_{i}\,,\\
		&\tau_{i}(h_{j}) = e^{{\rm ad}(e_{i})}e^{-{\rm ad}(f_{i})}(h_{j} - \CA_{ji}e_{i}) = e^{{\rm ad}(e_{i})}(h_{j} - \CA_{ji}e_{i} - \CA_{ji}h_{i})\,,\\
		&\tau_{i}(h_{j}) = h_{j} - \CA_{ji}h_{i} = h_{j} - \alpha_{i}(h_{j})h_{i}\,.
	\end{aligned}
\end{equation}
Now we are ready to analyze $\tau_{i}(e_{j})$. 
Let $h\in \fh$ be arbitrary. A quick observation shows that the roots change with respect to the Weyl group reads:
\begin{equation}
	\begin{aligned}
		\left[h, \tau_{i}(e_{j})\right] &= [\tau_{i}(h), \tau_{i}(e_{j})] + \alpha_{i}(h)[h_{i}, \tau_{i}(e_{j})] = \tau_{i}\bigl([h, e_{j}]\bigr) - \alpha_i(h)[\tau_{i}(h_{i}), \tau_{i}(e_{j})] = \,\\
		&= \alpha_{j}(h)\tau_{i}(e_{j}) - \alpha_{i}(h)\tau_{i}\bigl([h_{i}, e_{j}]\bigr) = \alpha_{j}(h)\tau_{i}(e_{j}) - \alpha_{i}(h)\alpha_{j}(h_{i})\tau_{i}(e_{j}) = \,\\
		&= \bigl(\alpha_{j}(h) - \alpha_{i}(h)\alpha_{j}(h_{i})\bigr)\tau_{i}(e_{j}) = \bigl(\alpha_{j} - \CA_{ji}\alpha_{i}\bigr)(h)\, \tau_{i}(e_{j}) = \,\\
		&= \underline{s_{i}(\alpha_j)(h)}\, \tau_{i}(e_{j})\,.
	\end{aligned}
\end{equation}
Here we used the fact that $\tau_{i}$ is a Lie algebra automorphism together with the definition of the Cartan matrix $\CA_{ji} = \alpha_{j}(h_{i})$. 
The result tells us that ${\rm ad}(h)$ has eigenvalue $s_{i}(\alpha_j)(h)$, which indicates essentially $\tau_{i}(e_{j})$ lives in the root space $\fg_{s_{i}(\alpha_{j})}$. 
Therefore, $\tau_{i}$ could be regarded as an automorphism induced by simple reflection $s_{i}$.

Furthermore we derive the explicit automorphisms for the algebras $\fs\fl_{n+1}$ whose Cartan matrices are given by \eqref{sl(n)_data}. 
First, we choose $|i - j| > 1$. In this case, the Serre relations are very simple:
\begin{equation}
	{\rm ad}_{e_{i}}(e_{j}) = 0\,, \quad {\rm ad}_{f_{i}}(f_{j}) = 0\,.
\end{equation}
Therefore:
\begin{equation}
	\begin{aligned}
		&\tau_{i}(e_{j}) = e^{{\rm ad}(e_{i})}e^{-{\rm ad}(f_{i})} (e_{j}) = e^{{\rm ad}(e_{i})}(e_{j}) = e_{j}\,\\
		&\tau_{i}(f_{j}) = f_{j}\,.
	\end{aligned}
\end{equation}
The non-trivial cases are $|i - j| = 1$. We start with the raising operators:
\begin{equation}
	\begin{aligned}
		\tau_{i}(e_{j}) &= e^{{\rm ad}(e_{i})}e^{-{\rm ad}(f_{i})}e^{{\rm ad}(e_{i})}(e_{j}) = e^{{\rm ad}(e_{i})}e^{-{\rm ad}(f_{i})}(e_{j} + [e_{i}, e_{j}]) = \,\\
		&= e^{{\rm ad}(e_{i})}(e_{j} + [e_{i}, e_{j}] - e_{j}) = \underline{[e_{i}, e_{j}]}\,.
	\end{aligned}
\end{equation}
One could perform similar calculations with the lowering operators:
\begin{equation}
	\tau_{i}(f_{j}) = e^{{\rm ad}(e_{i})}e^{-{\rm ad}(f_{i})}(f_{j}) = f_{j} - [f_{i}, f_{j}] - f_{j} = \underline{[f_{j}, f_{i}]}\,.
\end{equation}
The case $i = j$ was already covered above in detail.

For other Lie algebras the calculations are more involved. We will not cover them here. Instead we would argue the choice of automorphisms \eqref{Weyl_Tits_2} generates the following rule:
\begin{equation}
	\begin{aligned}
		&s_{\alpha}(\beta) = \beta - \CA_{\beta\alpha}\alpha\,,\\
		&\tau_{\alpha}(e_{\beta}) = \frac{1}{m!}[\underset{m}{\underbrace{e_{\alpha}, \dots, [e_{\alpha}}}, e_{\beta}]] = \frac{1}{m!}\bigl({\rm ad}(e_{\alpha})\bigr)^{m}(e_{\beta})\,,\\
		&\tau_{\alpha}(f_{\beta}) =\frac{(-1)^{m}}{m!} [\underset{m}{\underbrace{f_{\alpha}, \dots, [f_{\alpha}}, f_{\beta}]}] = \frac{(-1)^{m}}{m!}\bigl({\rm ad}(f_{\alpha})\bigr)^{m}(f_{\beta})\,,
	\end{aligned}
\end{equation}
where $m = -\CA_{\beta\alpha}$. This rule is also supported by the fact that $e_{\beta}$, as a vector in the adjoint representation of a chosen Lie algebra, is the lowest vector in $\fs\fl_{2}^{(\alpha)}$ sub-module with weight $-m$, what follows from the Serre relations. 

%%%%%%%%%%%%%%%%%%%%%%%%%%%%%%%%%%%%%%%%%%%%%%%%%%%%%%%%%%%%%%%%%%%%%%%%%%%%%%%%%%%%%%%%%%
%%%%%%%%%%%%%%%%%%%%%%%%%%%%%%%%%%%%%%%%%%%%%%%%%%%%%%%%%%%%%%%%%%%%%%%%%%%%%%%%%%%%%%%%%%
%%%%%%%%%%%%%%%%%%%%%%%%%%%%%%%%%%%%%%%%%%%%%%%%%%%%%%%%%%%%%%%%%%%%%%%%%%%%%%%%%%%%%%%%%%
%%%%%%%%%%%%%%%%%%%%%%%%%%%%%%%%%%%%%%%%%%%%%%%%%%%%%%%%%%%%%%%%%%%%%%%%%%%%%%%%%%%%%%%%%%

\section{Moduli space examples for \texorpdfstring{$\myA$}{A} series}\label{app:examples}

\subsection{Quiver \texorpdfstring{$\fs\fl_2$}{su2}}\label{sec:exp_sl_2}

The simplest example of quivers where the machinery discussed in this note is applicable is a family of quivers for $\fs\fl_2$.
They have only one available free parameter $h$:
\begin{equation}
	\fQ_{1,1,h}=\left\{\begin{array}{c}
		\begin{tikzpicture}
			%%%%%%%%%%%%%%%%%%%%%%%%%%%%%%%%
			\begin{scope}[rotate=-90]
				\draw[postaction=decorate, decoration={markings, mark= at position 0.7 with {\arrow{stealth}}}] (0,-1.2) to[out=100,in=260] node[pos=0.3, above] {$\scriptstyle R$} (0,0);
				\draw[postaction=decorate, decoration={markings, mark= at position 0.7 with {\arrow{stealth}}}] (0,0) to[out=280,in=80] node[pos=0.7, below] {$\scriptstyle S$} (0,-1.2);
				\begin{scope}[shift={(0,-1.2)}]
					\draw[fill=gray] (-0.08,-0.08) -- (-0.08,0.08) -- (0.08,0.08) -- (0.08,-0.08) -- cycle;
				\end{scope}
			\end{scope}
			%%%%%%%%%%%%%%%%%%%%%%%%%%%%%%%%%%%%%%%%%%%%%%%%%%%%5
			\begin{scope}[rotate=-90]
			\draw[postaction=decorate, decoration={markings, mark= at position 0.8 with {\arrow{stealth}}}] (0,0) to[out=60,in=0] (0,0.6) to[out=180,in=120] (0,0);
			\end{scope}
			\node[right] at (0.6,0) {$\scriptstyle C_1$};
			\draw[fill=\myblue] (0,0) circle (0.08);
			%%%%%%%%%%%%%%%%%%%%%%%%%%%%%%%%%%%%%%%%%%%%%%%%%%%%%%%%%%%%%%%
			\node[below] at (0,-0.08) {$\scriptstyle \zeta_1$};
		\end{tikzpicture}
	\end{array},\quad W=\Tr\left[{\color{burgundy}S C_1^{h}R}\right]\right\}\,.
\end{equation}

The moduli space is 1d and is spanned by $\zeta_1$.
There is a marginal stability wall at $\zeta_1^{\rm crit}=0$ and two respective phases ${\bf Ph}_+:\;\zeta_1>0$, ${\bf Ph}_-:\;\zeta_1<0$:
\begin{equation}
	\begin{array}{c}
		\begin{tikzpicture}
			\draw[draw=none, fill=white!70!green] (0,-0.1) -- (0,0.1) -- (1.8,0.1) -- (1.8,-0.1) -- cycle;
			\draw[draw=none, fill=white!70!orange] (0,-0.1) -- (0,0.1) -- (-1.8,0.1) -- (-1.8,-0.1) -- cycle;
			\draw[-stealth] (-2,0) -- (2,0);
			\node[right] at (2,0) {$\scriptstyle \zeta_1$};
			\draw[ultra thick] (0,-0.1) -- (0,0.1);
			\node[below] at (0,-0.1) {$\scriptstyle 0$};
			\node[above] at (1,0.1) {$\scriptstyle {\bf Ph}_+$};
			\node[above] at (-1,0.1) {$\scriptstyle {\bf Ph}_-$};
		\end{tikzpicture}
	\end{array}
\end{equation}

We construct immediately in general fixed points for both branches corresponding to $h+1$-dimensional representations of $\fs\fl_2$:
\begin{equation}
	\begin{aligned}
		&{\bf Ph}_+:\quad v_k^{(+)}\to\raisebox{0.1cm}{$\begin{array}{c}
			\begin{tikzpicture}[yscale=-1]
				\draw[postaction=decorate, decoration={markings, mark= at position 0.7 with {\arrow{stealth}}}] (0,0) -- (1,0) node[pos=0.5,below] {$\scriptstyle R$};
				\draw[postaction=decorate, decoration={markings, mark= at position 0.7 with {\arrow{stealth}}}] (1,0) -- (2,0) node[pos=0.5,below] {$\scriptstyle C_1$};
				\draw[postaction=decorate, decoration={markings, mark= at position 0.7 with {\arrow{stealth}}}] (3,0) -- (4,0) node[pos=0.5,below] {$\scriptstyle C_1$};
				\draw[dashed] (2,0) -- (3,0);
				\draw[fill=\myblue] (1,0) circle (0.08) (2,0) circle (0.08) (3,0) circle (0.08) (4,0) circle (0.08);
				\draw[fill=gray] (-0.08,-0.08) -- (-0.08,0.08) -- (0.08,0.08) -- (0.08,-0.08) -- cycle;
				\draw (0.9,-0.1) to[out=270,in=180] (1,-0.2) -- (2.4,-0.2) to[out=0, in=90] (2.5,-0.3) to[out=90,in=180] (2.6,-0.2) -- (4,-0.2) to[out=0,in=270] (4.1,-0.1);
				\node[above] at (2.5,-0.3) {\scriptsize $k$ nodes};
			\end{tikzpicture}
		\end{array}$},\quad d_1=k,\quad 0\leq k\leq h\,,\\
		&{\bf Ph}_-:\quad v_k^{(-)}\to\raisebox{0.1cm}{$\begin{array}{c}
				\begin{tikzpicture}[yscale=-1]
					\draw[postaction=decorate, decoration={markings, mark= at position 0.3 with {\arrowreversed{stealth}}}] (0,0) -- (1,0) node[pos=0.5,below] {$\scriptstyle S$};
					\draw[postaction=decorate, decoration={markings, mark= at position 0.3 with {\arrowreversed{stealth}}}] (1,0) -- (2,0) node[pos=0.5,below] {$\scriptstyle C_1$};
					\draw[postaction=decorate, decoration={markings, mark= at position 0.3 with {\arrowreversed{stealth}}}] (3,0) -- (4,0) node[pos=0.5,below] {$\scriptstyle C_1$};
					\draw[dashed] (2,0) -- (3,0);
					\draw[fill=\myblue] (1,0) circle (0.08) (2,0) circle (0.08) (3,0) circle (0.08) (4,0) circle (0.08);
					\draw[fill=gray] (-0.08,-0.08) -- (-0.08,0.08) -- (0.08,0.08) -- (0.08,-0.08) -- cycle;
					\draw (0.9,-0.1) to[out=270,in=180] (1,-0.2) -- (2.4,-0.2) to[out=0, in=90] (2.5,-0.3) to[out=90,in=180] (2.6,-0.2) -- (4,-0.2) to[out=0,in=270] (4.1,-0.1);
					\node[above] at (2.5,-0.3) {\scriptsize $k$ nodes};
				\end{tikzpicture}
			\end{array}$},\quad d_1=k,\quad 0\leq k\leq h\,.
	\end{aligned}
\end{equation}

The generating function in this case reads:
\begin{equation}
	{\bf DT}_{\fQ_{1,1,h}}=\frac{1-q^{h+1}}{1-q}\,.
\end{equation}

Weyl mutations in this case are generated by a single reflection $s_1$, $s_1^2=\bbone$.
Mutation $s_1$ maps ${\bf Ph}_+$ with $\zeta_1>0$ to ${\bf Ph}_-$ with $\check \zeta_1=-\zeta_1<0$ according to \eqref{non_Nak_du_sta}.
Respectively, for quiver dimensions following \eqref{non_Nak_du_dim} we obtain $\check d_1=h-d_1$, so on solutions to \eqref{fp_eq} mutation $s_1$ acts as:
\begin{equation}
	s_1:\quad v_k^{(+)}\;\mapsto \; v_{h-k}^{(-)}, \quad 0\leq k\leq h\,.
\end{equation}

Let us check how other mutation rules of Sec.~\ref{sec:Weyl_mut} work in this situation.
The only non-trivial rule in this case is item \ref{majoritem}.
For given $k$ we calculate:
\begin{equation}
	\begin{aligned}
		& \alpha=\left(\begin{array}{cccccc}
			\check S & \check S\check C_u &\ldots & \check S\check C_u^{h-1}
		\end{array}\right)=\underbrace{\left(\begin{array}{cccccc}
		1 & 0 & \ldots &0 & \ldots & 0\\
		0 & 1 & \ldots & 0 & \ldots & 0\\
		\ldots\\
		0 & 0 & \ldots & 1 & \ldots & 0\\
		\end{array}\right)^T}_{h \;\rm{columns}}\left.\rule{0cm}{1cm}\right\}{\scriptstyle h-k\; {\rm rows}}\,,\\
		&\beta=\left(\begin{array}{cccccc}
			C_u^{h-1}R & C_u^{h-2}R &\ldots & R
		\end{array}\right)^T=\underbrace{\left(\begin{array}{cccccc}
			0 & \ldots & 1 &\ldots & 0 & 0\\
			\ldots\\
			0 & \ldots & 0 & \ldots & 1 & 0\\
			0 &\ldots & 0 & \ldots & 0 & 1\\
		\end{array}\right)}_{h\;\rm{columns}}\left.\rule{0cm}{1cm}\right\}{\scriptstyle k\; {\rm rows}}\,,
	\end{aligned}
\end{equation}
so that clearly,
\begin{equation}
	0\longrightarrow \IC^{h-k}\overset{\alpha}{\longrightarrow}\IC^h\overset{\beta}{\longrightarrow}\IC^k\longrightarrow 0 \mbox{ is a short exact sequence}\,.
\end{equation}

\emph{However} the opposite is not true.
If we act by mutation $s_1=s_1^{-1}$ we should return from ${\bf Ph}_-$ back to ${\bf Ph}_+$.
Nevertheless if we think of ${\bf Ph}_-$ as a primary phase and of ${\bf Ph}_+$ as a dual one, then $R=0$ and $\check S=0$. Moreover $\alpha=0$, $\beta=0$, and these maps do not form a short exact sequence.
This statement is independent of $h$ and could be extended to the case $h=1$ when the quiver variety is of the Nakajima type and rules from Sec.~\ref{sec:Nak_quiv_Weyl} are applicable.
We believe this \emph{inconsistency} could be cured by selecting as $\fQ$ a theory with $\zeta>0$ and as $\check \fQ$ the one with $\zeta<0$, or by considering a resolution for complex moment map by complex stability parameters \cite{lusztig2000quiver}.
We will consider the latter approach elsewhere.

%%%%%%%%%%%%%%%%%%%%%%%%%%%%%%%%%%%%%%%%%%%%%%%%%%%%%%%%%%%%%%%%%%%%%%%%%%%%%%%%%%%%%%%%%%
%%%%%%%%%%%%%%%%%%%%%%%%%%%%%%%%%%%%%%%%%%%%%%%%%%%%%%%%%%%%%%%%%%%%%%%%%%%%%%%%%%%%%%%%%%
%%%%%%%%%%%%%%%%%%%%%%%%%%%%%%%%%%%%%%%%%%%%%%%%%%%%%%%%%%%%%%%%%%%%%%%%%%%%%%%%%%%%%%%%%%
%%%%%%%%%%%%%%%%%%%%%%%%%%%%%%%%%%%%%%%%%%%%%%%%%%%%%%%%%%%%%%%%%%%%%%%%%%%%%%%%%%%%%%%%%%

\subsection{Quiver \texorpdfstring{$\fs\fl_3$}{su3}}

The phase (moduli) space picture for $\fs\fl_3$ is depicted in Fig.\ref{fig:su(3)_phsp}.

\begin{figure}[ht!]
	\centering
	\begin{tikzpicture}
		\draw[thick, -stealth] (-1.5,0) -- (1.5,0);
		\draw[thick, -stealth] (0,-1.5) -- (0,1.5);
		\node[right] at (1.5,0) {$\scriptstyle \zeta_1$};
		\node[above] at (0,1.5) {$\scriptstyle \zeta_2$};
		\draw[draw=white, fill=palette2] (0,0) -- (1.2,0) -- (1.2,1.2) -- (0,1.2) -- cycle;
		\draw[draw=white, fill=palette3] (0,0) -- (0,1.2) -- (-1.2,1.2) -- cycle;
		\draw[draw=white, fill=palette4] (0,0) -- (-1.2,1.2) -- (-1.2,0) -- cycle;
		\draw[draw=white, fill=palette6] (0,0) -- (-1.2,0) -- (-1.2,-1.2) -- (0,-1.2) -- cycle;
		\draw[draw=white, fill=palette7] (0,0) -- (0,-1.2) -- (1.2,-1.2) -- cycle;
		\draw[draw=white, fill=palette5] (0,0) -- (1.2,0) -- (1.2,-1.2) -- cycle;
		\node[white] at (0.5,0.5) {\footnotesize I};
		\node[white] at (-0.3,0.8) {\footnotesize II};
		\node[white] at (-0.8,0.3) {\footnotesize III};
		\node[white] at (-0.5,-0.5) {\footnotesize IV};
		\node[white] at (0.3,-0.8) {\footnotesize V};
		\node[white] at (0.8,-0.3) {\footnotesize VI};
	\end{tikzpicture}
	\caption{Moduli space of an $\fs\fl_3$ quiver.}\label{fig:su(3)_phsp}
\end{figure}

Consider transformations:
\begin{equation}
	s_1=\left(
	\begin{array}{cc}
		-1 & 0 \\
		1 & 1 \\
	\end{array}
	\right),\quad s_2=\left(
	\begin{array}{cc}
		1 & 1 \\
		0 & -1 \\
	\end{array}
	\right),\quad s_1^2=s_2^2=\bbone,\quad s_1s_2s_1=s_2s_1s_2\,.
\end{equation}
Let us say that $\vec{\zeta}>0$ corresponds to a sequence of constraints $\zeta_a>0$ for all components $\zeta_a$ of vector $\vec\zeta$.
Then we characterize phases in the following way (cf. Fig.~\ref{fig:su(3)_phsp}):\footnote{Let us note here that if a phase solid sector in the moduli space is characterized by a sequence of inequalities $s_a s_b s_c s_d \vec\zeta>0$, then to arrive to a point in this phase starting with any point in the cyclic chamber one should apply mutations in the reverse order $\vec\zeta=s_d s_c s_b s_a\vec\zeta_{\rm cyc}$.}
\begin{equation}
	\mbox{\bf I}:\;\vec{\zeta}>0,\quad \mbox{\bf II}:\;s_1\vec{\zeta}>0,\quad \mbox{\bf III}:\;s_2s_1\vec{\zeta}>0,\quad \mbox{\bf IV}:\;(s_1s_2s_1=s_2s_1s_2)\vec{\zeta}>0,\quad \mbox{\bf V}:\;s_1s_2\vec{\zeta}>0, \quad \mbox{\bf VI}:\;s_2\vec{\zeta}>0\,.
\end{equation}

\subsubsection{Irrep  \texorpdfstring{$[2,2]$}{[2,2]}}

Here we list all the explicit solutions in all phases to theory \eqref{su(3), [2]}.
The fields that are not mentioned all have zero vevs.

{\bf Phase I} $(\zeta_1>0,\zeta_2>0)$ solutions:
\begin{subequations}
\begin{equation}
\begin{array}{c}
			\begin{tikzpicture}
				\begin{scope}[shift={(-0.8,0)}]
					\draw[fill=gray] (-0.08,-0.08) -- (-0.08,0.08) -- (0.08,0.08) -- (0.08,-0.08) -- cycle;
				\end{scope}
			\end{tikzpicture}
		\end{array}: \quad \varnothing\,,
\end{equation}
\begin{equation}
		\begin{array}{c}
			\begin{tikzpicture}
				\draw[postaction=decorate, decoration={markings, mark= at position 0.7 with {\arrow{stealth}}}] (-0.8,0) -- (0,0) node[pos=0.5,below] {$\scriptstyle R$};
				\draw[fill=black!40!red] (0,0) circle (0.08);
				\begin{scope}[shift={(-0.8,0)}]
					\draw[fill=gray] (-0.08,-0.08) -- (-0.08,0.08) -- (0.08,0.08) -- (0.08,-0.08) -- cycle;
				\end{scope}
			\end{tikzpicture}
		\end{array}:\quad R=\left(\sqrt{\zeta_1}\right),\; C_1=(0)\,,
\end{equation}
\begin{equation}
		\begin{array}{c}
			\begin{tikzpicture}
				\draw[postaction=decorate, decoration={markings, mark= at position 0.7 with {\arrow{stealth}}}] (-0.8,0) -- (0,0) node[pos=0.5,below] {$\scriptstyle R$};
				\draw[postaction=decorate, decoration={markings, mark= at position 0.7 with {\arrow{stealth}}}] (0,0) -- (0,0.8) node[pos=0.5,left] {$\scriptstyle C_1$};
				\draw[fill=black!40!red] (0,0) circle (0.08) (0,0.8) circle (0.08);
				\begin{scope}[shift={(-0.8,0)}]
					\draw[fill=gray] (-0.08,-0.08) -- (-0.08,0.08) -- (0.08,0.08) -- (0.08,-0.08) -- cycle;
				\end{scope}
			\end{tikzpicture}
		\end{array}:\quad R=\left(\begin{array}{c}
			\sqrt{2\zeta_1}\\ 0
		\end{array}\right),\; C_1=\left(\begin{array}{cc}
			0 & 0 \\
			\sqrt{\zeta_1} & 0\\
		\end{array}\right)\,,
\end{equation}
\begin{equation}
		\begin{array}{c}
			\begin{tikzpicture}
				\draw[postaction=decorate, decoration={markings, mark= at position 0.7 with {\arrow{stealth}}}] (-0.8,0) -- (0,0) node[pos=0.5,below] {$\scriptstyle R$};
				\draw[postaction=decorate, decoration={markings, mark= at position 0.7 with {\arrow{stealth}}}] (0,0) -- (0.8,0) node[pos=0.5,above] {$\scriptstyle A_1$};
				\draw[fill=black!40!red] (0,0) circle (0.08);
				\draw[fill=black!40!green] (0.8,0) circle (0.08);
				\begin{scope}[shift={(-0.8,0)}]
					\draw[fill=gray] (-0.08,-0.08) -- (-0.08,0.08) -- (0.08,0.08) -- (0.08,-0.08) -- cycle;
				\end{scope}
			\end{tikzpicture}
		\end{array}:\quad R=\left(\sqrt{\zeta_1+\zeta_2}\right),\; C_1=(0),\; A_1=\left(\sqrt{\zeta_2}\right),\;B_1=\left(0\right),\quad C_2=(0)\,,
\end{equation}
\begin{equation}
		\begin{array}{c}
			\begin{tikzpicture}
				\draw[postaction=decorate, decoration={markings, mark= at position 0.7 with {\arrow{stealth}}}] (-0.8,0) -- (0,0) node[pos=0.5,below] {$\scriptstyle R$};
				\draw[postaction=decorate, decoration={markings, mark= at position 0.7 with {\arrow{stealth}}}] (0,0) -- (0,0.8) node[pos=0.5,left] {$\scriptstyle C_1$};
				\draw[postaction=decorate, decoration={markings, mark= at position 0.7 with {\arrow{stealth}}}] (0,0) -- (0.8,0) node[pos=0.5,above] {$\scriptstyle A_1$};
				\draw[fill=black!40!red] (0,0) circle (0.08) (0,0.8) circle (0.08);
				\draw[fill=black!40!green] (0.8,0) circle (0.08);
				\begin{scope}[shift={(-0.8,0)}]
					\draw[fill=gray] (-0.08,-0.08) -- (-0.08,0.08) -- (0.08,0.08) -- (0.08,-0.08) -- cycle;
				\end{scope}
			\end{tikzpicture}
		\end{array}:\quad R=\left(\begin{array}{c}
			\sqrt{2\zeta_1+\zeta_2}\\0
		\end{array}\right),\; C_1=\left(\begin{array}{cc}
			0 & 0 \\
			\sqrt{\zeta_1} & 0
		\end{array}\right),\; A_1=\left(\begin{array}{cc}
			\sqrt{\zeta_2} & 0
		\end{array}\right),\;B_1=\left(\begin{array}{c}
			0 \\ 0
		\end{array}\right),\quad C_2=(0)\,,
	\end{equation}
\begin{equation}
	\begin{aligned}
		&\begin{array}{c}
			\begin{tikzpicture}
				\draw[postaction=decorate, decoration={markings, mark= at position 0.7 with {\arrow{stealth}}}] (-0.8,0) -- (0,0) node[pos=0.5,below] {$\scriptstyle R$};
				\draw[postaction=decorate, decoration={markings, mark= at position 0.7 with {\arrow{stealth}}}] (0,0) -- (0,0.8) node[pos=0.5,left] {$\scriptstyle C_1$};
				\draw[postaction=decorate, decoration={markings, mark= at position 0.7 with {\arrow{stealth}}}] (0,0) -- (0.8,0) node[pos=0.5,above] {$\scriptstyle A_1$};
				\draw[postaction=decorate, decoration={markings, mark= at position 0.7 with {\arrow{stealth}}}] (0,0.8) -- (0.8,0.8) node[pos=0.5,above] {$\scriptstyle A_1$};
				\draw[postaction=decorate, decoration={markings, mark= at position 0.7 with {\arrow{stealth}}}] (0.8,0) -- (0.8,0.8) node[pos=0.5,right] {$\scriptstyle C_2$};
				\draw[fill=black!40!red] (0,0) circle (0.08) (0,0.8) circle (0.08);
				\draw[fill=black!40!green] (0.8,0) circle (0.08) (0.8,0.8) circle (0.08);
				\begin{scope}[shift={(-0.8,0)}]
					\draw[fill=gray] (-0.08,-0.08) -- (-0.08,0.08) -- (0.08,0.08) -- (0.08,-0.08) -- cycle;
				\end{scope}
			\end{tikzpicture}
		\end{array}:\quad R=\left(\begin{array}{c}
			\sqrt{2\zeta_1+2\zeta_2}\\ 0
		\end{array}\right),\; C_1=\left(\begin{array}{cc}
			0 & 0 \\
			\sqrt{\frac{(\zeta_1+\zeta_2)(\zeta_1+2\zeta_2)}{\zeta_1+3\zeta_2}} & 0
		\end{array}\right)\,,\\
		& A_1=\left(\begin{array}{cc}
			\sqrt{\frac{2(\zeta_1+2\zeta_2)\zeta_2}{\zeta_1+3\zeta_2}} &0 \\ 0 & \sqrt{\frac{2\zeta_2^2}{\zeta_1+3\zeta_2}}
		\end{array}\right),\;
		B_1=\left(\begin{array}{cc}
			0  & 0\\ 0 & 0
		\end{array}\right),\quad C_2=\left(\begin{array}{cc}
			0 & 0 \\
			\sqrt{\frac{(\zeta_1+\zeta_2)\zeta_2}{\zeta_1+3\zeta_2}} & 0
		\end{array}\right)\,.
	\end{aligned}
\end{equation}
\end{subequations}

{\bf Phase II }$(\zeta_1<0,\zeta_1+\zeta_2>0)$ solutions:
\begin{subequations}
\begin{equation}
		\begin{array}{c}
			\begin{tikzpicture}
				\begin{scope}[shift={(-0.8,0)}]
					\draw[fill=gray] (-0.08,-0.08) -- (-0.08,0.08) -- (0.08,0.08) -- (0.08,-0.08) -- cycle;
				\end{scope}
			\end{tikzpicture}
		\end{array}: \quad \varnothing\,,
\end{equation}
\begin{equation}
		\begin{array}{c}
			\begin{tikzpicture}
				\draw[postaction=decorate, decoration={markings, mark= at position 0.3 with {\arrowreversed{stealth}}}] (-0.8,0) -- (0,0) node[pos=0.5,below] {$\scriptstyle S$};
				\draw[fill=black!40!red] (0,0) circle (0.08);
				\begin{scope}[shift={(-0.8,0)}]
					\draw[fill=gray] (-0.08,-0.08) -- (-0.08,0.08) -- (0.08,0.08) -- (0.08,-0.08) -- cycle;
				\end{scope}
			\end{tikzpicture}
		\end{array}:\quad S=\left(\sqrt{-\zeta_1}\right),\; C_1=(0)\,,
\end{equation}
\begin{equation}
		\begin{array}{c}
			\begin{tikzpicture}
				\draw[postaction=decorate, decoration={markings, mark= at position 0.3 with {\arrowreversed{stealth}}}] (-0.8,0) -- (0,0) node[pos=0.5,below] {$\scriptstyle S$};
				\draw[postaction=decorate, decoration={markings, mark= at position 0.3 with {\arrowreversed{stealth}}}] (0,0) -- (0,0.8) node[pos=0.5,left] {$\scriptstyle C_1$};
				\draw[fill=black!40!red] (0,0) circle (0.08) (0,0.8) circle (0.08);
				\begin{scope}[shift={(-0.8,0)}]
					\draw[fill=gray] (-0.08,-0.08) -- (-0.08,0.08) -- (0.08,0.08) -- (0.08,-0.08) -- cycle;
				\end{scope}
			\end{tikzpicture}
		\end{array}:\quad S=\left(\begin{array}{cc}
			\sqrt{-2\zeta_1}& 0
		\end{array}\right),\; C_1=\left(\begin{array}{cc}
			0 & \sqrt{-\zeta_1} \\
			0 & 0\\
		\end{array}\right)\,,
\end{equation}
\begin{equation}
		\begin{array}{c}
			\begin{tikzpicture}
				\draw[postaction=decorate, decoration={markings, mark= at position 0.7 with {\arrow{stealth}}}] (-0.8,0) -- (0,0) node[pos=0.5,below] {$\scriptstyle R$};
				\draw[postaction=decorate, decoration={markings, mark= at position 0.7 with {\arrow{stealth}}}] (0,0) -- (0.8,0) node[pos=0.5,above] {$\scriptstyle A_1$};
				\draw[fill=black!40!red] (0,0) circle (0.08);
				\draw[fill=black!40!green] (0.8,0) circle (0.08);
				\begin{scope}[shift={(-0.8,0)}]
					\draw[fill=gray] (-0.08,-0.08) -- (-0.08,0.08) -- (0.08,0.08) -- (0.08,-0.08) -- cycle;
				\end{scope}
			\end{tikzpicture}
		\end{array}:\quad R=\left(\sqrt{\zeta_1+\zeta_2}\right),\; C_1=(0),\; A_1=\left(\sqrt{\zeta_2}\right),\;B_1=\left(0\right),\quad C_2=(0)\,,
\end{equation}
\begin{equation}
		\begin{array}{c}
			\begin{tikzpicture}
				\draw[postaction=decorate, decoration={markings, mark= at position 0.7 with {\arrow{stealth}}}] (-0.8,0) -- (0,0.5) node[pos=0.5,above] {$\scriptstyle R$};
				\draw[postaction=decorate, decoration={markings, mark= at position 0.3 with {\arrowreversed{stealth}}}] (-0.8,0) -- (0,-0.5) node[pos=0.5,below] {$\scriptstyle S$};
				\draw[postaction=decorate, decoration={markings, mark= at position 0.7 with {\arrow{stealth}}}] (0,0.5) -- (0.8,0.5) node[pos=0.5,above] {$\scriptstyle A_1$};
				\draw[fill=black!40!red] (0,-0.5) circle (0.08) (0,0.5) circle (0.08);
				\draw[fill=black!40!green] (0.8,0.5) circle (0.08);
				\begin{scope}[shift={(-0.8,0)}]
					\draw[fill=gray] (-0.08,-0.08) -- (-0.08,0.08) -- (0.08,0.08) -- (0.08,-0.08) -- cycle;
				\end{scope}
			\end{tikzpicture}
		\end{array}:\quad R=\left(\begin{array}{c}
			\sqrt{\zeta_1+\zeta_2}\\0
		\end{array}\right),\; S=\left(\begin{array}{cc}
			0 & \sqrt{-\zeta_1}
		\end{array}\right),\; C_1=\left(\begin{array}{cc}
			0 & 0 \\
			0 & 0
		\end{array}\right),\; A_1=\left(\begin{array}{cc}
			\sqrt{\zeta_2} & 0
		\end{array}\right)\,,
\end{equation}
\begin{equation}
		\begin{aligned}
		&\begin{array}{c}
			\begin{tikzpicture}
				\draw[postaction=decorate, decoration={markings, mark= at position 0.7 with {\arrow{stealth}}}] (-0.8,0) -- (0,0) node[pos=0.5,below] {$\scriptstyle R$};
				\draw[postaction=decorate, decoration={markings, mark= at position 0.7 with {\arrow{stealth}}}] (0,0) -- (0,0.8) node[pos=0.5,left] {$\scriptstyle C_1$};
				\draw[postaction=decorate, decoration={markings, mark= at position 0.7 with {\arrow{stealth}}}] (0,0) -- (0.8,0) node[pos=0.5,above] {$\scriptstyle A_1$};
				\draw[postaction=decorate, decoration={markings, mark= at position 0.7 with {\arrow{stealth}}}] (0,0.8) -- (0.8,0.8) node[pos=0.5,above] {$\scriptstyle A_1$};
				\draw[postaction=decorate, decoration={markings, mark= at position 0.7 with {\arrow{stealth}}}] (0.8,0) -- (0.8,0.8) node[pos=0.5,right] {$\scriptstyle C_2$};
				\draw[fill=black!40!red] (0,0) circle (0.08) (0,0.8) circle (0.08);
				\draw[fill=black!40!green] (0.8,0) circle (0.08) (0.8,0.8) circle (0.08);
				\begin{scope}[shift={(-0.8,0)}]
					\draw[fill=gray] (-0.08,-0.08) -- (-0.08,0.08) -- (0.08,0.08) -- (0.08,-0.08) -- cycle;
				\end{scope}
			\end{tikzpicture}
		\end{array}:\quad R=\left(\begin{array}{c}
			\sqrt{2\zeta_1+2\zeta_2}\\ 0
		\end{array}\right),\; C_1=\left(\begin{array}{cc}
			0 & 0 \\
			\sqrt{\frac{(\zeta_1+\zeta_2)(\zeta_1+2\zeta_2)}{\zeta_1+3\zeta_2}} & 0
		\end{array}\right)\,,\\
		& A_1=\left(\begin{array}{cc}
			\sqrt{\frac{2(\zeta_1+2\zeta_2)\zeta_2}{\zeta_1+3\zeta_2}} &0 \\ 0 & \sqrt{\frac{2\zeta_2^2}{\zeta_1+3\zeta_2}}
		\end{array}\right),\;
		B_1=\left(\begin{array}{cc}
			0  & 0\\ 0 & 0
		\end{array}\right),\quad C_2=\left(\begin{array}{cc}
			0 & 0 \\
			\sqrt{\frac{(\zeta_1+\zeta_2)\zeta_2}{\zeta_1+3\zeta_2}} & 0
		\end{array}\right)\,.
	\end{aligned}
\end{equation}
\end{subequations}

{\bf Phase III }$(\zeta_1+\zeta_2<0,\zeta_2>0)$ solutions:
\begin{subequations}
	\begin{equation}
		\begin{array}{c}
			\begin{tikzpicture}
				\begin{scope}[shift={(-0.8,0)}]
					\draw[fill=gray] (-0.08,-0.08) -- (-0.08,0.08) -- (0.08,0.08) -- (0.08,-0.08) -- cycle;
				\end{scope}
			\end{tikzpicture}
		\end{array}: \quad \varnothing\,,
	\end{equation}
	\begin{equation}
		\begin{array}{c}
			\begin{tikzpicture}
				\draw[postaction=decorate, decoration={markings, mark= at position 0.3 with {\arrowreversed{stealth}}}] (-0.8,0) -- (0,0) node[pos=0.5,below] {$\scriptstyle S$};
				\draw[fill=black!40!red] (0,0) circle (0.08);
				\begin{scope}[shift={(-0.8,0)}]
					\draw[fill=gray] (-0.08,-0.08) -- (-0.08,0.08) -- (0.08,0.08) -- (0.08,-0.08) -- cycle;
				\end{scope}
			\end{tikzpicture}
		\end{array}:\quad S=\left(\sqrt{-\zeta_1}\right),\; C_1=(0)\,,
	\end{equation}
	\begin{equation}
		\begin{array}{c}
			\begin{tikzpicture}
				\draw[postaction=decorate, decoration={markings, mark= at position 0.3 with {\arrowreversed{stealth}}}] (-0.8,0) -- (0,0) node[pos=0.5,below] {$\scriptstyle S$};
				\draw[postaction=decorate, decoration={markings, mark= at position 0.3 with {\arrowreversed{stealth}}}] (0,0) -- (0,0.8) node[pos=0.5,left] {$\scriptstyle C_1$};
				\draw[fill=black!40!red] (0,0) circle (0.08) (0,0.8) circle (0.08);
				\begin{scope}[shift={(-0.8,0)}]
					\draw[fill=gray] (-0.08,-0.08) -- (-0.08,0.08) -- (0.08,0.08) -- (0.08,-0.08) -- cycle;
				\end{scope}
			\end{tikzpicture}
		\end{array}:\quad S=\left(\begin{array}{cc}
			\sqrt{-2\zeta_1}& 0
		\end{array}\right),\; C_1=\left(\begin{array}{cc}
			0 & \sqrt{-\zeta_1} \\
			0 & 0\\
		\end{array}\right)\,,
	\end{equation}
	\begin{equation}
		\begin{array}{c}
			\begin{tikzpicture}
				\draw[postaction=decorate, decoration={markings, mark= at position 0.3 with {\arrowreversed{stealth}}}] (-0.8,0) -- (0,0) node[pos=0.5,below] {$\scriptstyle S$};
				\draw[postaction=decorate, decoration={markings, mark= at position 0.7 with {\arrow{stealth}}}] (0,0) -- (0.8,0) node[pos=0.5,above] {$\scriptstyle A_1$};
				\draw[fill=black!40!red] (0,0) circle (0.08);
				\draw[fill=black!40!green] (0.8,0) circle (0.08);
				\begin{scope}[shift={(-0.8,0)}]
					\draw[fill=gray] (-0.08,-0.08) -- (-0.08,0.08) -- (0.08,0.08) -- (0.08,-0.08) -- cycle;
				\end{scope}
			\end{tikzpicture}
		\end{array}:\quad S=\left(\sqrt{-\zeta_1-\zeta_2}\right),\; C_1=(0),\; A_1=\left(\sqrt{\zeta_2}\right),\;B_1=\left(0\right),\quad C_2=(0)\,,
	\end{equation}
	\begin{equation}
		\begin{array}{c}
			\begin{tikzpicture}
				\draw[postaction=decorate, decoration={markings, mark= at position 0.3 with {\arrowreversed{stealth}}}] (-0.8,0) -- (0,0) node[pos=0.5,below] {$\scriptstyle S$};
				\draw[postaction=decorate, decoration={markings, mark= at position 0.3 with {\arrowreversed{stealth}}}] (0,0) -- (0,0.8) node[pos=0.5,left] {$\scriptstyle C_1$};
				\draw[postaction=decorate, decoration={markings, mark= at position 0.7 with {\arrow{stealth}}}] (0,0.8) -- (0.8,0.8) node[pos=0.5,below] {$\scriptstyle A_1$};
				\draw[fill=black!40!red] (0,0) circle (0.08) (0,0.8) circle (0.08);
				\draw[fill=black!40!green] (0.8,0.8) circle (0.08);
				\begin{scope}[shift={(-0.8,0)}]
					\draw[fill=gray] (-0.08,-0.08) -- (-0.08,0.08) -- (0.08,0.08) -- (0.08,-0.08) -- cycle;
				\end{scope}
			\end{tikzpicture}
		\end{array}:\quad S=\left(\begin{array}{cc}
			\sqrt{-2\zeta_1-\zeta_2}& 0
		\end{array}\right),\; C_1=\left(\begin{array}{cc}
			0 & \sqrt{-\zeta_1-\zeta_2} \\
			0 & 0
		\end{array}\right),\; A_1=\left(\begin{array}{cc}
			0 & \sqrt{\zeta_2}
		\end{array}\right)\,,
	\end{equation}
	\begin{equation}
		\begin{aligned}
			&\begin{array}{c}
				\begin{tikzpicture}
					\draw[postaction=decorate, decoration={markings, mark= at position 0.3 with {\arrowreversed{stealth}}}] (-0.8,0) -- (0,0) node[pos=0.5,below] {$\scriptstyle S$};
					\draw[postaction=decorate, decoration={markings, mark= at position 0.3 with {\arrowreversed{stealth}}}] (0,0) -- (0,0.8) node[pos=0.5,left] {$\scriptstyle C_1$};
					\draw[postaction=decorate, decoration={markings, mark= at position 0.7 with {\arrow{stealth}}}] (0,0) -- (0.8,0) node[pos=0.5,above] {$\scriptstyle A_1$};
					\draw[postaction=decorate, decoration={markings, mark= at position 0.7 with {\arrow{stealth}}}] (0,0.8) -- (0.8,0.8) node[pos=0.5,above] {$\scriptstyle A_1$};
					\draw[postaction=decorate, decoration={markings, mark= at position 0.3 with {\arrowreversed{stealth}}}] (0.8,0) -- (0.8,0.8) node[pos=0.5,right] {$\scriptstyle C_2$};
					\draw[fill=black!40!red] (0,0) circle (0.08) (0,0.8) circle (0.08);
					\draw[fill=black!40!green] (0.8,0) circle (0.08) (0.8,0.8) circle (0.08);
					\begin{scope}[shift={(-0.8,0)}]
						\draw[fill=gray] (-0.08,-0.08) -- (-0.08,0.08) -- (0.08,0.08) -- (0.08,-0.08) -- cycle;
					\end{scope}
				\end{tikzpicture}
			\end{array}:\quad S=\left(\begin{array}{cc}
				\sqrt{-2\zeta_1-2\zeta_2} & 0
			\end{array}\right),\; C_1=\left(\begin{array}{cc}
				0 & \sqrt{\frac{\zeta_1(\zeta_1+\zeta_2)}{-\zeta_1+\zeta_2}} \\
				0 & 0
			\end{array}\right),\; B_1=\left(\begin{array}{cc}
				0  & 0\\ 0 & 0
			\end{array}\right)\,,\\
			& A_1=\left(\begin{array}{cc}
				\sqrt{\frac{2\zeta_2^2}{-\zeta_1+\zeta_2}} &0 \\ 0 & \sqrt{\frac{-2\zeta_1\zeta_2}{-\zeta_1+\zeta_2}}
			\end{array}\right),\; C_2=\left(\begin{array}{cc}
				0 & \sqrt{\frac{-\zeta_2(\zeta_1+\zeta_2)}{-\zeta_1+\zeta_2}}\\
				0 & 0
			\end{array}\right)\,.
		\end{aligned}
	\end{equation}
\end{subequations}

{\bf Phase IV} $(\zeta_1<0,\zeta_2<0)$ solutions:
\begin{subequations}
	\begin{equation}
		\begin{array}{c}
			\begin{tikzpicture}
				\begin{scope}[shift={(-0.8,0)}]
					\draw[fill=gray] (-0.08,-0.08) -- (-0.08,0.08) -- (0.08,0.08) -- (0.08,-0.08) -- cycle;
				\end{scope}
			\end{tikzpicture}
		\end{array}: \quad \varnothing\,,
	\end{equation}
	\begin{equation}
		\begin{array}{c}
			\begin{tikzpicture}
				\draw[postaction=decorate, decoration={markings, mark= at position 0.3 with {\arrowreversed{stealth}}}] (-0.8,0) -- (0,0) node[pos=0.5,below] {$\scriptstyle S$};
				\draw[fill=black!40!red] (0,0) circle (0.08);
				\begin{scope}[shift={(-0.8,0)}]
					\draw[fill=gray] (-0.08,-0.08) -- (-0.08,0.08) -- (0.08,0.08) -- (0.08,-0.08) -- cycle;
				\end{scope}
			\end{tikzpicture}
		\end{array}:\quad S=\left(\sqrt{-\zeta_1}\right),\; C_1=(0)\,,
	\end{equation}
	\begin{equation}
		\begin{array}{c}
			\begin{tikzpicture}
				\draw[postaction=decorate, decoration={markings, mark= at position 0.3 with {\arrowreversed{stealth}}}] (-0.8,0) -- (0,0) node[pos=0.5,below] {$\scriptstyle S$};
				\draw[postaction=decorate, decoration={markings, mark= at position 0.3 with {\arrowreversed{stealth}}}] (0,0) -- (0,0.8) node[pos=0.5,left] {$\scriptstyle C_1$};
				\draw[fill=black!40!red] (0,0) circle (0.08) (0,0.8) circle (0.08);
				\begin{scope}[shift={(-0.8,0)}]
					\draw[fill=gray] (-0.08,-0.08) -- (-0.08,0.08) -- (0.08,0.08) -- (0.08,-0.08) -- cycle;
				\end{scope}
			\end{tikzpicture}
		\end{array}:\quad S=\left(\begin{array}{cc}
			\sqrt{-2\zeta_1}& 0
		\end{array}\right),\; C_1=\left(\begin{array}{cc}
			0 & \sqrt{-\zeta_1} \\
			0 & 0\\
		\end{array}\right)\,,
	\end{equation}
	\begin{equation}
		\begin{array}{c}
			\begin{tikzpicture}
				\draw[postaction=decorate, decoration={markings, mark= at position 0.3 with {\arrowreversed{stealth}}}] (-0.8,0) -- (0,0) node[pos=0.5,below] {$\scriptstyle S$};
				\draw[postaction=decorate, decoration={markings, mark= at position 0.3 with {\arrowreversed{stealth}}}] (0,0) -- (0.8,0) node[pos=0.5,above] {$\scriptstyle B_1$};
				\draw[fill=black!40!red] (0,0) circle (0.08);
				\draw[fill=black!40!green] (0.8,0) circle (0.08);
				\begin{scope}[shift={(-0.8,0)}]
					\draw[fill=gray] (-0.08,-0.08) -- (-0.08,0.08) -- (0.08,0.08) -- (0.08,-0.08) -- cycle;
				\end{scope}
			\end{tikzpicture}
		\end{array}:\quad S=\left(\sqrt{-\zeta_1-\zeta_2}\right),\; C_1=(0),\; A_1=\left(0\right),\;B_1=\left(\sqrt{-\zeta_2}\right),\quad C_2=(0)\,,
	\end{equation}
	\begin{equation}
		\begin{array}{c}
			\begin{tikzpicture}
				\draw[postaction=decorate, decoration={markings, mark= at position 0.3 with {\arrowreversed{stealth}}}] (-0.8,0) -- (0,0) node[pos=0.5,below] {$\scriptstyle S$};
				\draw[postaction=decorate, decoration={markings, mark= at position 0.3 with {\arrowreversed{stealth}}}] (0,0) -- (0,0.8) node[pos=0.5,left] {$\scriptstyle C_1$};
				\draw[postaction=decorate, decoration={markings, mark= at position 0.3 with {\arrowreversed{stealth}}}] (0,0) -- (0.8,0) node[pos=0.5,above] {$\scriptstyle B_1$};
				\draw[fill=black!40!red] (0,0) circle (0.08) (0,0.8) circle (0.08);
				\draw[fill=black!40!green] (0.8,0) circle (0.08);
				\begin{scope}[shift={(-0.8,0)}]
					\draw[fill=gray] (-0.08,-0.08) -- (-0.08,0.08) -- (0.08,0.08) -- (0.08,-0.08) -- cycle;
				\end{scope}
			\end{tikzpicture}
		\end{array}:\quad S=\left(\begin{array}{cc}
			\sqrt{-2\zeta_1-\zeta_2} & 0
		\end{array}\right),\; C_1=\left(\begin{array}{cc}
			0 & \sqrt{-\zeta_1} \\
			0 & 0
		\end{array}\right),\; A_1=\left(\begin{array}{cc}
			0 & 0
		\end{array}\right),\;B_1=\left(\begin{array}{c}
			\sqrt{-\zeta_2} \\ 0
		\end{array}\right)\,,
	\end{equation}
	\begin{equation}
		\begin{aligned}
			&\begin{array}{c}
				\begin{tikzpicture}
					\draw[postaction=decorate, decoration={markings, mark= at position 0.3 with {\arrowreversed{stealth}}}] (-0.8,0) -- (0,0) node[pos=0.5,below] {$\scriptstyle S$};
					\draw[postaction=decorate, decoration={markings, mark= at position 0.3 with {\arrowreversed{stealth}}}] (0,0) -- (0,0.8) node[pos=0.5,left] {$\scriptstyle C_1$};
					\draw[postaction=decorate, decoration={markings, mark= at position 0.3 with {\arrowreversed{stealth}}}] (0,0) -- (0.8,0) node[pos=0.5,above] {$\scriptstyle B_1$};
					\draw[postaction=decorate, decoration={markings, mark= at position 0.3 with {\arrowreversed{stealth}}}] (0,0.8) -- (0.8,0.8) node[pos=0.5,above] {$\scriptstyle B_1$};
					\draw[postaction=decorate, decoration={markings, mark= at position 0.3 with {\arrowreversed{stealth}}}] (0.8,0) -- (0.8,0.8) node[pos=0.5,right] {$\scriptstyle C_2$};
					\draw[fill=black!40!red] (0,0) circle (0.08) (0,0.8) circle (0.08);
					\draw[fill=black!40!green] (0.8,0) circle (0.08) (0.8,0.8) circle (0.08);
					\begin{scope}[shift={(-0.8,0)}]
						\draw[fill=gray] (-0.08,-0.08) -- (-0.08,0.08) -- (0.08,0.08) -- (0.08,-0.08) -- cycle;
					\end{scope}
				\end{tikzpicture}
			\end{array}:\quad S=\left(\begin{array}{cc}
				\sqrt{-2\zeta_1-2\zeta_2}& 0
			\end{array}\right),\; C_1=\left(\begin{array}{cc}
				0 & \sqrt{-\frac{(\zeta_1+\zeta_2)(\zeta_1+2\zeta_2)}{\zeta_1+3\zeta_2}} \\
				0 & 0
			\end{array}\right)\,,\\
			& B_1=\left(\begin{array}{cc}
				\sqrt{-\frac{2(\zeta_1+2\zeta_2)\zeta_2}{\zeta_1+3\zeta_2}} &0 \\ 0 & \sqrt{-\frac{2\zeta_2^2}{\zeta_1+3\zeta_2}}
			\end{array}\right),\;
			A_1=\left(\begin{array}{cc}
				0  & 0\\ 0 & 0
			\end{array}\right),\quad C_2=\left(\begin{array}{cc}
				0 & \sqrt{-\frac{(\zeta_1+\zeta_2)\zeta_2}{\zeta_1+3\zeta_2}}  \\
				0& 0
			\end{array}\right)\,.
		\end{aligned}
	\end{equation}
\end{subequations}

{\bf Phase V }$(\zeta_1>0,\zeta_1+\zeta_2<0)$ solutions:
\begin{subequations}
	\begin{equation}
		\begin{array}{c}
			\begin{tikzpicture}
				\begin{scope}[shift={(-0.8,0)}]
					\draw[fill=gray] (-0.08,-0.08) -- (-0.08,0.08) -- (0.08,0.08) -- (0.08,-0.08) -- cycle;
				\end{scope}
			\end{tikzpicture}
		\end{array}: \quad \varnothing\,,
	\end{equation}
	\begin{equation}
		\begin{array}{c}
			\begin{tikzpicture}
				\draw[postaction=decorate, decoration={markings, mark= at position 0.7 with {\arrow{stealth}}}] (-0.8,0) -- (0,0) node[pos=0.5,below] {$\scriptstyle R$};
				\draw[fill=black!40!red] (0,0) circle (0.08);
				\begin{scope}[shift={(-0.8,0)}]
					\draw[fill=gray] (-0.08,-0.08) -- (-0.08,0.08) -- (0.08,0.08) -- (0.08,-0.08) -- cycle;
				\end{scope}
			\end{tikzpicture}
		\end{array}:\quad R=\left(\sqrt{\zeta_1}\right),\; C_1=(0)\,,
	\end{equation}
	\begin{equation}
		\begin{array}{c}
			\begin{tikzpicture}
				\draw[postaction=decorate, decoration={markings, mark= at position 0.7 with {\arrow{stealth}}}] (-0.8,0) -- (0,0) node[pos=0.5,below] {$\scriptstyle R$};
				\draw[postaction=decorate, decoration={markings, mark= at position 0.7 with {\arrow{stealth}}}] (0,0) -- (0,0.8) node[pos=0.5,left] {$\scriptstyle C_1$};
				\draw[fill=black!40!red] (0,0) circle (0.08) (0,0.8) circle (0.08);
				\begin{scope}[shift={(-0.8,0)}]
					\draw[fill=gray] (-0.08,-0.08) -- (-0.08,0.08) -- (0.08,0.08) -- (0.08,-0.08) -- cycle;
				\end{scope}
			\end{tikzpicture}
		\end{array}:\quad R=\left(\begin{array}{c}
			\sqrt{2\zeta_1}\\ 0
		\end{array}\right),\; C_1=\left(\begin{array}{cc}
			0 & 0 \\
			\sqrt{\zeta_1} & 0\\
		\end{array}\right)\,,
	\end{equation}
	\begin{equation}
		\begin{array}{c}
			\begin{tikzpicture}
				\draw[postaction=decorate, decoration={markings, mark= at position 0.3 with {\arrowreversed{stealth}}}] (-0.8,0) -- (0,0) node[pos=0.5,below] {$\scriptstyle S$};
				\draw[postaction=decorate, decoration={markings, mark= at position 0.3 with {\arrowreversed{stealth}}}] (0,0) -- (0.8,0) node[pos=0.5,above] {$\scriptstyle B_1$};
				\draw[fill=black!40!red] (0,0) circle (0.08);
				\draw[fill=black!40!green] (0.8,0) circle (0.08);
				\begin{scope}[shift={(-0.8,0)}]
					\draw[fill=gray] (-0.08,-0.08) -- (-0.08,0.08) -- (0.08,0.08) -- (0.08,-0.08) -- cycle;
				\end{scope}
			\end{tikzpicture}
		\end{array}:\quad S=\left(\sqrt{-\zeta_1-\zeta_2}\right),\; C_1=(0),\; A_1=\left(0\right),\;B_1=\left(\sqrt{-\zeta_2}\right),\quad C_2=(0)\,,
	\end{equation}
	\begin{equation}
		\begin{array}{c}
			\begin{tikzpicture}
				\draw[postaction=decorate, decoration={markings, mark= at position 0.3 with {\arrowreversed{stealth}}}] (-0.8,0) -- (0,0.5) node[pos=0.5,above] {$\scriptstyle S$};
				\draw[postaction=decorate, decoration={markings, mark= at position 0.7 with {\arrow{stealth}}}] (-0.8,0) -- (0,-0.5) node[pos=0.5,below] {$\scriptstyle R$};
				\draw[postaction=decorate, decoration={markings, mark= at position 0.3 with {\arrowreversed{stealth}}}] (0,0.5) -- (0.8,0.5) node[pos=0.5,above] {$\scriptstyle B_1$};
				\draw[fill=black!40!red] (0,-0.5) circle (0.08) (0,0.5) circle (0.08);
				\draw[fill=black!40!green] (0.8,0.5) circle (0.08);
				\begin{scope}[shift={(-0.8,0)}]
					\draw[fill=gray] (-0.08,-0.08) -- (-0.08,0.08) -- (0.08,0.08) -- (0.08,-0.08) -- cycle;
				\end{scope}
			\end{tikzpicture}
		\end{array}:\quad S=\left(\begin{array}{c}
			\sqrt{-\zeta_1-\zeta_2}\\0
		\end{array}\right),\; R=\left(\begin{array}{cc}
			0 & \sqrt{\zeta_1}
		\end{array}\right),\; C_1=\left(\begin{array}{cc}
			0 & 0 \\
			0 & 0
		\end{array}\right),\; B_1=\left(\begin{array}{cc}
			\sqrt{-\zeta_2} & 0
		\end{array}\right)\,,
	\end{equation}
	\begin{equation}
		\begin{aligned}
			&\begin{array}{c}
				\begin{tikzpicture}
					\draw[postaction=decorate, decoration={markings, mark= at position 0.3 with {\arrowreversed{stealth}}}] (-0.8,0) -- (0,0) node[pos=0.5,below] {$\scriptstyle S$};
					\draw[postaction=decorate, decoration={markings, mark= at position 0.3 with {\arrowreversed{stealth}}}] (0,0) -- (0,0.8) node[pos=0.5,left] {$\scriptstyle C_1$};
					\draw[postaction=decorate, decoration={markings, mark= at position 0.3 with {\arrowreversed{stealth}}}] (0,0) -- (0.8,0) node[pos=0.5,above] {$\scriptstyle B_1$};
					\draw[postaction=decorate, decoration={markings, mark= at position 0.3 with {\arrowreversed{stealth}}}] (0,0.8) -- (0.8,0.8) node[pos=0.5,above] {$\scriptstyle B_1$};
					\draw[postaction=decorate, decoration={markings, mark= at position 0.3 with {\arrowreversed{stealth}}}] (0.8,0) -- (0.8,0.8) node[pos=0.5,right] {$\scriptstyle C_2$};
					\draw[fill=black!40!red] (0,0) circle (0.08) (0,0.8) circle (0.08);
					\draw[fill=black!40!green] (0.8,0) circle (0.08) (0.8,0.8) circle (0.08);
					\begin{scope}[shift={(-0.8,0)}]
						\draw[fill=gray] (-0.08,-0.08) -- (-0.08,0.08) -- (0.08,0.08) -- (0.08,-0.08) -- cycle;
					\end{scope}
				\end{tikzpicture}
			\end{array}:\quad S=\left(\begin{array}{cc}
				\sqrt{-2\zeta_1-2\zeta_2}& 0
			\end{array}\right),\; C_1=\left(\begin{array}{cc}
				0 & \sqrt{-\frac{(\zeta_1+\zeta_2)(\zeta_1+2\zeta_2)}{\zeta_1+3\zeta_2}} \\
				0 & 0
			\end{array}\right)\,,\\
			& B_1=\left(\begin{array}{cc}
				\sqrt{-\frac{2(\zeta_1+2\zeta_2)\zeta_2}{\zeta_1+3\zeta_2}} &0 \\ 0 & \sqrt{-\frac{2\zeta_2^2}{\zeta_1+3\zeta_2}}
			\end{array}\right),\;
			A_1=\left(\begin{array}{cc}
				0  & 0\\ 0 & 0
			\end{array}\right),\quad C_2=\left(\begin{array}{cc}
				0 & \sqrt{-\frac{(\zeta_1+\zeta_2)\zeta_2}{\zeta_1+3\zeta_2}} \\
				0 & 0
			\end{array}\right)\,.
		\end{aligned}
	\end{equation}
\end{subequations}

{\bf Phase VI }$(\zeta_1+\zeta_2>0,\zeta_2<0)$ solutions:
\begin{subequations}
\begin{equation}
		\begin{array}{c}
			\begin{tikzpicture}
				\begin{scope}[shift={(-0.8,0)}]
					\draw[fill=gray] (-0.08,-0.08) -- (-0.08,0.08) -- (0.08,0.08) -- (0.08,-0.08) -- cycle;
				\end{scope}
			\end{tikzpicture}
		\end{array}: \quad \varnothing\,,
\end{equation}
\begin{equation}
		\begin{array}{c}
			\begin{tikzpicture}
				\draw[postaction=decorate, decoration={markings, mark= at position 0.7 with {\arrow{stealth}}}] (-0.8,0) -- (0,0) node[pos=0.5,below] {$\scriptstyle R$};
				\draw[fill=black!40!red] (0,0) circle (0.08);
				\begin{scope}[shift={(-0.8,0)}]
					\draw[fill=gray] (-0.08,-0.08) -- (-0.08,0.08) -- (0.08,0.08) -- (0.08,-0.08) -- cycle;
				\end{scope}
			\end{tikzpicture}
		\end{array}:\quad R=\left(\sqrt{\zeta_1}\right),\; C_1=(0)\,,
\end{equation}
\begin{equation}
		\begin{array}{c}
			\begin{tikzpicture}
				\draw[postaction=decorate, decoration={markings, mark= at position 0.7 with {\arrow{stealth}}}] (-0.8,0) -- (0,0) node[pos=0.5,below] {$\scriptstyle R$};
				\draw[postaction=decorate, decoration={markings, mark= at position 0.7 with {\arrow{stealth}}}] (0,0) -- (0,0.8) node[pos=0.5,left] {$\scriptstyle C_1$};
				\draw[fill=black!40!red] (0,0) circle (0.08) (0,0.8) circle (0.08);
				\begin{scope}[shift={(-0.8,0)}]
					\draw[fill=gray] (-0.08,-0.08) -- (-0.08,0.08) -- (0.08,0.08) -- (0.08,-0.08) -- cycle;
				\end{scope}
			\end{tikzpicture}
		\end{array}:\quad R=\left(\begin{array}{c}
			\sqrt{2\zeta_1}\\ 0
		\end{array}\right),\; C_1=\left(\begin{array}{cc}
			0 & 0 \\
			\sqrt{\zeta_1} & 0\\
		\end{array}\right)\,,
\end{equation}
\begin{equation}
		\begin{array}{c}
			\begin{tikzpicture}
				\draw[postaction=decorate, decoration={markings, mark= at position 0.7 with {\arrow{stealth}}}] (-0.8,0) -- (0,0) node[pos=0.5,below] {$\scriptstyle R$};
				\draw[postaction=decorate, decoration={markings, mark= at position 0.3 with {\arrowreversed{stealth}}}] (0,0) -- (0.8,0) node[pos=0.5,above] {$\scriptstyle B_1$};
				\draw[fill=black!40!red] (0,0) circle (0.08);
				\draw[fill=black!40!green] (0.8,0) circle (0.08);
				\begin{scope}[shift={(-0.8,0)}]
					\draw[fill=gray] (-0.08,-0.08) -- (-0.08,0.08) -- (0.08,0.08) -- (0.08,-0.08) -- cycle;
				\end{scope}
			\end{tikzpicture}
		\end{array}:\quad R=\left(\sqrt{\zeta_1+\zeta_2}\right),\; C_1=(0),\; A_1=\left(0\right),\;B_1=\left(\sqrt{-\zeta_2}\right),\quad C_2=(0)\,,
\end{equation}
\begin{equation}
		\begin{array}{c}
			\begin{tikzpicture}
				\draw[postaction=decorate, decoration={markings, mark= at position 0.7 with {\arrow{stealth}}}] (-0.8,0) -- (0,0) node[pos=0.5,below] {$\scriptstyle R$};
				\draw[postaction=decorate, decoration={markings, mark= at position 0.7 with {\arrow{stealth}}}] (0,0) -- (0,0.8) node[pos=0.5,left] {$\scriptstyle C_1$};
				\draw[postaction=decorate, decoration={markings, mark= at position 0.3 with {\arrowreversed{stealth}}}] (0,0.8) -- (0.8,0.8) node[pos=0.5,below] {$\scriptstyle B_1$};
				\draw[fill=black!40!red] (0,0) circle (0.08) (0,0.8) circle (0.08);
				\draw[fill=black!40!green] (0.8,0.8) circle (0.08);
				\begin{scope}[shift={(-0.8,0)}]
					\draw[fill=gray] (-0.08,-0.08) -- (-0.08,0.08) -- (0.08,0.08) -- (0.08,-0.08) -- cycle;
				\end{scope}
			\end{tikzpicture}
		\end{array}:\quad R=\left(\begin{array}{c}
			\sqrt{2\zeta_1+\zeta_2}\\0
		\end{array}\right),\; C_1=\left(\begin{array}{cc}
			0 & 0 \\
			\sqrt{\zeta_1+\zeta_2} & 0
		\end{array}\right),\; B_1=\left(\begin{array}{c}
			0 \\ \sqrt{-\zeta_2}
		\end{array}\right),\quad C_2=(0)\,,
\end{equation}
\begin{equation}
	\begin{aligned}
		&\begin{array}{c}
			\begin{tikzpicture}
				\draw[postaction=decorate, decoration={markings, mark= at position 0.7 with {\arrow{stealth}}}] (-0.8,0) -- (0,0) node[pos=0.5,below] {$\scriptstyle R$};
				\draw[postaction=decorate, decoration={markings, mark= at position 0.7 with {\arrow{stealth}}}] (0,0) -- (0,0.8) node[pos=0.5,left] {$\scriptstyle C_1$};
				\draw[postaction=decorate, decoration={markings, mark= at position 0.3 with {\arrowreversed{stealth}}}] (0,0) -- (0.8,0) node[pos=0.5,above] {$\scriptstyle B_1$};
				\draw[postaction=decorate, decoration={markings, mark= at position 0.3 with {\arrowreversed{stealth}}}] (0,0.8) -- (0.8,0.8) node[pos=0.5,above] {$\scriptstyle B_1$};
				\draw[postaction=decorate, decoration={markings, mark= at position 0.7 with {\arrow{stealth}}}] (0.8,0) -- (0.8,0.8) node[pos=0.5,right] {$\scriptstyle C_2$};
				\draw[fill=black!40!red] (0,0) circle (0.08) (0,0.8) circle (0.08);
				\draw[fill=black!40!green] (0.8,0) circle (0.08) (0.8,0.8) circle (0.08);
				\begin{scope}[shift={(-0.8,0)}]
					\draw[fill=gray] (-0.08,-0.08) -- (-0.08,0.08) -- (0.08,0.08) -- (0.08,-0.08) -- cycle;
				\end{scope}
			\end{tikzpicture}
		\end{array}:\quad R=\left(\begin{array}{c}
			\sqrt{2\zeta_1+2\zeta_2}\\ 0
		\end{array}\right),\; C_1=\left(\begin{array}{cc}
			0 & 0 \\
			\sqrt{\frac{\zeta_1(\zeta_1+\zeta_2)}{\zeta_1-\zeta_2}} & 0
		\end{array}\right),\; A_1=\left(\begin{array}{cc}
			0  & 0\\ 0 & 0
		\end{array}\right)\,,\\
		& B_1=\left(\begin{array}{cc}
			\sqrt{\frac{2\zeta_2^2}{\zeta_1-\zeta_2}} &0 \\ 0 & \sqrt{\frac{-2\zeta_1\zeta_2}{\zeta_1-\zeta_2}}
		\end{array}\right),\; C_2=\left(\begin{array}{cc}
			0 & 0 \\
			\sqrt{\frac{-\zeta_2(\zeta_1+\zeta_2)}{\zeta_1-\zeta_2}} & 0
		\end{array}\right)\,.
	\end{aligned}
\end{equation}
\end{subequations}

%%%%%%%%%%%%%%%%%%%%%%%%%%%%%%%%%%%%%%%%%%%%%%%%%%%%%%%%%%%%%%%%%%%%%%%%%%%
%%%%%%%%%%%%%%%%%%%%%%%%%%%%%%%%%%%%%%%%%%%%%%%%%%%%%%%%%%%%%%%%%%%%%%%%%%%
%%%%%%%%%%%%%%%%%%%%%%%%%%%%%%%%%%%%%%%%%%%%%%%%%%%%%%%%%%%%%%%%%%%%%%%%%%%

\subsubsection{Irrep  \texorpdfstring{$[3,3]$}{[3,3]}}

Here we consider a theory analogous to \eqref{su(3), [2]}, yet with term $SC_1^3R$ in the superpotential.

Generating function for solutions reads:
\begin{equation}
	{\bf DT}_{\fQ_{2,1,3}}=1+q_1+q_1q_2+q_1^2+q_1^2q_2+q_1^2q_2^2+q_1^3+q_1^3q_2+q_1^3q_2^2+q_1^3q_2^3\,.
\end{equation}

For this quiver we find the following solutions to \eqref{fp_eq} in various phases.
Solutions are presented in the form of atomic structure plots for which it is checked numerically that respective ans\"atze flow to valid $G_{\IR}(\fQ)$-elements in the steepest descend method.

{\bf Phase I} $(\zeta_1>0,\zeta_2>0)$ solutions:
\begin{equation}
	\begin{aligned}
	&1\to\!\!\!
\begin{array}{c}
		\begin{tikzpicture}[scale=0.8]
			\begin{scope}[shift={(0,0)}]
				\draw[fill=gray] (-0.1,-0.1) -- (-0.1,0.1) -- (0.1,0.1) -- (0.1,-0.1) -- cycle;
			\end{scope}
		\end{tikzpicture}
	\end{array}\!\!,\;
	q_1\to\!\!\!\begin{array}{c}
	\begin{tikzpicture}[scale=0.8]
		\draw[postaction=decorate, decoration={markings, mark= at position 0.7 with {\arrow{stealth}}}] (0,0) -- (1,0) node[pos=0.5,above] {$\scriptstyle R$};
		\draw[fill=black!40!red] (1,0) circle (0.1);
		\begin{scope}[shift={(0,0)}]
			\draw[fill=gray] (-0.1,-0.1) -- (-0.1,0.1) -- (0.1,0.1) -- (0.1,-0.1) -- cycle;
		\end{scope}
	\end{tikzpicture}
	\end{array}\!\!,\;
	q_1q_2\to\!\!\!\begin{array}{c}
	\begin{tikzpicture}[scale=0.8]
		\draw[postaction=decorate, decoration={markings, mark= at position 0.7 with {\arrow{stealth}}}] (0,0) -- (1,0) node[pos=0.5,above] {$\scriptstyle R$};
		\draw[postaction=decorate, decoration={markings, mark= at position 0.7 with {\arrow{stealth}}}] (1,0) -- (1,-1) node[pos=0.5,left] {$\scriptstyle A_1$};
		\draw[fill=black!40!red] (1,0) circle (0.1);
		\draw[fill=black!40!green] (1,-1) circle (0.1);
		\begin{scope}[shift={(0,0)}]
			\draw[fill=gray] (-0.1,-0.1) -- (-0.1,0.1) -- (0.1,0.1) -- (0.1,-0.1) -- cycle;
		\end{scope}
	\end{tikzpicture}
	\end{array}\!\!,\;
	q_1^2\to\!\!\!\begin{array}{c}
		\begin{tikzpicture}[scale=0.8]
			\draw[postaction=decorate, decoration={markings, mark= at position 0.7 with {\arrow{stealth}}}] (0,0) -- (1,0) node[pos=0.5,above] {$\scriptstyle R$};
			\draw[postaction=decorate, decoration={markings, mark= at position 0.7 with {\arrow{stealth}}}] (1,0) -- (2,0) node[pos=0.5,above] {$\scriptstyle C_1$};
			\draw[fill=black!40!red] (1,0) circle (0.1) (2,0) circle (0.1);
			\begin{scope}[shift={(0,0)}]
				\draw[fill=gray] (-0.1,-0.1) -- (-0.1,0.1) -- (0.1,0.1) -- (0.1,-0.1) -- cycle;
			\end{scope}
		\end{tikzpicture}
	\end{array}\!\!,\;
	q_1^2q_2\to\!\!\!\begin{array}{c}
	\begin{tikzpicture}[scale=0.8]
		\draw[postaction=decorate, decoration={markings, mark= at position 0.7 with {\arrow{stealth}}}] (0,0) -- (1,0) node[pos=0.5,above] {$\scriptstyle R$};
		\draw[postaction=decorate, decoration={markings, mark= at position 0.7 with {\arrow{stealth}}}] (1,0) -- (2,0) node[pos=0.5,above] {$\scriptstyle C_1$};
		\draw[postaction=decorate, decoration={markings, mark= at position 0.7 with {\arrow{stealth}}}] (1,0) -- (1,-1) node[pos=0.5,left] {$\scriptstyle A_1$};
		\draw[fill=black!40!red] (1,0) circle (0.1) (2,0) circle (0.1);
		\draw[fill=black!40!green] (1,-1) circle (0.1);
		\begin{scope}[shift={(0,0)}]
			\draw[fill=gray] (-0.1,-0.1) -- (-0.1,0.1) -- (0.1,0.1) -- (0.1,-0.1) -- cycle;
		\end{scope}
	\end{tikzpicture}
	\end{array}\!\!,\;
	q_1^2q_2^2\to\!\!\!\begin{array}{c}
	\begin{tikzpicture}[scale=0.8]
		\draw[postaction=decorate, decoration={markings, mark= at position 0.7 with {\arrow{stealth}}}] (0,0) -- (1,0) node[pos=0.5,above] {$\scriptstyle R$};
		\draw[postaction=decorate, decoration={markings, mark= at position 0.7 with {\arrow{stealth}}}] (1,0) -- (2,0) node[pos=0.5,above] {$\scriptstyle C_1$};
		\draw[postaction=decorate, decoration={markings, mark= at position 0.7 with {\arrow{stealth}}}] (1,-1) -- (2,-1) node[pos=0.5,above] {$\scriptstyle C_2$};
		\draw[postaction=decorate, decoration={markings, mark= at position 0.7 with {\arrow{stealth}}}] (1,0) -- (1,-1) node[pos=0.5,left] {$\scriptstyle A_1$};
		\draw[postaction=decorate, decoration={markings, mark= at position 0.7 with {\arrow{stealth}}}] (2,0) -- (2,-1) node[pos=0.5,right] {$\scriptstyle A_1$};
		\draw[fill=black!40!red] (1,0) circle (0.1) (2,0) circle (0.1);
		\draw[fill=black!40!green] (1,-1) circle (0.1) (2,-1) circle (0.1);
		\begin{scope}[shift={(0,0)}]
			\draw[fill=gray] (-0.1,-0.1) -- (-0.1,0.1) -- (0.1,0.1) -- (0.1,-0.1) -- cycle;
		\end{scope}
	\end{tikzpicture}
	\end{array}\,,\\
	&q_1^3\to\!\!\!\begin{array}{c}
		\begin{tikzpicture}[scale=0.8]
			\draw[postaction=decorate, decoration={markings, mark= at position 0.7 with {\arrow{stealth}}}] (0,0) -- (1,0) node[pos=0.5,above] {$\scriptstyle R$};
			\draw[postaction=decorate, decoration={markings, mark= at position 0.7 with {\arrow{stealth}}}] (1,0) -- (2,0) node[pos=0.5,above] {$\scriptstyle C_1$};
			\draw[postaction=decorate, decoration={markings, mark= at position 0.7 with {\arrow{stealth}}}] (2,0) -- (3,0) node[pos=0.5,above] {$\scriptstyle C_1$};
			\draw[fill=black!40!red] (1,0) circle (0.1) (2,0) circle (0.1) (3,0) circle (0.1);
			\begin{scope}[shift={(0,0)}]
				\draw[fill=gray] (-0.1,-0.1) -- (-0.1,0.1) -- (0.1,0.1) -- (0.1,-0.1) -- cycle;
			\end{scope}
		\end{tikzpicture}
	\end{array}\!\!,\;q_1^3q_2\to\!\!\!\begin{array}{c}
	\begin{tikzpicture}[scale=0.8]
		\draw[postaction=decorate, decoration={markings, mark= at position 0.7 with {\arrow{stealth}}}] (0,0) -- (1,0) node[pos=0.5,above] {$\scriptstyle R$};
		\draw[postaction=decorate, decoration={markings, mark= at position 0.7 with {\arrow{stealth}}}] (1,0) -- (2,0) node[pos=0.5,above] {$\scriptstyle C_1$};
		\draw[postaction=decorate, decoration={markings, mark= at position 0.7 with {\arrow{stealth}}}] (2,0) -- (3,0) node[pos=0.5,above] {$\scriptstyle C_1$};
		\draw[postaction=decorate, decoration={markings, mark= at position 0.7 with {\arrow{stealth}}}] (1,0) -- (1,-1) node[pos=0.5,left] {$\scriptstyle A_1$};
		\draw[fill=black!40!red] (1,0) circle (0.1) (2,0) circle (0.1) (3,0) circle (0.1);
		\draw[fill=black!40!green] (1,-1) circle (0.1);
		\begin{scope}[shift={(0,0)}]
			\draw[fill=gray] (-0.1,-0.1) -- (-0.1,0.1) -- (0.1,0.1) -- (0.1,-0.1) -- cycle;
		\end{scope}
	\end{tikzpicture}
	\end{array}\!\!,\;q_1^3q_2^2\to\!\!\!\begin{array}{c}
	\begin{tikzpicture}[scale=0.8]
		\draw[postaction=decorate, decoration={markings, mark= at position 0.7 with {\arrow{stealth}}}] (0,0) -- (1,0) node[pos=0.5,above] {$\scriptstyle R$};
		\draw[postaction=decorate, decoration={markings, mark= at position 0.7 with {\arrow{stealth}}}] (1,0) -- (2,0) node[pos=0.5,above] {$\scriptstyle C_1$};
		\draw[postaction=decorate, decoration={markings, mark= at position 0.7 with {\arrow{stealth}}}] (2,0) -- (3,0) node[pos=0.5,above] {$\scriptstyle C_1$};
		\draw[postaction=decorate, decoration={markings, mark= at position 0.7 with {\arrow{stealth}}}] (1,0) -- (1,-1) node[pos=0.5,left] {$\scriptstyle A_1$};
		\draw[postaction=decorate, decoration={markings, mark= at position 0.7 with {\arrow{stealth}}}] (2,0) -- (2,-1) node[pos=0.5,right] {$\scriptstyle A_1$};
		\draw[postaction=decorate, decoration={markings, mark= at position 0.7 with {\arrow{stealth}}}] (1,-1) -- (2,-1) node[pos=0.5,above] {$\scriptstyle C_2$};
		\draw[fill=black!40!red] (1,0) circle (0.1) (2,0) circle (0.1) (3,0) circle (0.1);
		\draw[fill=black!40!green] (1,-1) circle (0.1) (2,-1) circle (0.1);
		\begin{scope}[shift={(0,0)}]
			\draw[fill=gray] (-0.1,-0.1) -- (-0.1,0.1) -- (0.1,0.1) -- (0.1,-0.1) -- cycle;
		\end{scope}
	\end{tikzpicture}
	\end{array}\!\!,\;q_1^3q_2^3\to\!\!\!\begin{array}{c}
	\begin{tikzpicture}[scale=0.8]
		\draw[postaction=decorate, decoration={markings, mark= at position 0.7 with {\arrow{stealth}}}] (0,0) -- (1,0) node[pos=0.5,above] {$\scriptstyle R$};
		\draw[postaction=decorate, decoration={markings, mark= at position 0.7 with {\arrow{stealth}}}] (1,0) -- (2,0) node[pos=0.5,above] {$\scriptstyle C_1$};
		\draw[postaction=decorate, decoration={markings, mark= at position 0.7 with {\arrow{stealth}}}] (2,0) -- (3,0) node[pos=0.5,above] {$\scriptstyle C_1$};
		\draw[postaction=decorate, decoration={markings, mark= at position 0.7 with {\arrow{stealth}}}] (1,0) -- (1,-1) node[pos=0.5,left] {$\scriptstyle A_1$};
		\draw[postaction=decorate, decoration={markings, mark= at position 0.7 with {\arrow{stealth}}}] (2,0) -- (2,-1) node[pos=0.3,right] {$\scriptstyle A_1$};
		\draw[postaction=decorate, decoration={markings, mark= at position 0.7 with {\arrow{stealth}}}] (3,0) -- (3,-1) node[pos=0.5,right] {$\scriptstyle A_1$};
		\draw[postaction=decorate, decoration={markings, mark= at position 0.7 with {\arrow{stealth}}}] (1,-1) -- (2,-1) node[pos=0.5,above] {$\scriptstyle C_2$};
		\draw[postaction=decorate, decoration={markings, mark= at position 0.7 with {\arrow{stealth}}}] (2,-1) -- (3,-1) node[pos=0.5,above] {$\scriptstyle C_2$};
		\draw[fill=black!40!red] (1,0) circle (0.1) (2,0) circle (0.1) (3,0) circle (0.1);
		\draw[fill=black!40!green] (1,-1) circle (0.1) (2,-1) circle (0.1) (3,-1) circle (0.1);
		\begin{scope}[shift={(0,0)}]
			\draw[fill=gray] (-0.1,-0.1) -- (-0.1,0.1) -- (0.1,0.1) -- (0.1,-0.1) -- cycle;
		\end{scope}
	\end{tikzpicture}
	\end{array}\!.
	\end{aligned}
\end{equation}

{\bf Phase II }$(\zeta_1<0,\zeta_1+\zeta_2>0)$ solutions:
\begin{equation}
	\begin{aligned}
		&1\to\!\!\!\begin{array}{c}
			\begin{tikzpicture}[scale=0.8]
				\begin{scope}[shift={(0,0)}]
					\draw[fill=gray] (-0.1,-0.1) -- (-0.1,0.1) -- (0.1,0.1) -- (0.1,-0.1) -- cycle;
				\end{scope}
			\end{tikzpicture}
		\end{array}\!\!,\;
		q_1\to\!\!\!\begin{array}{c}
			\begin{tikzpicture}[scale=0.8]
				\draw[postaction=decorate, decoration={markings, mark= at position 0.7 with {\arrow{stealth}}}] (-1,0) -- (0,0) node[pos=0.5,above] {$\scriptstyle S$};
				\draw[fill=black!40!red] (-1,0) circle (0.1);
				\begin{scope}[shift={(0,0)}]
					\draw[fill=gray] (-0.1,-0.1) -- (-0.1,0.1) -- (0.1,0.1) -- (0.1,-0.1) -- cycle;
				\end{scope}
			\end{tikzpicture}
		\end{array}\!\!,\;
		q_1q_2\to\!\!\!\begin{array}{c}
			\begin{tikzpicture}[scale=0.8]
				\draw[postaction=decorate, decoration={markings, mark= at position 0.7 with {\arrow{stealth}}}] (0,0) -- (1,0) node[pos=0.5,above] {$\scriptstyle R$};
				\draw[postaction=decorate, decoration={markings, mark= at position 0.7 with {\arrow{stealth}}}] (1,0) -- (1,-1) node[pos=0.5,left] {$\scriptstyle A_1$};
				\draw[fill=black!40!red] (1,0) circle (0.1);
				\draw[fill=black!40!green] (1,-1) circle (0.1);
				\begin{scope}[shift={(0,0)}]
					\draw[fill=gray] (-0.1,-0.1) -- (-0.1,0.1) -- (0.1,0.1) -- (0.1,-0.1) -- cycle;
				\end{scope}
			\end{tikzpicture}
		\end{array}\!\!,\;
		q_1^2\to\!\!\!\begin{array}{c}
			\begin{tikzpicture}[scale=0.8]
				\draw[postaction=decorate, decoration={markings, mark= at position 0.7 with {\arrow{stealth}}}] (-1,0) -- (0,0) node[pos=0.5,above] {$\scriptstyle S$};
				\draw[postaction=decorate, decoration={markings, mark= at position 0.7 with {\arrow{stealth}}}] (-2,0) -- (-1,0) node[pos=0.5,above] {$\scriptstyle C_1$};
				\draw[fill=black!40!red] (-1,0) circle (0.1) (-2,0) circle (0.1);
				\begin{scope}[shift={(0,0)}]
					\draw[fill=gray] (-0.1,-0.1) -- (-0.1,0.1) -- (0.1,0.1) -- (0.1,-0.1) -- cycle;
				\end{scope}
			\end{tikzpicture}
		\end{array}\!\!,\;
		q_1^2q_2\to\!\!\!\begin{array}{c}
			\begin{tikzpicture}[scale=0.8]
				\draw[postaction=decorate, decoration={markings, mark= at position 0.7 with {\arrow{stealth}}}] (0,0) -- (1,0) node[pos=0.5,above] {$\scriptstyle R$};
				\draw[postaction=decorate, decoration={markings, mark= at position 0.7 with {\arrow{stealth}}}] (1,0) -- (1,-1) node[pos=0.5,left] {$\scriptstyle A_1$};
				\draw[postaction=decorate, decoration={markings, mark= at position 0.7 with {\arrow{stealth}}}] (-1,0) -- (0,0) node[pos=0.5,above] {$\scriptstyle S$};
				\draw[fill=black!40!red] (1,0) circle (0.1) (-1,0) circle (0.1);
				\draw[fill=black!40!green] (1,-1) circle (0.1);
				\begin{scope}[shift={(0,0)}]
					\draw[fill=gray] (-0.1,-0.1) -- (-0.1,0.1) -- (0.1,0.1) -- (0.1,-0.1) -- cycle;
				\end{scope}
			\end{tikzpicture}
		\end{array}\!\!,\;
		q_1^2q_2^2\to\!\!\!\begin{array}{c}
			\begin{tikzpicture}[scale=0.8]
				\draw[postaction=decorate, decoration={markings, mark= at position 0.7 with {\arrow{stealth}}}] (0,0) -- (1,0) node[pos=0.5,above] {$\scriptstyle R$};
				\draw[postaction=decorate, decoration={markings, mark= at position 0.7 with {\arrow{stealth}}}] (1,0) -- (2,0) node[pos=0.5,above] {$\scriptstyle C_1$};
				\draw[postaction=decorate, decoration={markings, mark= at position 0.7 with {\arrow{stealth}}}] (1,-1) -- (2,-1) node[pos=0.5,above] {$\scriptstyle C_2$};
				\draw[postaction=decorate, decoration={markings, mark= at position 0.7 with {\arrow{stealth}}}] (1,0) -- (1,-1) node[pos=0.5,left] {$\scriptstyle A_1$};
				\draw[postaction=decorate, decoration={markings, mark= at position 0.7 with {\arrow{stealth}}}] (2,0) -- (2,-1) node[pos=0.5,right] {$\scriptstyle A_1$};
				\draw[fill=black!40!red] (1,0) circle (0.1) (2,0) circle (0.1);
				\draw[fill=black!40!green] (1,-1) circle (0.1) (2,-1) circle (0.1);
				\begin{scope}[shift={(0,0)}]
					\draw[fill=gray] (-0.1,-0.1) -- (-0.1,0.1) -- (0.1,0.1) -- (0.1,-0.1) -- cycle;
				\end{scope}
			\end{tikzpicture}
		\end{array}\,,\\
		&q_1^3\to\!\!\!\begin{array}{c}
			\begin{tikzpicture}[scale=0.8]
				\draw[postaction=decorate, decoration={markings, mark= at position 0.7 with {\arrow{stealth}}}] (-1,0) -- (0,0) node[pos=0.5,above] {$\scriptstyle S$};
				\draw[postaction=decorate, decoration={markings, mark= at position 0.7 with {\arrow{stealth}}}] (-2,0) -- (-1,0) node[pos=0.5,above] {$\scriptstyle C_1$};
				\draw[postaction=decorate, decoration={markings, mark= at position 0.7 with {\arrow{stealth}}}] (-3,0) -- (-2,0) node[pos=0.5,above] {$\scriptstyle C_1$};
				\draw[fill=black!40!red] (-1,0) circle (0.1) (-2,0) circle (0.1) (-3,0) circle (0.1);
				\begin{scope}[shift={(0,0)}]
					\draw[fill=gray] (-0.1,-0.1) -- (-0.1,0.1) -- (0.1,0.1) -- (0.1,-0.1) -- cycle;
				\end{scope}
			\end{tikzpicture}
		\end{array}\!\!,\;
		q_1^3q_2\to\!\!\!\begin{array}{c}
			\begin{tikzpicture}[scale=0.8]
				\draw[postaction=decorate, decoration={markings, mark= at position 0.7 with {\arrow{stealth}}}] (0,0) -- (1,0) node[pos=0.5,above] {$\scriptstyle R$};
				\draw[postaction=decorate, decoration={markings, mark= at position 0.7 with {\arrow{stealth}}}] (1,0) -- (1,-1) node[pos=0.5,left] {$\scriptstyle A_1$};
				\draw[postaction=decorate, decoration={markings, mark= at position 0.7 with {\arrow{stealth}}}] (-1,0) -- (0,0) node[pos=0.5,above] {$\scriptstyle S$};
				\draw[postaction=decorate, decoration={markings, mark= at position 0.7 with {\arrow{stealth}}}] (-2,0) -- (-1,0) node[pos=0.5,above] {$\scriptstyle C_1$};
				\draw[fill=black!40!red] (1,0) circle (0.1) (-1,0) circle (0.1) (-2,0) circle (0.1);
				\draw[fill=black!40!green] (1,-1) circle (0.1);
				\begin{scope}[shift={(0,0)}]
					\draw[fill=gray] (-0.1,-0.1) -- (-0.1,0.1) -- (0.1,0.1) -- (0.1,-0.1) -- cycle;
				\end{scope}
			\end{tikzpicture}
		\end{array}\!\!,\;
		q_1^3q_2^2\to\!\!\!\begin{array}{c}
			\begin{tikzpicture}[scale=0.8]
				\draw[postaction=decorate, decoration={markings, mark= at position 0.7 with {\arrow{stealth}}}] (0,0) -- (1,0) node[pos=0.5,above] {$\scriptstyle R$};
				\draw[postaction=decorate, decoration={markings, mark= at position 0.7 with {\arrow{stealth}}}] (1,0) -- (2,0) node[pos=0.5,above] {$\scriptstyle C_1$};
				\draw[postaction=decorate, decoration={markings, mark= at position 0.7 with {\arrow{stealth}}}] (1,-1) -- (2,-1) node[pos=0.5,above] {$\scriptstyle C_2$};
				\draw[postaction=decorate, decoration={markings, mark= at position 0.7 with {\arrow{stealth}}}] (1,0) -- (1,-1) node[pos=0.5,left] {$\scriptstyle A_1$};
				\draw[postaction=decorate, decoration={markings, mark= at position 0.7 with {\arrow{stealth}}}] (2,0) -- (2,-1) node[pos=0.3,left] {$\scriptstyle A_1$};
				\draw[postaction=decorate, decoration={markings, mark= at position 0.7 with {\arrow{stealth}}}] (-1,0) -- (0,0) node[pos=0.5,above] {$\scriptstyle S$};
				\draw[fill=black!40!red] (1,0) circle (0.1) (2,0) circle (0.1) (-1,0) circle (0.1);
				\draw[fill=black!40!green] (1,-1) circle (0.1) (2,-1) circle (0.1);
				\begin{scope}[shift={(0,0)}]
					\draw[fill=gray] (-0.1,-0.1) -- (-0.1,0.1) -- (0.1,0.1) -- (0.1,-0.1) -- cycle;
				\end{scope}
			\end{tikzpicture}
		\end{array}%
\!.
	\end{aligned}
\end{equation}

Duality pattern for vectors reads:
\begin{equation}
	%
\right)
				\end{array}$};
			\draw[stealth-stealth] (-0.2,0) to[out=120,in=60] node[pos=0.5,above] {$\scriptstyle s_1$} (-1.8,0);
			\draw[stealth-stealth] (-2.2,0) to[out=120,in=60] node[pos=0.5,above] {$\scriptstyle s_2$} (-3.8,0);
			\draw[stealth-stealth] (-4.2,0) to[out=120,in=60] node[pos=0.5,above] {$\scriptstyle s_1$} (-5.8,0);
			\draw[stealth-stealth] (0.2,0) to[out=60,in=120] node[pos=0.5,above] {$\scriptstyle s_2$} (1.8,0);
			\draw[stealth-stealth] (2.2,0) to[out=60,in=120] node[pos=0.5,above] {$\scriptstyle s_1$} (3.8,0);
			\draw[stealth-stealth] (-6.2,0) to[out=120,in=180] (-5,1) -- (3,1) node[pos=0.5,above]{$\scriptstyle s_2$} to[out=0,in=60] (4.2,0);
		\end{tikzpicture}
	\end{array}
\end{equation}

According to this duality pattern, for example, $s_2$ connects the following dual solutions in phases II and V:
\begin{equation}
	v_1^{(\zeta_2>0)}=\begin{array}{c}
		\begin{tikzpicture}[scale=0.8]
			\draw[postaction=decorate, decoration={markings, mark= at position 0.7 with {\arrow{stealth}}}] (0,0) -- (1,0) node[pos=0.5,above] {$\scriptstyle R$};
			\draw[postaction=decorate, decoration={markings, mark= at position 0.7 with {\arrow{stealth}}}] (1,0) -- (1,-1) node[pos=0.5,left] {$\scriptstyle A_1$};
			\draw[postaction=decorate, decoration={markings, mark= at position 0.7 with {\arrow{stealth}}}] (-1,0) -- (0,0) node[pos=0.5,above] {$\scriptstyle S$};
			\draw[postaction=decorate, decoration={markings, mark= at position 0.7 with {\arrow{stealth}}}] (-2,0) -- (-1,0) node[pos=0.5,above] {$\scriptstyle C_1$};
			\draw[fill=black!40!red] (1,0) circle (0.1) (-1,0) circle (0.1) (-2,0) circle (0.1);
			\draw[fill=black!40!green] (1,-1) circle (0.1);
			\begin{scope}[shift={(0,0)}]
				\draw[fill=gray] (-0.1,-0.1) -- (-0.1,0.1) -- (0.1,0.1) -- (0.1,-0.1) -- cycle;
			\end{scope}
		\end{tikzpicture}
	\end{array}\quad\longleftrightarrow\quad v_2^{(\zeta_2<0)}=\begin{array}{c}
	\begin{tikzpicture}[scale=0.8, xscale=-1]
		\draw[postaction=decorate, decoration={markings, mark= at position 0.3 with {\arrowreversed{stealth}}}] (0,0) -- (1,0) node[pos=0.5,above] {$\scriptstyle S$};
		\draw[postaction=decorate, decoration={markings, mark= at position 0.3 with {\arrowreversed{stealth}}}] (1,0) -- (2,0) node[pos=0.5,above] {$\scriptstyle C_1$};
		\draw[postaction=decorate, decoration={markings, mark= at position 0.3 with {\arrowreversed{stealth}}}] (-1,0) -- (0,0) node[pos=0.5,above] {$\scriptstyle R$};
		\draw[postaction=decorate, decoration={markings, mark= at position 0.3 with {\arrowreversed{stealth}}}] (1,0) -- (1,-1) node[pos=0.5,right] {$\scriptstyle B_1$};
		\draw[postaction=decorate, decoration={markings, mark= at position 0.3 with {\arrowreversed{stealth}}}] (2,0) -- (2,-1) node[pos=0.3,right] {$\scriptstyle B_1$};
		\draw[postaction=decorate, decoration={markings, mark= at position 0.3 with {\arrowreversed{stealth}}}] (1,-1) -- (2,-1) node[pos=0.5,above] {$\scriptstyle C_2$};
		\draw[fill=black!40!red] (1,0) circle (0.1) (2,0) circle (0.1) (-1,0) circle (0.1);
		\draw[fill=black!40!green] (1,-1) circle (0.1) (2,-1) circle (0.1);
		\begin{scope}[shift={(0,0)}]
			\draw[fill=gray] (-0.1,-0.1) -- (-0.1,0.1) -- (0.1,0.1) -- (0.1,-0.1) -- cycle;
		\end{scope}
	\end{tikzpicture}
	\end{array}\,.
\end{equation}

In this case we observe:
\begin{equation}
	\begin{aligned}
	&\alpha\left(v_2^{(\zeta_2<0)}\right)=\left(
	\begin{array}{cc}
		0 & 0 \\
		1 & 0 \\
		0 & 1 \\
	\end{array}
	\right),\quad \beta\left(v_1^{(\zeta_2>0)}\right)=\left(
	\begin{array}{ccc}
		-1 & 0 & 0 \\
	\end{array}
	\right)\,,\\
	&0\longrightarrow \IC^{2}\overset{\alpha}{\longrightarrow}\IC^3\overset{\beta}{\longrightarrow}\IC\longrightarrow 0 \mbox{ is a short exact sequence}\,.
	\end{aligned}
\end{equation}

Let us consider one more example of $s_1$ connecting the following solutions in phases V and IV respectively:
\begin{equation}
	v_3^{(\zeta_1>0)}=\begin{array}{c}
		\begin{tikzpicture}[scale=0.8, xscale=-1]
			\draw[postaction=decorate, decoration={markings, mark= at position 0.3 with {\arrowreversed{stealth}}}] (0,0) -- (1,0) node[pos=0.5,above] {$\scriptstyle S$};
			\draw[postaction=decorate, decoration={markings, mark= at position 0.3 with {\arrowreversed{stealth}}}] (-1,0) -- (0,0) node[pos=0.5,above] {$\scriptstyle R$};
			\draw[postaction=decorate, decoration={markings, mark= at position 0.3 with {\arrowreversed{stealth}}}] (1,0) -- (1,-1) node[pos=0.5,right] {$\scriptstyle B_1$};
			\draw[fill=black!40!red] (1,0) circle (0.1) (-1,0) circle (0.1);
			\draw[fill=black!40!green] (1,-1) circle (0.1);
			\begin{scope}[shift={(0,0)}]
				\draw[fill=gray] (-0.1,-0.1) -- (-0.1,0.1) -- (0.1,0.1) -- (0.1,-0.1) -- cycle;
			\end{scope}
		\end{tikzpicture}
	\end{array}\quad\longleftrightarrow\quad v_4^{(\zeta_1<0)}=\begin{array}{c}
	\begin{tikzpicture}[scale=0.8, xscale=-1]
		\draw[postaction=decorate, decoration={markings, mark= at position 0.3 with {\arrowreversed{stealth}}}] (0,0) -- (1,0) node[pos=0.5,above] {$\scriptstyle S$};
		\draw[postaction=decorate, decoration={markings, mark= at position 0.3 with {\arrowreversed{stealth}}}] (1,0) -- (2,0) node[pos=0.5,above] {$\scriptstyle C_1$};
		\draw[postaction=decorate, decoration={markings, mark= at position 0.3 with {\arrowreversed{stealth}}}] (1,0) -- (1,-1) node[pos=0.5,right] {$\scriptstyle B_1$};
		\draw[fill=black!40!red] (1,0) circle (0.1) (2,0) circle (0.1);
		\draw[fill=black!40!green] (1,-1) circle (0.1);
		\begin{scope}[shift={(0,0)}]
			\draw[fill=gray] (-0.1,-0.1) -- (-0.1,0.1) -- (0.1,0.1) -- (0.1,-0.1) -- cycle;
		\end{scope}
	\end{tikzpicture}
	\end{array}\,.
\end{equation}

In this case we observe:
\begin{equation}
	\begin{aligned}
		&\alpha\left(v_4^{(\zeta_2<0)}\right)=\left(
		\begin{array}{cc}
			1 & 0 \\
			0 & 1 \\
			0 & 0 \\
			0 & 0 \\
		\end{array}
		\right),\quad \beta\left(v_3^{(\zeta_2>0)}\right)=\left(
		\begin{array}{cccc}
			0 & 0 & 1 & 0 \\
			0 & 0 & 0 & 1 \\
		\end{array}
		\right)\,,\\
		&0\longrightarrow \IC^{2}\overset{\alpha}{\longrightarrow}\IC^4\overset{\beta}{\longrightarrow}\IC^2\longrightarrow 0 \mbox{ is a short exact sequence}\,.
	\end{aligned}
\end{equation}

%%%%%%%%%%%%%%%%%%%%%%%%%%%%%%%%%%%%%%%%%%%%%%%%%%%%%%%%%%%%%%%%%%%%%%%%%%%
%%%%%%%%%%%%%%%%%%%%%%%%%%%%%%%%%%%%%%%%%%%%%%%%%%%%%%%%%%%%%%%%%%%%%%%%%%%
%%%%%%%%%%%%%%%%%%%%%%%%%%%%%%%%%%%%%%%%%%%%%%%%%%%%%%%%%%%%%%%%%%%%%%%%%%%

\subsection{Quiver \texorpdfstring{$\fs\fl_4$}{su4}}

\begin{figure}[ht!]
	\centering
	\begin{tikzpicture}
		\node at (0,0) {$\begin{array}{c}
				\begin{tikzpicture}[scale=2.3]
					\tikzset{sty1/.style={thick}}
					\tikzset{sty2/.style={gray, thin}}
					%%%%%%%%%%%%%%%%%%%%%%%%%%%%%%%%%%%%%%%
					\draw[sty2](0.99985,0.01749) to[out=91.2566,in=300.] (0.87174,0.48997) to[out=120.,in=325.87] (0.55331,0.83297);
					\draw[sty2](0.50752,-0.86164) to[out=32.3132,in=241.644] (0.84203,-0.53942) to[out=61.6443,in=266.037] (0.99996,-0.0087);
					\draw[sty2](-0.48477,-0.87464) to[out=-28.7434,in=180.] (-0.01155,-0.99993) to[out=0.,in=205.87] (0.44472,-0.89567);
					\draw[sty2](-0.49244,0.87034) to[out=27.6868,in=178.356] (-0.04614,0.99894) to[out=-1.64428,in=153.963] (0.49244,0.87034);
					\draw[sty2](0.99999,0.0029) to[out=90.7505,in=310.148] (0.901,0.25046) to[out=130.148,in=329.751] (0.70711,0.40825);
					\draw[sty2](0.50251,0.86456) to[out=-30.7505,in=109.852] (0.66741,0.65506) to[out=-70.1478,in=90.2487] (0.70711,0.40825);
					\draw[sty2](0.99998,-0.00577) to[out=-90.7442,in=72.4585] (0.96909,-0.20142) to[out=-107.542,in=49.2489] (0.8165,-0.4714);
					\draw[sty2](0.50499,-0.86312) to[out=30.7442,in=227.542] (0.65898,-0.73855) to[out=47.5415,in=250.751] (0.8165,-0.4714);
					\draw[sty2](-0.99985,0.01749) to[out=88.7434,in=240.] (-0.87174,0.48997) to[out=60.,in=214.13] (-0.55331,0.83297);
					\draw[sty2](-0.50752,-0.86164) to[out=147.687,in=298.356] (-0.84203,-0.53942) to[out=118.356,in=273.963] (-0.99996,-0.0087);
					\draw[sty2](-0.49748,-0.86747) to[out=-29.2495,in=190.148] (-0.23359,-0.90552) to[out=10.1478,in=209.751] (0.,-0.8165);
					\draw[sty2](0.49748,-0.86747) to[out=-150.751,in=349.852] (0.23359,-0.90552) to[out=169.852,in=330.249] (0.,-0.8165);
					\draw[sty2](-0.49494,0.86892) to[out=29.2431,in=149.073] (0.29444,0.82998) to[out=-30.9273,in=120.334] (0.70711,0.40825);
					\draw[sty2](0.8165,-0.4714) to[out=71.0361,in=277.711] (0.85699,-0.0291) to[out=97.7112,in=299.669] (0.70711,0.40825);
					\draw[sty2](0.,-0.8165) to[out=0.33078,in=202.853] (0.4537,-0.72762) to[out=22.8529,in=228.964] (0.8165,-0.4714);
					\draw[sty2](-0.50251,0.86456) to[out=-149.249,in=70.1478] (-0.66741,0.65506) to[out=-109.852,in=89.7513] (-0.70711,0.40825);
					\draw[sty2](-0.99999,0.0029) to[out=89.2495,in=229.852] (-0.901,0.25046) to[out=49.8522,in=210.249] (-0.70711,0.40825);
					\draw[sty2](-0.99997,-0.00583) to[out=-89.2431,in=150.927] (-0.57156,-0.66998) to[out=-29.0727,in=179.666] (0.,-0.8165);
					\draw[sty2](-0.70711,0.40825) to[out=29.75,in=191.423] (-0.33872,0.54322) to[out=11.4231,in=150.25] (0.70711,0.40825);
					\draw[sty2](0.,-0.8165) to[out=30.25,in=228.577] (0.30108,-0.56495) to[out=48.5769,in=269.75] (0.70711,0.40825);
					\draw[sty2](0.,-0.8165) to[out=149.75,in=311.423] (-0.30108,-0.56495) to[out=131.423,in=270.25] (-0.70711,0.40825);
					%%%%%%%%%%%%%%%%%%%%%%%%%%%%%%%%%%%%%%%
					\draw[-stealth] (0,0) -- (-1.06066, -0.612372) node[pos=1,left] {$\scriptstyle \zeta_1$};
					\draw[-stealth] (0,0) -- (1.06066, -0.612372) node[pos=1,right] {$\scriptstyle \zeta_2$};
					\draw[-stealth] (0,0) -- (0., 1.22474) node[pos=1,left] {$\scriptstyle \zeta_3$};
					\draw[fill=palette4] (0,0) -- (-0.8165,0.4714) to[out=-108.964,in=97.7112] (-0.85699,0.0291) to[out=-82.2888,in=119.669] (-0.70711,-0.40825) -- cycle;
					\draw[fill=palette2] (0,0) -- (0.,0.8165) to[out=-179.669,in=22.8529] (-0.4537,0.72762) to[out=-157.147,in=48.9639] (-0.8165,0.4714) -- cycle;
					\draw[fill=palette3] (0,0) -- (0.,0.8165) to[out=-149.75,in=48.5769] (-0.30108,0.56495) to[out=-131.423,in=89.75] (-0.70711,-0.40825) -- cycle;
					%%%%%%%%%%%%%%%%%%%%%%%%%%%%%%%%%%%%%%%
					\draw[sty1](-0.70711,-0.40825) to[out=-29.75,in=168.577] (-0.33872,-0.54322) to[out=-11.4231,in=209.75] (0.70711,-0.40825);
					\draw[sty1](-0.70711,-0.40825) to[out=89.75,in=227.943] (-0.30108,0.56495) to[out=47.9428,in=210.25] (0.,0.8165);
					\draw[sty1](0.70711,-0.40825) to[out=90.25,in=312.057] (0.30108,0.56495) to[out=132.057,in=329.75] (0.,0.8165);
					\draw[sty1](0.70711,-0.40825) to[out=30.2487,in=230.495] (0.901,-0.25046) to[out=50.4945,in=269.249] (0.99999,-0.0029);
					\draw[sty1](0.,0.8165) to[out=-0.33412,in=149.814] (0.57156,0.66998) to[out=-30.1862,in=90.7569] (0.99997,0.00583);
					\draw[sty1](-0.70711,-0.40825) to[out=119.669,in=277.147] (-0.85699,0.0291) to[out=97.1471,in=251.036] (-0.8165,0.4714);
					\draw[sty1](-0.8165,0.4714) to[out=48.964,in=202.289] (-0.4537,0.72762) to[out=22.2888,in=180.331] (0.,0.8165);
					\draw[sty1](-0.70711,-0.40825) to[out=-59.6659,in=150.186] (-0.29444,-0.82998) to[out=-29.8138,in=209.243] (0.49494,-0.86892);
					\draw[sty1](0.70711,-0.40825) to[out=-90.2487,in=69.5055] (0.66741,-0.65506) to[out=-110.495,in=30.7505] (0.50251,-0.86456);
					\draw[sty1](0.99996,0.0087) to[out=92.3132,in=301.644] (0.88817,0.45951) to[out=121.644,in=326.037] (0.50752,0.86164);
					\draw[sty1](0.,0.8165) to[out=29.7513,in=189.505] (0.23359,0.90552) to[out=9.50547,in=150.751] (0.49748,0.86747);
					\draw[sty1](0.51507,-0.85715) to[out=31.2566,in=240.] (0.86019,-0.50997) to[out=60.,in=265.87] (0.99803,-0.0627);
					\draw[sty1](-0.8165,0.4714) to[out=70.7511,in=227.089] (-0.65898,0.73855) to[out=47.0893,in=210.744] (-0.50499,0.86312);
					\draw[sty1](0.,0.8165) to[out=150.249,in=350.495] (-0.23359,0.90552) to[out=170.495,in=29.2495] (-0.49748,0.86747);
					\draw[sty1](-0.70711,-0.40825) to[out=149.751,in=309.505] (-0.901,-0.25046) to[out=129.505,in=270.751] (-0.99999,-0.0029);
					\draw[sty1](-0.8165,0.4714) to[out=-130.751,in=72.9107] (-0.96909,0.20142) to[out=-107.089,in=89.2558] (-0.99998,0.00577);
					\draw[sty1](-0.70711,-0.40825) to[out=-89.7513,in=110.495] (-0.66741,-0.65506) to[out=-69.5055,in=149.249] (-0.50251,-0.86456);
					\draw[sty1](-0.49244,-0.87034) to[out=-27.6868,in=181.644] (-0.04614,-0.99894) to[out=1.64428,in=206.037] (0.49244,-0.87034);
					\draw[sty1](-0.48477,0.87464) to[out=28.7434,in=180.] (-0.01155,0.99993) to[out=0.,in=154.13] (0.44472,0.89567);
					\draw[sty1](-0.99996,0.0087) to[out=87.6868,in=238.356] (-0.88817,0.45951) to[out=58.3557,in=213.963] (-0.50752,0.86164);
					\draw[sty1](-0.51507,-0.85715) to[out=148.743,in=300.] (-0.86019,-0.50997) to[out=120.,in=274.13] (-0.99803,-0.0627);
					%%%%%%%%%%%%%%%%%%%%%%%%%%%%%%%%%%%%%%%
					\draw[ultra thick] (0,0) circle (1);
					\begin{scope}[scale=0.965]
					\foreach \t/\x/\y in {1/0.3/0.3,2/0.866025/0.166667,4/0.288675/-0.833333,5/0.707107/0.680414,6/0.852803/-0.492366,7/-0.566947/0.763763,8/-0.944911/0.109109,9/-0.235702/-0.952579,10/0./0.984732,14/-0.852803/-0.492366}
					{
						\node[burgundy] at (\x,\y) {\footnotesize\bf \t};
					}
					\node[white] at (-0.5,0.5) {\footnotesize\bf 3};
					\end{scope}
				\end{tikzpicture}
			\end{array}$};
		\node at (8,0) {$\begin{array}{c}
				\begin{tikzpicture}
					\draw (-3.2,-2.5) -- (3.2,-2.5) -- (3.2,3.2) -- (-3.2,3.2) -- cycle;
					\foreach \a/\b/\r in {0./1.41421/1.73205,1.22474/-0.707107/1.73205,-1.22474/-0.707107/1.73205,0./-0.707107/1.22474,0.612372/0.353553/1.22474,0./0./1.}
					{
						\draw (\a,\b) circle (\r);
					}
					\foreach \t/\x/\y in {1/0/0,2/0.588571/0.11327,3/-0.448288/0.258819,4/0.19619/-0.566352,5/0.592986/0.570602,6/0.72636/-0.419364,7/-0.433245/0.583646,8/-0.722074/0.083378,9/-0.197662/-0.798842,10/0./0.838728,11/1.03255/0.596141,12/0.997117/-0.575686,13/-0.752168/0.434265,14/-0.72636/-0.419364,15/0./-1.19228,16/0.291873/1.17959,17/1.36668/-0.15781,18/0.820006/-1.10467,19/-1.03255/0.596141,20/-0.87562/-0.842566,21/-0.546117/1.5765,22/1.67303/-0.965926,23/-1.63835/-0.315301,24/2/2}
					{
						\node at (\x,\y) {\tiny \t};
					}
				\end{tikzpicture}
			\end{array}$};
		\node at (0,-3.5) {(a)};
		\node at (8,-3.5) {(b)};
	\end{tikzpicture}
	\caption{Projected moduli space of the $\fs\fu_4$ quiver: (a) on a unit 2-sphere, (b) on a stereographic plane}\label{fig:su_4_moduli}
\end{figure}
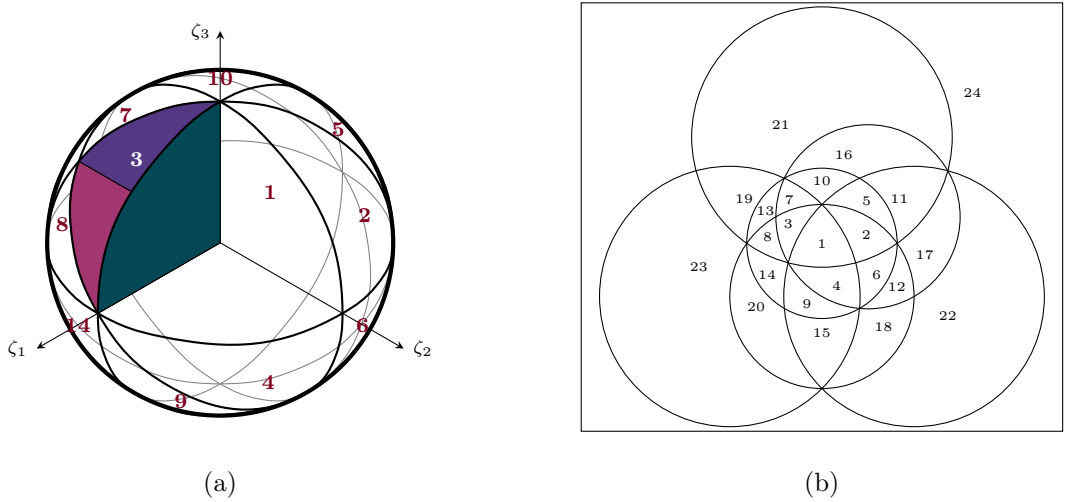

An $\fs\fl_4$ quiver has 3 nodes, and, respectively, the FI parameter space is 3d.
Apparently, starting with a collection of quiver morphism solutions $M$ at given $\vec \zeta$ we can easily produce solutions for the whole ray $\lambda\vec \zeta$, $\lambda>0$ by simple rescaling $\sqrt{\lambda}M$.
So the marginal stability walls could appear only on planes passing through 0.
It is natural to work with projective coordinates modulo $\lambda$, then the whole moduli space is projected to $S^2$ (see Fig.~\ref{fig:su_4_moduli}).

The action on the Weyl mutations on FI parameters is generated by three reflections:
\begin{equation}
	s_1=\left(
	\begin{array}{ccc}
		-1 & 0 & 0 \\
		1 & 1 & 0 \\
		0 & 0 & 1 \\
	\end{array}
	\right),\quad s_2=\left(
	\begin{array}{ccc}
		1 & 1 & 0 \\
		0 & -1 & 0 \\
		0 & 1 & 1 \\
	\end{array}
	\right),\quad s_3=\left(
	\begin{array}{ccc}
		1 & 0 & 0 \\
		0 & 1 & 1 \\
		0 & 0 & -1 \\
	\end{array}
	\right)\,,
\end{equation}
satisfying permutation group rules:
\begin{equation}
	s_1^2=s_2^2=s_3^2=\bbone, \; s_1s_3=s_3s_1,\; s_1s_2s_1=s_2s_1s_2,\; s_3s_2s_3=s_2s_3s_2\,.
\end{equation}
So there are $4!=24$ phases in Fig.~\ref{fig:su_4_moduli} marked by numbers in the following way:
\begin{equation}
	\begin{aligned}
		& {\bf Ph}_{1}:\;\vec \zeta>0,\quad {\bf Ph}_{2}:\; s_1\vec \zeta>0,\quad {\bf Ph}_{3}:\; s_2\vec \zeta>0,\quad {\bf Ph}_{4}:\; s_3\vec \zeta>0,\quad {\bf Ph}_{5}:\; s_2 s_1\vec \zeta>0,\quad {\bf Ph}_{6}:\; s_3 s_1\vec \zeta>0\,,\\
		& {\bf Ph}_{7}:\; s_1 s_2\vec \zeta>0,\quad {\bf Ph}_{8}:\; s_3 s_2\vec \zeta>0,\quad {\bf Ph}_{9}:\; s_2 s_3\vec \zeta>0,\quad {\bf Ph}_{10}:\; s_1 s_2 s_1\vec \zeta>0,\quad {\bf Ph}_{11}:\; s_3 s_2 s_1\vec \zeta>0\,,\\
		& {\bf Ph}_{12}:\; s_2 s_3 s_1\vec \zeta>0,\quad {\bf Ph}_{13}:\; s_3 s_1 s_2\vec \zeta>0,\quad {\bf Ph}_{14}:\; s_2 s_3 s_2\vec \zeta>0,\quad {\bf Ph}_{15}:\; s_1 s_2 s_3\vec \zeta>0\,,\\
		& {\bf Ph}_{16}:\; s_3 s_1 s_2 s_1\vec \zeta>0,\quad {\bf Ph}_{17}:\; s_2 s_3 s_2 s_1\vec \zeta>0,\quad {\bf Ph}_{18}:\; s_1 s_2 s_3 s_1\vec \zeta>0,\quad {\bf Ph}_{19}:\; s_2 s_3 s_1 s_2\vec \zeta>0\,,\\
		& {\bf Ph}_{20}:\; s_1 s_2 s_3 s_2\vec \zeta>0,\quad {\bf Ph}_{21}:\; s_2 s_3 s_1 s_2 s_1\vec \zeta>0,\quad {\bf Ph}_{22}:\; s_1 s_2 s_3 s_2 s_1\vec \zeta>0,\quad {\bf Ph}_{23}:\; s_1 s_2 s_3 s_1 s_2\vec \zeta>0\,,\\
		& {\bf Ph}_{24}:\; s_1 s_2 s_3 s_1 s_2 s_1\vec \zeta>0\,.
	\end{aligned}
\end{equation}

\subsubsection{Irrep [2,2]}

This theory is described by the following quiver:
\begin{equation}\label{sl(4), [2,2], quiv}
	\fQ_{3,2,2}=\left\{\begin{array}{c}
		\begin{tikzpicture}
			%%%%%%%%%%%%%%%%%%%%%%%%%%%%%%%%%%%%%%%%%%%%%%%%%%%%%%%%%%%%%
			\begin{scope}[shift = {(4,0)}]
				\draw[postaction=decorate, decoration={markings, mark= at position 0.7 with {\arrow{stealth}}}] (0,-1.2) to[out=100,in=260] node[pos=0.3, left] {$\scriptstyle R$} (0,0);
				\draw[postaction=decorate, decoration={markings, mark= at position 0.7 with {\arrow{stealth}}}] (0,0) to[out=280,in=80] node[pos=0.7, right] {$\scriptstyle S$} (0,-1.2);
				%%%%%%%%%%%%%%%%%%%%%%%%%%%%%%%%%%%%%%%%%%%%%%%%%%%%5
				\draw[postaction=decorate, decoration={markings, mark= at position 0.7 with {\arrow{stealth}}}] (0,0) to[out=20,in=160] node[pos=0.5,above] {$\scriptstyle A_2$} (1.5,0);
				\draw[postaction=decorate, decoration={markings, mark= at position 0.7 with {\arrow{stealth}}}] (1.5,0) to[out=200,in=340] node[pos=0.5,below] {$\scriptstyle B_2$} (0,0);
				\begin{scope}[shift={(-1.5,0)}]
					\draw[postaction=decorate, decoration={markings, mark= at position 0.7 with {\arrow{stealth}}}] (0,0) to[out=20,in=160] node[pos=0.5,above] {$\scriptstyle A_{1}$} (1.5,0);
					\draw[postaction=decorate, decoration={markings, mark= at position 0.7 with {\arrow{stealth}}}] (1.5,0) to[out=200,in=340] node[pos=0.5,below] {$\scriptstyle B_{1}$} (0,0);
				\end{scope}
				\draw[postaction=decorate, decoration={markings, mark= at position 0.8 with {\arrow{stealth}}}] (0,0) to[out=60,in=0] (0,0.6) to[out=180,in=120] (0,0);
				\node[above] at (0,0.6) {$\scriptstyle C_2$};
				\begin{scope}[shift={(1.5,0)}]
					\draw[postaction=decorate, decoration={markings, mark= at position 0.8 with {\arrow{stealth}}}] (0,0) to[out=60,in=0] (0,0.6) to[out=180,in=120] (0,0);
					\node[above] at (0,0.6) {$\scriptstyle C_{3}$};
				\end{scope}
				\begin{scope}[shift={(-1.5,0)}]
					\draw[postaction=decorate, decoration={markings, mark= at position 0.8 with {\arrow{stealth}}}] (0,0) to[out=60,in=0] (0,0.6) to[out=180,in=120] (0,0);
					\node[above] at (0,0.6) {$\scriptstyle C_{1}$};
				\end{scope}
				\draw[fill=black!40!red] (-1.5,0) circle (0.08);
				\draw[fill=black!40!green] (0,0) circle (0.08);
				\draw[fill=\myblue] (1.5,0) circle (0.08);
				%%%%%%%%%%%%%%%%%%%%%%%%%%%%%%%%
				\begin{scope}[shift={(0,-1.2)}]
					\draw[fill=gray] (-0.08,-0.08) -- (-0.08,0.08) -- (0.08,0.08) -- (0.08,-0.08) -- cycle;
				\end{scope}
			\end{scope}
			%%%%%%%%%%%%%%%%%%%%%%%%%%%%%%%%%%%%%%%%%%%%%%%%%%%%%%%%%%%%%%%
		\end{tikzpicture}
	\end{array},\;W=\Tr\left[A_1C_1B_1+A_2C_2B_2-B_{1}C_2A_{1}-B_{2}C_{3}A_{2}+{\color{burgundy}S C_2^{2}R}\right]\right\}\,.
\end{equation}

The generating function for solution numbers reads:
\begin{equation}\label{sl(4), [2,2], DT}
	\begin{aligned}
	&{\bf DT}_{3,2,2}=1+q_{\bf\color{black!40!green} 2}+q_{\bf\color{black!40!red} 1} q_{\bf\color{black!40!green} 2}+q_{\bf\color{black!40!green} 2}^2+q_{\bf\color{black!40!red} 1} q_{\bf\color{black!40!green} 2}^2+q_{\bf\color{black!40!red} 1}^2 q_{\bf\color{black!40!green} 2}^2+q_{\bf\color{black!40!green} 2} q_{\bf\color{\myblue} 3}+q_{\bf\color{black!40!red} 1} q_{\bf\color{black!40!green} 2} q_{\bf\color{\myblue} 3}+q_{\bf\color{black!40!green} 2}^2 q_{\bf\color{\myblue} 3}+\underline{2 q_{\bf\color{black!40!red} 1} q_{\bf\color{black!40!green} 2}^2 q_{\bf\color{\myblue} 3}}+q_{\bf\color{black!40!red} 1}^2 q_{\bf\color{black!40!green} 2}^2 q_{\bf\color{\myblue} 3}+q_{\bf\color{black!40!red} 1} q_{\bf\color{black!40!green} 2}^3 q_{\bf\color{\myblue} 3}+\\
	&+q_{\bf\color{black!40!red} 1}^2 q_{\bf\color{black!40!green} 2}^3 q_{\bf\color{\myblue} 3}+q_{\bf\color{black!40!green} 2}^2 q_{\bf\color{\myblue} 3}^2+q_{\bf\color{black!40!red} 1} q_{\bf\color{black!40!green} 2}^2 q_{\bf\color{\myblue} 3}^2+q_{\bf\color{black!40!red} 1}^2 q_{\bf\color{black!40!green} 2}^2 q_{\bf\color{\myblue} 3}^2+q_{\bf\color{black!40!red} 1} q_{\bf\color{black!40!green} 2}^3 q_{\bf\color{\myblue} 3}^2+q_{\bf\color{black!40!red} 1}^2 q_{\bf\color{black!40!green} 2}^3 q_{\bf\color{\myblue} 3}^2+q_{\bf\color{black!40!red} 1}^2 q_{\bf\color{black!40!green} 2}^4 q_{\bf\color{\myblue} 3}^2\,.
	\end{aligned}
\end{equation}

We would like to construct explicitly all the solutions in the cyclic chamber and follow the action of few mutations on them.
Here we present results of this computation with the help of atomic structure plots according to Sec.~\ref{sec:atomic}.
To lighten up notations we omit edge labels.
Those labels could be re-constructed uniquely by comparing endpoint atom colors and arrow direction with those of \eqref{sl(4), [2,2], quiv}.
Also to warn the reader we mark specifically edges that are not co-directed with naturally chosen edge directions $\vec A$, $\vec B$, $\vec C$ as in Fig.~\ref{fig:boxedYoung} by the violet color.
For all the atomic structure plots it is checked numerically that respective ans\"atze flow to valid $G_{\IR}(\fQ)$-elements in the steepest descend method.
\begin{subequations}
%1
\begin{equation}
	\#1:\;\varnothing\;\to\;{\bf Ph}_1:\,
\begin{array}{c}
		\begin{tikzpicture}[scale=0.7]
				\draw[fill=gray] (-0.1,-0.1) -- (-0.1,0.1) -- (0.1,0.1) -- (0.1,-0.1) -- cycle;
		\end{tikzpicture}
	\end{array}%
\,.
\end{equation}

\end{subequations}

%%%%%%%%%%%%%%%%%%%%%%%%%%%%%%%%%%%%%%%%%%%%%%%%%%%%%%%%%%%%%%%%%%%%%%%%%%%
%%%%%%%%%%%%%%%%%%%%%%%%%%%%%%%%%%%%%%%%%%%%%%%%%%%%%%%%%%%%%%%%%%%%%%%%%%%
%%%%%%%%%%%%%%%%%%%%%%%%%%%%%%%%%%%%%%%%%%%%%%%%%%%%%%%%%%%%%%%%%%%%%%%%%%%

\section{Moduli space examples beyond \texorpdfstring{$\fs\fl_{n+1}$}{sl(n+1)}}\label{app:examples2}

\subsection{Quiver \texorpdfstring{$\fs\fo_5 \simeq \fs\fp_4$}{so5}}\label{app:B2}

Algebras $\fs\fo_{5} \simeq \fs\fp_4$ have corresponding Dynkin diagram $\myB_2$.
We choose here conventions that the short root is on the right, so the Cartan matrix reads:
\begin{equation}\label{B2 Cartan}
	\CA = \begin{pmatrix}
		2 & -2\\
		-1 & 2
	\end{pmatrix} \,.
\end{equation}
For simplicity, we will denote irreducible representations by  Dynkin labels of the highest weight.

The aim of this section is to argue that the phase space of $\fs\fo_5$ quiver could be depicted as in Fig.~\ref{fig:B2_phsp}.
\begin{figure}[ht!]
	\centering
	\begin{tikzpicture}[scale=1.5]
		\draw[thick, -stealth] (-1.5,0) -- (1.5,0);
		\draw[thick, -stealth] (0,-1.5) -- (0,1.5);
		\node[right] at (1.5,0) {$\scriptstyle \zeta_1$};
		\node[above] at (0,1.5) {$\scriptstyle \zeta_2$};
		\draw[draw=white, fill=Xpurple!40!black] (0,0) -- (1.2,0) -- (1.2,1.2) -- (0,1.2) -- cycle; % Phase I
		\draw[draw=white, fill=Xmagenta!40!black] (0,0) -- (0,1.2) -- (-1.2,1.2) -- (0, 0) -- cycle; % Phase II
		\draw[draw=white, fill=Xmagenta!60!black] (0,0) -- (-1.2,1.2) -- (-1.2, 0.6) -- (0, 0) -- cycle; % Phase III
		\draw[draw=white, fill=Xmagenta!80!black] (0,0) -- (-1.2,0.6) -- (-1.2,0) -- cycle; % Phase IV
		\draw[draw=white, fill=Xpurple!80!black] (0,0) -- (-1.2,0) -- (-1.2,-1.2) -- (0,-1.2) -- cycle; % Phase V
		\draw[draw=white, fill=Xblue!80!black] (0,0) -- (0,-1.2) -- (1.2,-1.2) -- (0, 0) -- cycle; % Phase VI
		\draw[draw=white, fill=Xblue!60!black] (0,0) -- (1.2,-1.2) -- (1.2,-0.6) -- (0, 0) -- cycle; % Phase VII
		\draw[draw=white, fill=Xblue!40!black] (0,0) -- (1.2,-0.6) -- (1.2,0) -- (0, 0) -- cycle; % Phase VIII
		\node[white] at (0.9,0.9) {\scriptsize I};
		\node[white] at (-0.5,0.9) {\scriptsize II};
		\node[white] at (-0.85, 0.6) {\scriptsize III};
		\node[white] at (-0.9, 0.2) {\scriptsize IV};
		\node[white] at (-0.9,-0.9) {\scriptsize V};
		\node[white] at (0.5,-0.9) {\scriptsize VI};
		\node[white] at (0.85, -0.6) {\scriptsize VII};
		\node[white] at (0.9, -0.2) {\scriptsize VIII};
	\end{tikzpicture}
	\caption{Moduli space of a $\fs\fo_5$ quiver.}\label{fig:B2_phsp}
\end{figure}

The Weyl reflections acting on the stability parameters have a form of simple matrices:
\begin{equation}
	s_{1} = \begin{pmatrix}
		-1 & 0 \\
		1 & 1
	\end{pmatrix}\,,\quad
	s_{2} = \begin{pmatrix}
		1 & 2\\
		0 & -1
	\end{pmatrix}\,, \qquad s_{1}^{2} = s_{2}^{2} = \bbone\,,\quad s_{1}s_{2}s_{1}s_{2} = s_{2}s_{1}s_{2}s_{1}\,.
\end{equation}
We define the phases in the following way:
\begin{equation}
	\begin{aligned}
		&\mbox{\bf I}:\;\vec{\zeta}>0,\quad \mbox{\bf II}:\;s_1\vec{\zeta}>0,\quad \mbox{\bf III}:\;s_2s_1\vec{\zeta}>0,\quad \mbox{\bf IV}:\;s_1s_2s_1\vec{\zeta}>0\,,\\
		&\mbox{\bf V}:\;(s_1s_2s_{1}s_{2} = s_{2}s_{1}s_{2}s_{1})\vec{\zeta}>0,\quad\mbox{\bf VI}:\;s_2\vec{\zeta}>0,\quad \mbox{\bf VII}:\;s_{1}s_{2}\vec{\zeta} > 0,\quad\mbox{\bf VIII}:\; s_{2}\vec{\zeta} > 0\,.
	\end{aligned}
\end{equation}

\subsubsection{Irrep $(2, 0)$}

The corresponding quiver reads:
\begin{equation}\label{so(5), (2, 0)}
	\fQ_{\fs\fo_5, (2, 0)} = \left\{\begin{array}{c}
		\begin{tikzpicture}
			%%%%%%%%%%%%%%%%%%%%%%%%%%%%%%%%
			\begin{scope}[rotate=-90]
				\draw[postaction=decorate, decoration={markings, mark= at position 0.7 with {\arrow{stealth}}}] (0,-1.2) to[out=100,in=260] node[pos=0.3, above] {$\scriptstyle R_{1}$} (0,0);
				\draw[postaction=decorate, decoration={markings, mark= at position 0.7 with {\arrow{stealth}}}] (0,0) to[out=280,in=80] node[pos=0.7, below] {$\scriptstyle S_{1}$} (0,-1.2);
				\begin{scope}[shift={(0,-1.2)}]
					\draw[fill=gray] (-0.08,-0.08) -- (-0.08,0.08) -- (0.08,0.08) -- (0.08,-0.08) -- cycle;
				\end{scope}
			\end{scope}
			%%%%%%%%%%%%%%%%%%%%%%%%%%%%%%%%%%%%%%%%%%%%%%%%%%%%5
			\draw[postaction=decorate, decoration={markings, mark= at position 0.7 with {\arrow{stealth}}}] (0,0) to[out=20,in=160] node[pos=0.5,above] {$\scriptstyle A_1$} (1.5,0);
			\draw[postaction=decorate, decoration={markings, mark= at position 0.7 with {\arrow{stealth}}}] (1.5,0) to[out=200,in=340] node[pos=0.5,below] {$\scriptstyle B_1$} (0,0);
			\draw[postaction=decorate, decoration={markings, mark= at position 0.8 with {\arrow{stealth}}}] (0,0) to[out=60,in=0] (0,0.6) to[out=180,in=120] (0,0);
			\node[above] at (0,0.6) {$\scriptstyle C_1$};
			\begin{scope}[shift={(1.5,0)}]
				\draw[postaction=decorate, decoration={markings, mark= at position 0.8 with {\arrow{stealth}}}] (0,0) to[out=60,in=0] (0,0.6) to[out=180,in=120] (0,0);
				\node[above] at (0,0.6) {$\scriptstyle C_{2}$};
			\end{scope}
			\draw[fill=\myred] (0,0) circle (0.08);
			\draw[fill=\mygreen] (1.5,0) circle (0.08);
			%%%%%%%%%%%%%%%%%%%%%%%%%%%%%%%%%%%%%%%%%%%%%%%%%%%%%%%%%%%%%%%
			\node[below] at (0,-0.08) {$\scriptstyle \zeta_1$};
			\node[below] at (1.5,-0.08) {$\scriptstyle \zeta_2$};
		\end{tikzpicture}
	\end{array},\quad W=\Tr\left[A_1C_1B_1-B_{1}C_{2}^{2}A_{1}+{\color{burgundy}S_{1} C_1^{2}R_{1}}\right]\right\}\,.
\end{equation}

We will argue that this representation resembles the 14-dimensional irreducible representation of $\fs\fo_5$:
\begin{equation}
	\dim_{\fs\fo_5}|(2, 0)| = 14\,.
\end{equation}

The DT generating function reads in this case:
\begin{equation}
	{\bf DT}_{\fQ_{\fs\fo_{5}, (2, 0)}} = 1 + q_{1} + q_{1}^{2} + q_{1}q_{2} + q_{1}^{2}q_{2} + q_{1}q_{2}^{2} + \underline{2q_{1}^{2}q_{2}^{2}} + q_{1}^{3}q_{2}^{2} + q_{1}^{2}q_{2}^{3} + q_{1}^{3}q_{2}^{3} + q_{1}^{2}q_{2}^{4} + q_{1}^{3}q_{2}^{4} + q_{1}^{4}q_{2}^{4}\,.
\end{equation}

The D- and F-term conditions for \eqref{so(5), (2, 0)} read:
\begin{equation}\label{B2 F-relations}
	\begin{aligned}
		&[C_{1}, C_{1}^{\dagger}] + R_{1}R_{1}^{\dagger} + B_{1}B_{1}^{\dagger} - A_{1}^{\dagger}A_{1} - S_{1}^{\dagger}S_{1} =\zeta_{1} \bbone_{d_{1}\times d_{1}}\,,\\
		&[C_{2}, C_{2}^{\dagger}] + A_{1}A_{1}^{\dagger} - B_{1}^{\dagger}B_{1} = \zeta_{2} \bbone_{d_{2}\times d_{2}},\;B_{1}A_{1} + C_{1}R_{1}S_{1} + R_{1}S_{1}C_{1} = 0\,,\\
		&A_{1}B_{1}C_{2} + C_{2}A_{1}B_{1} = 0,\; C_{1}B_{1} - B_{1}C_{2}^{2} = 0,\; A_{1}C_{1} - C_{2}^{2}A_{1} = 0,\; C_{1}^{2}R_{1} = 0,\; S_{1}C_{1}^{2} = 0\,.
	\end{aligned}
\end{equation}
One assigns to the fields the following flavor charges:
\begin{equation}\label{B2 fcharges}
	\begin{array}{c|c|c|c|c|c|c}
		\mbox{Field} & A_1 & B_1 & C_1 & C_{2} & R_{1} & S_{1}\\
		\hline
		\mbox{Equiv. ch.} & \epsilon_1 & \epsilon_2 & -\epsilon_1-\epsilon_2 & -\frac{1}{2}(\epsilon_{1} + \epsilon_{2}) & 0 & 2(\epsilon_1+\epsilon_2)
	\end{array}\,.
\end{equation}

As before, we expect $S_{1} = 0$ in the cyclic phase. 
Relations
\begin{equation}
	C_{1}B_{1} = B_{1}C_{2}^{2}\,, \quad A_{1}C_{1} = C_{2}^{2}A_{1}\,,
\end{equation} 
highlight another difference from the case of $Y(\fs\fl_{3})$. 
Fields $C_{1}$ and $C_{2}$ are essentially non-equivalent and play slightly different roles in a construction of representations. 
For example, $C_{1}R_{1} = 0$ does not imply $C_{2}A_{1}R_{1} = 0$.

Some analogs of molten crtystal structures were presented in \cite{Bao:2025hfu} in similar cases of non-simply laced quivers.

{\bf Phase I} $(\zeta_1>0,\zeta_2>0)$ solutions:
\begin{equation}
	\scalebox{0.9}{$
$}
\end{equation}

In this case we would like to justify our suggestion \eqref{non-simp1} to generalize Weyl mutations of Sec.~\ref{sec:Weyl_mut} in an example.
Let us consider the following fixed point:
\begin{equation}
	v_{1}^{(\zeta_{2} > 0)} = %
\,.
\end{equation}
Based on this atomic structure plot we could set the following ansatz for quiver morphisms:
\begin{equation}
	\begin{aligned}
		&R_{1} = %
\,.
	\end{aligned}
\end{equation}
Substituting this ansatz into equations \eqref{B2 F-relations} we fix non-zero vevs:
\begin{equation}
	x_{1} = \sqrt{2(\zeta_{1} + \zeta_{2})}\,, \quad x_{2} = \sqrt{\zeta_{1}}\,,\quad x_{3} = \sqrt{2\zeta_{2}}\,,\quad x_{4} = \sqrt{\zeta_{2}}\,.
\end{equation}

To construct a dual fixed point $\check v_{1}^{(\zeta_{2} < 0)}$ in  phase VIII consider respective maps $\alpha$ and $\beta$. 
We construct the map $\beta$ according to \eqref{non-simp1}:
\begin{equation}
	\beta\Bigl(v_{1}^{(\zeta_{2} > 0)}\Bigr) = %
\,.
\end{equation}
Then the null space of this map suggest the following alleged map $\alpha$:
\begin{equation}
	\alpha\Bigl(\check v_{1}^{(\zeta_{2} < 0)}\Bigr) = %
\,.
\end{equation}
Chosen in this way these maps form a short exact sequence:
\begin{equation}
	\begin{array}{c}
		\begin{tikzpicture}
			\node (A) at (-3.5,0) {$0$};
			\node (B) at (-2,0) {$\IC^{2}$};
			\node (C) at (0,0) {$\smash{\underset{V_{1}^{\oplus 2}}{\underbrace{\IC^{2}\oplus \IC^{2}}}}$};
			\node (D) at (2,0) {$\IC^{2}$};
			\node (E) at (3.5,0) {$0$};
			\path (A) edge[->] (B) (B) edge[->] node[above] {$\scriptstyle \alpha$} (C) (C) edge[->] node[above] {$\scriptstyle \beta$} (D) (D) edge[->] (E);
		\end{tikzpicture}
	\end{array}\,.
\end{equation}
Map $\alpha$ suggests that meson field $\check{B}_{1}\check{C}_{2}$ in the dual phase obtains a non-zero vev.
So we guess that the dual fixed point correspond to the following atomic structure plot:
\begin{equation}
	\check v_{1}^{(\zeta_{2} < 0)} = %
\,.
\end{equation}
And indeed we check that this plot describes the following explicit expectation values for quiver morphisms in the dual phase:
\begin{equation}
	\begin{aligned}
		\check{R}_{1} = %
\,.
	\end{aligned}
\end{equation}

\subsubsection{Irrep $(0, 2)$}

The corresponding quiver reads:
\begin{equation}\label{so(5), (0, 2)}
	\fQ_{\fs\fo_5, (2, 0)} = \left\{%
,\quad W=\Tr\left[A_1C_1B_1-B_{1}C_{2}^{2}A_{1}+{\color{green!40!black}S_{2} C_2^{2}R_{2}}\right]\right\}\,.
\end{equation}

Corresponding fixed points are described by the following equations:
\begin{equation}\label{so(5) (0,2) ADHM}
	\begin{aligned}
		&[C_{2}, C_{2}^{\dagger}] + R_{2}R_{2}^{\dagger} + A_{1}A_{1}^{\dagger} - B_{1}^{\dagger}B_{1} - S_{2}^{\dagger}S_{2}  = \zeta_{2} \bbone_{d_{2}\times d_{2}}\,,\\
		&[C_{1}, C_{1}^{\dagger}] + B_{1}B_{1}^{\dagger} - A_{1}^{\dagger}A_{1} =\zeta_{1} \bbone_{d_{1}\times d_{1}}\,, \quad B_{1}A_{1} = 0\,,\\
		&A_{1}B_{1}C_{2} + C_{2}A_{1}B_{1} + C_{2}R_{2}S_{2} + R_{2}S_{2}C_{2} = 0\,, \quad C_{2}^{2}R_{2} = 0\,, \quad S_{1}C_{2}^{2} = 0\,.
	\end{aligned}
\end{equation}

Again, we expect the quiver \eqref{so(5), (0, 2)} resemble 10-dimensional representation of the algebra $\fs\fo_5$ and the corresponding Yangian:
\begin{equation}
	\dim_{\fs\fo_5}|(0, 2)| = 10\,.
\end{equation}

The generating function takes the form:
\begin{equation}
	{\bf DT}_{\fQ_{\fs\fo_5, (0, 2)}}= 1 + q_{2} + q_{1}q_{2} + q_{2}^{2} + \underline{2q_{1}q_{2}^{2}} + q_{1}^{2}q_{2}^{2} + q_{1}q_{2}^{3} + q_{1}^{2}q_{2}^{3} + q_{1}^{2}q_{2}^{4}\,.
\end{equation}

One of the differences with the previous case that the path $A_{1}B_{1}R_{2}$ does not acquire zero vev, therefore, changes the number of states accordingly.

{\bf Phase I} $(\zeta_1>0,\zeta_2>0)$ solutions:
\begin{equation}\label{so(5) (0,2) ph I}
	\scalebox{0.9}{$\begin{array}{l}
		1\to\!\!\!\begin{array}{c}
			\begin{tikzpicture}[scale=0.8]
				\begin{scope}[shift={(0,0)}]
					\draw[fill=gray] (-0.1,-0.1) -- (-0.1,0.1) -- (0.1,0.1) -- (0.1,-0.1) -- cycle;
				\end{scope}
			\end{tikzpicture}
		\end{array}\!\!,\;
		q_2\to\!\!\!\begin{array}{c}
			\begin{tikzpicture}[scale=0.8]
				\draw[postaction=decorate, decoration={markings, mark= at position 0.7 with {\arrow{stealth}}}] (1,0) -- (0,0) node[pos=0.5,above] {$\scriptstyle R_{2}$};
				\draw[fill=\mygreen] (0,0) circle (0.1);
				\begin{scope}[shift={(1,0)}]
					\draw[fill=gray] (-0.1,-0.1) -- (-0.1,0.1) -- (0.1,0.1) -- (0.1,-0.1) -- cycle;
				\end{scope}
			\end{tikzpicture}
		\end{array}\!\!,\;
		q_1q_2\to\!\!\!\begin{array}{c}
			\begin{tikzpicture}[scale=0.8]
				\draw[postaction=decorate, decoration={markings, mark= at position 0.7 with {\arrow{stealth}}}] (1,-1) -- (0,-1) node[pos=0.5,below] {$\scriptstyle R_{2}$};
				\draw[postaction=decorate, decoration={markings, mark= at position 0.7 with {\arrow{stealth}}}] (0,-1) -- (0,0) node[pos=0.5,left] {$\scriptstyle B_1$};
				\draw[fill=\myred] (0,0) circle (0.1);
				\draw[fill=\mygreen] (0,-1) circle (0.1);
				\begin{scope}[shift={(1,-1)}]
					\draw[fill=gray] (-0.1,-0.1) -- (-0.1,0.1) -- (0.1,0.1) -- (0.1,-0.1) -- cycle;
				\end{scope}
			\end{tikzpicture}
		\end{array}\!\!,\;
		q_1^2\to\!\!\!\begin{array}{c}
			\begin{tikzpicture}[scale=0.8]
				\draw[postaction=decorate, decoration={markings, mark= at position 0.7 with {\arrow{stealth}}}] (1,0) -- (0,0) node[pos=0.5,above] {$\scriptstyle C_{2}$};
				\draw[postaction=decorate, decoration={markings, mark= at position 0.7 with {\arrow{stealth}}}] (2,0) -- (1,0) node[pos=0.5,below] {$\scriptstyle R_2$};
				\draw[fill=\mygreen] (0,0) circle (0.1) (1,0) circle (0.1);
				\begin{scope}[shift={(2,0)}]
					\draw[fill=gray] (-0.1,-0.1) -- (-0.1,0.1) -- (0.1,0.1) -- (0.1,-0.1) -- cycle;
				\end{scope}
			\end{tikzpicture}
		\end{array}\!\!,\;
		\underline{q_1q_2^{2}}\to\!\!\!\begin{array}{c}
			\begin{tikzpicture}[scale=0.8]
				\draw[postaction=decorate, decoration={markings, mark= at position 0.7 with {\arrow{stealth}}}] (1,0) -- (0,0) node[pos=0.5,below] {$\scriptstyle C_{2}$};
				\draw[postaction=decorate, decoration={markings, mark= at position 0.7 with {\arrow{stealth}}}] (2,0) -- (1,0) node[pos=0.5,below] {$\scriptstyle R_{2}$};
				\draw[postaction=decorate, decoration={markings, mark= at position 0.7 with {\arrow{stealth}}}] (1,0) -- (1,1) node[pos=0.5,left] {$\scriptstyle B_1$};
				\draw[fill=\mygreen] (1,0) circle (0.1) (0,0) circle (0.1);
				\draw[fill=\myred] (1,1) circle (0.1);
				\begin{scope}[shift={(2,0)}]
					\draw[fill=gray] (-0.1,-0.1) -- (-0.1,0.1) -- (0.1,0.1) -- (0.1,-0.1) -- cycle;
				\end{scope}
			\end{tikzpicture}
		\end{array}\!\!,\;
		\underline{q_1q_2^2}\to\!\!\!\begin{array}{c}
			\begin{tikzpicture}[scale=0.8]
				\draw[postaction=decorate, decoration={markings, mark= at position 0.7 with {\arrow{stealth}}}] (2,0) -- (1,0) node[pos=0.5,below] {$\scriptstyle R_{2}$};
				\draw[postaction=decorate, decoration={markings, mark= at position 0.7 with {\arrow{stealth}}}] (1,0) -- (0,1) node[pos=0.5,right] {$\scriptstyle B_1$};
				\draw[postaction=decorate, decoration={markings, mark= at position 0.7 with {\arrow{stealth}}}] (0,1) -- (-1,0) node[pos=0.5,left] {$\scriptstyle A_1$};
				\draw[fill=\mygreen] (1,0) circle (0.1) (-1,0) circle (0.1);
				\draw[fill=\myred] (0,1) circle (0.1);
				\begin{scope}[shift={(2,0)}]
					\draw[fill=gray] (-0.1,-0.1) -- (-0.1,0.1) -- (0.1,0.1) -- (0.1,-0.1) -- cycle;
				\end{scope}
			\end{tikzpicture}
		\end{array}\,,\\
		q_1q_{2}^3\to\!\!\!\begin{array}{c}
			\begin{tikzpicture}[scale=0.8]
				\draw[postaction=decorate, decoration={markings, mark= at position 0.7 with {\arrow{stealth}}}] (2,0) -- (1,0) node[pos=0.5,below] {$\scriptstyle R_{2}$};
				\draw[postaction=decorate, decoration={markings, mark= at position 0.7 with {\arrow{stealth}}}] (1,0) -- (0,1) node[pos=0.5,right] {$\scriptstyle B_1$};
				\draw[postaction=decorate, decoration={markings, mark= at position 0.7 with {\arrow{stealth}}}] (0,1) -- (-1,0) node[pos=0.5,left] {$\scriptstyle A_1$};
				\draw[postaction=decorate, decoration={markings, mark= at position 0.7 with {\arrow{stealth}}}] (1,0) -- (0,0) node[pos=0.5,below] {$\scriptstyle C_2$};
				\draw[fill=\mygreen] (1,0) circle (0.1) (0, 0) circle (0.1) (-1,0) circle (0.1);
				\draw[fill=\myred] (0,1) circle (0.1);
				\begin{scope}[shift={(2,0)}]
					\draw[fill=gray] (-0.1,-0.1) -- (-0.1,0.1) -- (0.1,0.1) -- (0.1,-0.1) -- cycle;
				\end{scope}
			\end{tikzpicture}
		\end{array}\!\!,\;q_1^{2}q_2^{2}\to\!\!\!\begin{array}{c}
			\begin{tikzpicture}[scale=0.8]
				\draw[postaction=decorate, decoration={markings, mark= at position 0.7 with {\arrow{stealth}}}] (1,0) -- (0,0) node[pos=0.5,below] {$\scriptstyle C_{2}$};
				\draw[postaction=decorate, decoration={markings, mark= at position 0.7 with {\arrow{stealth}}}] (2,0) -- (1,0) node[pos=0.5,below] {$\scriptstyle R_{2}$};
				\draw[postaction=decorate, decoration={markings, mark= at position 0.7 with {\arrow{stealth}}}] (1,0) -- (1,1) node[pos=0.5,left] {$\scriptstyle B_1$};
				\draw[postaction=decorate, decoration={markings, mark= at position 0.7 with {\arrow{stealth}}}] (0,0) -- (0,1) node[pos=0.5,left] {$\scriptstyle B_1$};
				\draw[fill=\mygreen] (1,0) circle (0.1) (0,0) circle (0.1);
				\draw[fill=\myred] (1,1) circle (0.1) (0, 1) circle (0.1);
				\begin{scope}[shift={(2,0)}]
					\draw[fill=gray] (-0.1,-0.1) -- (-0.1,0.1) -- (0.1,0.1) -- (0.1,-0.1) -- cycle;
				\end{scope}
			\end{tikzpicture}
		\end{array}\!\!,\;q_1^2q_2^3\to\!\!\!\begin{array}{c}
			\begin{tikzpicture}[scale=0.8]
				\draw[postaction=decorate, decoration={markings,  mark= at position 0.7 with {\arrow{stealth}}}] (2,0) -- (1,0) node[pos=0.5,below] {$\scriptstyle R_{2}$};
				\draw[postaction=decorate, decoration={markings,  mark= at position 0.7 with {\arrow{stealth}}}] (1,0) -- (0,1) node[pos=0.5,right] {$\scriptstyle B_1$};
				\draw[postaction=decorate, decoration={markings,  mark= at position 0.7 with {\arrow{stealth}}}] (0,0) -- (-1,1) node[pos=0.5,right] {$\scriptstyle B_1$};
				\draw[postaction=decorate, decoration={markings,  mark= at position 0.7 with {\arrow{stealth}}}] (0,1) -- (-1,0) node[pos=0.5,left] {$\scriptstyle A_1$};
				\draw[postaction=decorate, decoration={markings,  mark= at position 0.7 with {\arrow{stealth}}}] (1,0) -- (0,0) node[pos=0.5,below] {$\scriptstyle C_2$};
				\draw[fill=\mygreen] (1,0) circle (0.1) (0, 0)  circle (0.1) (-1,0) circle (0.1);
				\draw[fill=\myred] (0,1) circle (0.1) (-1, 1) circle (0.1);
				\begin{scope}[shift={(2,0)}]
					\draw[fill=gray] (-0.1,-0.1) -- (-0.1,0.1) --  (0.1,0.1) -- (0.1,-0.1) -- cycle;
				\end{scope}
			\end{tikzpicture}
		\end{array}\!\!,\;q_1^2q_2^4\to\!\!\!\begin{array}{c}
			\begin{tikzpicture}[scale=0.8]
				\draw[postaction=decorate, decoration={markings,  	mark= at position 0.7 with {\arrow{stealth}}}] (2,0) -- (1,0) node[pos=0.5,below] {$\scriptstyle R_{2}$};
				\draw[postaction=decorate, decoration={markings,  	mark= at position 0.7 with {\arrow{stealth}}}] (1,0) -- (0,1) node[pos=0.5,right] {$\scriptstyle B_1$};
				\draw[postaction=decorate, decoration={markings,  	mark= at position 0.7 with {\arrow{stealth}}}] (0,0) -- (-1,1) node[pos=0.5,right] {$\scriptstyle B_1$};
				\draw[postaction=decorate, decoration={markings,  	mark= at position 0.7 with {\arrow{stealth}}}] (0,1) -- (-1,0) node[pos=0.5,left] {$\scriptstyle A_1$};
				\draw[postaction=decorate, decoration={markings,  	mark= at position 0.7 with {\arrow{stealth}}}] (-1,1) -- (-2,0) node[pos=0.5,left] {$\scriptstyle A_1$};
				\draw[postaction=decorate, decoration={markings,  	mark= at position 0.7 with {\arrow{stealth}}}] (1,0) -- (0,0) node[pos=0.5,below] {$\scriptstyle C_2$};
				\draw[postaction=decorate, decoration={markings,  	mark= at position 0.7 with {\arrow{stealth}}}] (-1,0) -- (-2,0) node[pos=0.5,below] {$\scriptstyle C_2$};
				\draw[fill=\mygreen] (1,0) circle (0.1) (0, 0)  	circle (0.1) (-1,0) circle (0.1) (-2, 0) circle (0.1);
				\draw[fill=\myred] (0,1) circle (0.1) (-1, 1) circle 	(0.1);
				\begin{scope}[shift={(2,0)}]
					\draw[fill=gray] (-0.1,-0.1) -- (-0.1,0.1) --  	(0.1,0.1) -- (0.1,-0.1) -- cycle;
				\end{scope}
			\end{tikzpicture}
		\end{array}\!.
	\end{array}$}
\end{equation}

Her we should draw reader's attention to certain peculiarities appearing novel in this case in comparison to the $\myA$ Dynkin series.
Let us concentrate on a fixed point corresponding to $q_1^2q_2^4$ in \eqref{so(5) (0,2) ph I}.
This atomic structure plot sets the following ansatz for the fields:
\begin{equation}
	R_2=\left(
	\begin{array}{c}
		x_1 \\
		0 \\
		0 \\
		0 \\
	\end{array}
	\right),\quad C_2=\left(
	\begin{array}{cccc}
		0 & 0 & 0 & 0 \\
		x_2 & 0 & 0 & 0 \\
		0 & 0 & 0 & 0 \\
		0 & 0 & x_3 & 0 \\
	\end{array}
	\right),\quad A_1=\left(
	\begin{array}{cc}
		0 & 0 \\
		0 & 0 \\
		x_4 & 0 \\
		0 & x_5 \\
	\end{array}
	\right),\quad B_1=\left(
	\begin{array}{cccc}
		x_6 & 0 & 0 & 0 \\
		0 & x_7 & 0 & 0 \\
	\end{array}
	\right)\,,
\end{equation}
the rest of fields is zero.
Equations in the last line of \eqref{so(5) (0,2) ADHM} lead to the following constraints on $x_i$:
\begin{equation}\label{cubic}
	x_3 x_4 x_6+x_2 x_5 x_7=0\,.
\end{equation}
First of all this constraint is cubic that did not happen in the case of $\myA$ series when the atomic structure plot contained just a single loop.
Also we note that a solution when all $x_a>0$ is not available in this case anymore, what happened in the simply laced case only outside the cyclic chamber when atomic structure plots stop being molten crystals.
And here this effect occurs already in the cyclic chamber.
In comparison in \cite{Bao:2025hfu} these non-simply laced quivers were referred to as \emph{non-toric}.

Since \eqref{cubic} is cubic, expressions for vevs of $x_a$ are rather bulky in terms of radicals, so here we present a solution in a special point $\zeta_1=\zeta_2=\zeta>0$:
\begin{equation}
	x_1=\sqrt{ 6 \zeta},\; x_2=\sqrt{\frac{5 \zeta }{2}},\; x_3=-\sqrt{ \frac{\zeta }{2}},\; x_4=\sqrt{ \frac{3 \zeta }{2}},\; x_5=\sqrt{ \frac{\zeta }{2}},\; x_6=\sqrt{ \frac{5 \zeta }{2}},\; x_7=\sqrt{ \frac{3 \zeta }{2}}\,.
\end{equation}

\subsection{Quiver \texorpdfstring{$\fg_2$}{g2}}\label{app:G2}

Further we would like to consider an example of a $\fg_2$ quiver. 
Here we assume that the short root is on the right that corresponds to the following Cartan matrix:
\begin{equation}\label{G2 Cartan}
	\CA = \begin{pmatrix}
		2 & -3\\
		-1 & 2
	\end{pmatrix}\,.
\end{equation}

In this case the moduli space is divided into 12 phases depicted in  Fig.~\ref{fig:G2_phsp}.
\begin{figure}[ht!]
	\centering
	\begin{tikzpicture}[scale=2]
		\draw[thick, -stealth] (-1.5,0) -- (1.5,0);
		\draw[thick, -stealth] (0,-1.5) -- (0,1.5);
		\node[right] at (1.5,0) {$\scriptstyle \zeta_1$};
		\node[above] at (0,1.5) {$\scriptstyle \zeta_2$};
		\draw[draw=white, fill=Xpurple!40!black] (0,0) -- (1.2,0) -- (1.2,1.2) -- (0,1.2) -- cycle; % Phase I
		\draw[draw=white, fill=Xmagenta!40!black] (0,0) -- (0,1.2) -- (-1.2,1.2) -- (0, 0) -- cycle; % Phase II
		\draw[draw=white, fill=Xmagenta!50!black] (0,0) -- (-1.2,1.2) -- (-1.2, 0.8) -- (0, 0) -- cycle; % Phase III
		\draw[draw=white, fill=Xmagenta!60!black] (0,0) -- (-1.2,0.8) -- (-1.2, 0.6) -- (0, 0) -- cycle; % Phase IV
		\draw[draw=white, fill=Xmagenta!70!black] (0,0) -- (-1.2,0.6) -- (-1.2, 0.4) -- cycle; % Phase V
		\draw[draw=white, fill=Xmagenta!80!black] (0,0) -- (-1.2,0.4) -- (-1.2, 0) -- (0, 0) -- cycle; % Phase VI
		\draw[draw=white, fill=Xpurple!80!black] (0,0) -- (-1.2,0) -- (-1.2,-1.2) -- (0,-1.2) -- cycle; % Phase VII
		\draw[draw=white, fill=Xblue!80!black] (0,0) -- (0,-1.2) -- (1.2,-1.2) -- (0, 0) -- cycle; % Phase VIII
		\draw[draw=white, fill=Xblue!70!black] (0,0) -- (1.2,-1.2) -- (1.2,-0.8) -- (0, 0) -- cycle; % Phase IX
		\draw[draw=white, fill=Xblue!60!black] (0,0) -- (1.2,-0.8) -- (1.2,-0.6) -- (0, 0) -- cycle; % Phase X
		\draw[draw=white, fill=Xblue!50!black] (0,0) -- (1.2,-0.6) -- (1.2,-0.4) -- (0, 0) -- cycle; % Phase XI
		\draw[draw=white, fill=Xblue!40!black] (0,0) -- (1.2,-0.4) -- (1.2,0) -- (0, 0) -- cycle; % Phase XII
		\node[white] at (0.9,0.9) { I};
		\node[white] at (-0.5,0.9) { II};
		\node[white] at (-0.9, 0.75) {\scalebox{0.8}{III}};
		\node[white] at (-1, 0.58) {\scalebox{0.5}{IV}};
		\node[white] at (-1.1, 0.43) {\scalebox{0.5}{ V}};
		\node[white] at (-0.9,0.15) {\footnotesize VI};
		\node[white] at (-0.9, -0.9) {VII};
		\node[white] at (0.5, -0.9) {VIII};
		\node[white] at (0.9, -0.75) {\scalebox{0.8}{XI}};
		\node[white] at (1, -0.58) {\scalebox{0.5}{X}};
		\node[white] at (1.1, -0.43) {\scalebox{0.5}{XI}};
		\node[white] at (0.9, -0.15) {\footnotesize XII};
	\end{tikzpicture}
	\caption{Moduli space of a $\fg_2$ quiver.}\label{fig:G2_phsp}
\end{figure}

The Weyl reflections acting on the FI parameters take the following form:
\begin{equation}
	s_{1} = \begin{pmatrix}
		-1 & 0 \\
		1 & 1
	\end{pmatrix}\,,\quad
	s_{2} = \begin{pmatrix}
		1 & 3\\
		0 & -1
	\end{pmatrix}\,, \qquad s_{1}^{2} = s_{2}^{2} = \bbone\,,\quad s_{1}s_{2}s_{1}s_{2}s_{1}s_{2} = s_{2}s_{1}s_{2}s_{1}s_{2}s_{1}\,.
\end{equation}
We describe loci of phases depicted in Fig.~\ref{fig:G2_phsp} by the following inequalities:
\begin{equation}
	\begin{aligned}
		&\mbox{\bf I}:\;\vec{\zeta}>0,\quad \mbox{\bf II}:\;s_1\vec{\zeta}>0,\quad \mbox{\bf III}:\;s_2s_1\vec{\zeta}>0,\quad \mbox{\bf IV}:\;s_1s_2s_1\vec{\zeta}>0,,\quad \mbox{\bf V}:\;(s_2s_1)^{2}\vec{\zeta}>0,\quad\mbox{\bf VI}:\;s_{1}(s_{2}s_1)^{2}\vec{\zeta}>0\,,\\
		&\mbox{\bf VII}:\;(s_{1}s_{2})^{3}\vec{\zeta}>0,\quad\mbox{\bf VIII}:\;s_2(s_{1}s_{2})^{2}\vec{\zeta}>0,\quad \mbox{\bf IX}:\;(s_{1}s_{2})^{2}\vec{\zeta} > 0,\quad\mbox{\bf X}:\; s_{2}s_{1}s_{2}\vec{\zeta} > 0,\quad\mbox{\bf XI}:\;s_1s_2\vec{\zeta}>0\,,\\
		&\mbox{\bf XII}:\;s_1\vec{\zeta}>0\,.
	\end{aligned}
\end{equation}

\subsubsection{Irrep $(0, 1)$}

Consider the following quiver with a superpotential:
\begin{equation}\label{g(2), (0, 1)}
	\fQ_{\fg_2, (0, 1)} = \left\{\begin{array}{c}
		\begin{tikzpicture}
			%%%%%%%%%%%%%%%%%%%%%%%%%%%%%%%%
			\begin{scope}[shift={(1.5, 0)}, rotate=90]
				\draw[postaction=decorate, decoration={markings, mark= at position 0.7 with {\arrow{stealth}}}] (0,-1.2) to[out=100,in=260] node[pos=0.3, above] {$\scriptstyle S_{2}$} (0,0);
				\draw[postaction=decorate, decoration={markings, mark= at position 0.7 with {\arrow{stealth}}}] (0,0) to[out=280,in=80] node[pos=0.7, below] {$\scriptstyle R_{2}$} (0,-1.2);
				\begin{scope}[shift={(0,-1.2)}]
					\draw[fill=gray] (-0.08,-0.08) -- (-0.08,0.08) -- (0.08,0.08) -- (0.08,-0.08) -- cycle;
				\end{scope}
			\end{scope}
			%%%%%%%%%%%%%%%%%%%%%%%%%%%%%%%%%%%%%%%%%%%%%%%%%%%%5
			\draw[postaction=decorate, decoration={markings, mark= at position 0.7 with {\arrow{stealth}}}] (0,0) to[out=20,in=160] node[pos=0.5,above] {$\scriptstyle A_1$} (1.5,0);
			\draw[postaction=decorate, decoration={markings, mark= at position 0.7 with {\arrow{stealth}}}] (1.5,0) to[out=200,in=340] node[pos=0.5,below] {$\scriptstyle B_1$} (0,0);
			\draw[postaction=decorate, decoration={markings, mark= at position 0.8 with {\arrow{stealth}}}] (0,0) to[out=60,in=0] (0,0.6) to[out=180,in=120] (0,0);
			\node[above] at (0,0.6) {$\scriptstyle C_1$};
			\begin{scope}[shift={(1.5,0)}]
				\draw[postaction=decorate, decoration={markings, mark= at position 0.8 with {\arrow{stealth}}}] (0,0) to[out=60,in=0] (0,0.6) to[out=180,in=120] (0,0);
				\node[above] at (0,0.6) {$\scriptstyle C_{2}$};
			\end{scope}
			\draw[fill=\myred] (0,0) circle (0.08);
			\draw[fill=\mygreen] (1.5,0) circle (0.08);
			%%%%%%%%%%%%%%%%%%%%%%%%%%%%%%%%%%%%%%%%%%%%%%%%%%%%%%%%%%%%%%%
			\node[below] at (0,-0.08) {$\scriptstyle \zeta_1$};
			\node[below] at (1.5,-0.08) {$\scriptstyle \zeta_2$};
		\end{tikzpicture}
	\end{array},\quad W=\Tr\left[A_1C_1B_1-B_{1}C_{2}^{3}A_{1}+{\color{green!40!black}S_{2} C_2R_{2}}\right]\right\}\,.
\end{equation}

According to our classification fixed points on the corresponding quiver variety should correspond to vectors of a 7-dimensional representation of Lie algebra $\fg_2$ with Dynkin labels $(0,1)$ for the highest weight.

The generating function reads in this case:
\begin{equation}
	{\bf DT}_{\fQ_{\fg_{2}, (0,1)}}= 1 + q_{2} + q_{1}q_{2} + q_{1}q_{2}^{2} + q_{1}q_{2}^{3} + q_{1}^{2}q_{2}^{3} + q_{1}^{2}q_{2}^{4}\,.
\end{equation}

The F-term and D-term constraints on the quiver variety are of the following form:
\begin{equation}\label{G2 F-relations}
	\begin{aligned}
		&[C_{2}, C_{2}^{\dagger}] + R_{2}R_{2}^{\dagger} + A_{1}A_{1}^{\dagger} - B_{1}^{\dagger}B_{1} - S_{2}^{\dagger}S_{2} = \zeta_{2} \bbone_{d_{2}\times d_{2}}\,,\\
		&[C_{1}, C_{1}^{\dagger}] + B_{1}B_{1}^{\dagger} - A_{1}^{\dagger}A_{1} =\zeta_{1} \bbone_{d_{1}\times d_{1}}\,,\quad B_{1}A_{1} = 0\,,\\
		&A_{1}B_{1}C_{2}^{2} + C_{2}A_{1}B_{1}C_{2} + C_{2}^{2}A_{1}B_{1} = 0\,,\quad C_{1}B_{1} - B_{1}C_{2}^{3} = 0\,,\\
		&A_{1}C_{1} - C_{2}^{3}A_{1} = 0\,,\quad C_{2}R_{2} = 0\,, \quad S_{2}C_{2} = 0\,.
	\end{aligned}
\end{equation}
The fixed point conditions are restored from equivariant weights of the fields:
\begin{equation}\label{G2 fcharges}
\begin{array}{c|c|c|c|c|c|c}
		\mbox{Field} & A_1 & B_1 & C_1 & C_{2} & R_{2} & S_{2}\\
		\hline
		\mbox{Equiv. ch.} & \epsilon_1 & \epsilon_2 & -\epsilon_1-\epsilon_2 & -\frac{1}{3}(\epsilon_{1} + \epsilon_{2}) & 0 & \frac{1}{3}(\epsilon_1+\epsilon_2)
	\end{array}\,.
\end{equation}

We list the fixed points in the Phases I, II, and XII in terms of atomic structure plots below.

{\bf Phase I} $(\zeta_1>0,\zeta_2>0)$ solutions:
\begin{equation}
	\begin{aligned}
		&1\to\!\!\!%
\!.
	\end{aligned}
\end{equation}

Here we would like to demonstrate again as a modified rule \eqref{non-simp2} works for the following two dual fixed points in phases I and XII:
\begin{equation}
	v_{1}^{(\zeta_{2} > 0)} = %
\,.
\end{equation}

In this case we observe the maps:
\begin{equation}
	\alpha\Bigl(\check v_{1}^{(\zeta_{2}<0)}\Bigr) = %
\,,
\end{equation}
form a short exact sequence:
\begin{equation}
	%
\,.
\end{equation}
Then we could have fixed the vevs of morphisms for $\check v_{1}^{(\zeta_{2} < 0)}$ using \eqref{non-simp2} and \eqref{G2 F-relations}:
\begin{equation}
	\begin{aligned}
		{R}_{2} = %
\,.
\end{equation}

%%%%%%%%%%%%%%%%%%%%%%%%%%%%%%%%%%%%%%%%%%%%%%%%%%%%%%%%%%%%%%%%%%%%%%%%%%%
%%%%%%%%%%%%%%%%%%%%%%%%%%%%%%%%%%%%%%%%%%%%%%%%%%%%%%%%%%%%%%%%%%%%%%%%%%%
%%%%%%%%%%%%%%%%%%%%%%%%%%%%%%%%%%%%%%%%%%%%%%%%%%%%%%%%%%%%%%%%%%%%%%%%%%%

\bibliographystyle{utphys}
\bibliography{biblio}

\end{document}